\documentclass[10pt,oneside,a4paper,onecolumn]{elsarticle}
\pdfoutput=1
\usepackage{times}
\usepackage{fullpage}

\usepackage{gensymb}
\usepackage{tikz}
\usepackage{alltt}
\usepackage{upquote}
\usepackage{float}
\usepackage{array}
\usepackage{pgfplots}
\usetikzlibrary{positioning}
\usepackage{graphicx}
\usepackage[toc,page]{appendix}
\usepackage{amsmath,mathtools}
\usetikzlibrary{calc,decorations.markings,fit,shapes,chains,arrows}
\usepackage{amsthm}
\usepackage{amsfonts}
\usepackage{enumitem}
\usepackage{subcaption}
\usepackage{caption}
\usepackage{xcolor}
\usepackage{scalerel}
\usepackage[hidelinks=true]{hyperref}
\usepackage{algorithm}
\usepackage{multicol}
\usepackage{mathtools}

\tikzset{
  block/.style={rectangle, draw, rounded corners, text centered,text width = 16em, minimum height = 2em},
  line/.style={draw, -latex'}
  }
\tikzset{
  block2/.style={text centered,text width = 22em, minimum height = 2em},
  line/.style={draw, -latex'}
  }
\tikzset{
  block3/.style={rectangle, draw, rounded corners, text centered,text width = 19em, minimum height = 1em},
  line/.style={draw, -latex'}
  }
\tikzset{
  block4/.style={rectangle, draw, rounded corners, text centered,text width = 15em, minimum height = 1em},
  line/.style={draw, -latex'}
  }
\tikzset{
  blockNS/.style={rectangle, draw, fill=black!20, rounded corners, text centered,text width = 20em, minimum height = 2em, label={center:Navier-Stokes solver}},
  line/.style={draw, -latex'}
  }
\tikzset{
  blockAC/.style={rectangle, draw, fill=black!20, rounded corners, text centered,text width = 20em, minimum height = 2em, label={center:Allen-Cahn solver}},
  line/.style={draw, -latex'}
  }
\tikzset{  
  decision/.style = {diamond, draw, minimum width=4cm, minimum height=0.2cm},
  line/.style={draw, -latex'}
  }

\newcolumntype{M}[1]{>{\centering\arraybackslash}m{#1}}
\newcolumntype{N}{@{}m{0pt}@{}}

\tikzset{>=latex}
\graphicspath{ }
\def\@author#1{\g@addto@macro\elsauthors{\normalsize%
    \def\baselinestretch{1}%
    \upshape\authorsep#1\unskip\textsuperscript{%
      \ifx\@fnmark\@empty\else\unskip\sep\@fnmark\let\sep=,\fi
      \ifx\@corref\@empty\else\unskip\sep\@corref\let\sep=,\fi
      }%
    \def\authorsep{\unskip,\space}%
    \global\let\@fnmark\@empty
    \global\let\@corref\@empty
    \global\let\sep\@empty}%
    \@eadauthor={#1}
}

\graphicspath{{Pictures/}}

\begin{document}
\begin{frontmatter}
\title{A Positivity Preserving and Conservative Variational Scheme for Phase-Field Modeling of Two-Phase Flows}
\author[nus]{Vaibhav Joshi}
\ead{vaibhav.joshi16@u.nus.edu}

\author[nus]{Rajeev K. Jaiman\corref{cor1}}
\ead{mperkj@nus.edu.sg}
\cortext[cor1]{Corresponding author}
\address[nus]{Department of Mechanical Engineering, National University of Singapore, Singapore 119077}

\begin{abstract}
\noindent We present a positivity preserving variational scheme for the phase-field modeling of incompressible two-phase flows with high density ratio and using meshes of arbitrary topology. The variational finite element technique relies on the Allen-Cahn phase-field equation for capturing the phase interface on a fixed mesh with a mass conservative and energy-stable discretization. Mass is conserved by enforcing a Lagrange multiplier which has both temporal and spatial dependence on the solution of the phase-field equation. The spatial part of the Lagrange multiplier is written as a mid-point approximation to make the scheme energy-stable. This enables us to form a conservative, energy-stable and positivity preserving scheme. The proposed variational technique reduces spurious and unphysical oscillations in the solution while maintaining second-order spatial accuracy. To model a generic two-phase free-surface flow, we couple the Allen-Cahn phase-field equation with the Navier-Stokes equations. Comparison of results between standard linear stabilized finite element method and the present variational formulation shows a remarkable reduction of oscillations in the solution while retaining the boundedness of the phase-indicator field. We perform a standalone test to verify the accuracy and stability of the Allen-Cahn two-phase solver. Standard two-phase flow benchmarks such as Laplace-Young law and sloshing tank problem are carried out to assess the convergence and accuracy of the coupled Navier-Stokes and Allen-Cahn solver. Two- and three-dimensional dam break problem are then solved to assess the scheme for the problem with topological changes of the air-water interface on unstructured meshes. Finally, we demonstrate the phase-field solver for a practical problem of wave-structure interaction in offshore engineering using general three-dimensional unstructured meshes.

\end{abstract}

\end{frontmatter}

\section{Introduction}
Multiphase flows of two immiscible fluids  are encountered in many industrial applications ranging from small scale droplet interactions \cite{Khatavkar}, phase separation and microstructural evolution in materials \cite{Fried}, bubbly and slug flows in oil and natural gas pipelines to  free-surface ocean waves in the marine environment \cite{Sorensen, Chakrabarti}. 
The effects of multiphase flows are crucial from industrial point of view since they directly affect the design and optimization of the structures subjected to the interfacial flow conditions. Therefore it is essential to understand these effects physically and numerically to have improved structural designs and safer operational conditions. Of particular interest to the present work is the robust and efficient two-phase modeling of free-surface ocean waves to predict the air-water interface and the hydrodynamic forces on submerged offshore structures.

The numerical treatment of two-phase flow involving immiscible fluids poses certain challenges owing to the physical complexity in the representation and evolution of the phase interface \cite{Multiphase_Tryg}.  The representation of the continuum interface between the two phases can be considered by either interface tracking  or interface capturing techniques.  For interface tracking, a boundary between the two fluid sub-domains is explicitly tracked, for example, front tracking \cite{Unverdi} and arbitrary Lagrangian-Eulerian \cite{donea} methods. 
While these methods can accurately locate the interface position by tracking the moving boundary or markers on the interface, they may lead to difficulties for relatively larger interface motion and topological changes.  During the topological changes, re-meshing can be computationally expensive for interface tracking methods in three-dimensions. In interface capturing, no explicit representation of the interface is considered, instead the interface is represented implicitly using a field function throughout the computational domain on an Eulerian grid, such as level-set, volume-of-fluid and phase-field approaches. Through an implicit evolution of the interface field function, topological changes such as merging and breaking of interfaces can be naturally handled by interface capturing methods.
During the evolution of interface, the discontinuous nature of physical quantities such as density, viscosity and pressure can also pose difficulties during the numerical treatment of the interface. In particular, these discontinuities in the properties can lead to unphysical and spurious oscillations in the solution which can lead to unbounded behavior. The background fixed mesh has to be sufficiently resolved to capture these discontinuities leading to high computational cost. Furthermore, surface tension or capillary forces along the interface have to be accurately modeled. Several models have been discussed in the literature to evaluate the surface tension, one of which is the continuum surface force method \cite{Brackbill}. The conservation of mass is another challenge which is crucial to obtain accurate solution of the multiphase flow.

Among the types of interface capturing methods, the level-set and volume-of-fluid (VOF) are the most popular methods. The VOF method utilizes a volume of fluid function to extract the volume fraction of each fluid within every computational cell.   While the VOF method conserves mass accurately for incompressible flows, the calculation of interface curvature and normal from volume fractions can be quite complex due to interface reconstruction  \cite{Scardovelli, Unverdi, Rider}.  In addition, the smearing of the interface by numerical diffusion causes additional difficulty. On the other hand, the  level-set approach allows to have a non-smeared interface by constructing a signed-distance function (level-set function) of a discretization point to the interface \cite{Sethian}.  The zero-level-set of the distance function provides a sharp interface description and the curvature of the interface can be approximated with high accuracy. However, the level-set method does not conserve mass. The reasons behind the inability of level-set to conserve mass are numerical dissipation introduced due to discretization and re-initialization process to keep the level-set function as a signed distance function \cite{Lanhao}. Although methods including high order discretization scheme \cite{Sussman_1}, improved re-initialization \cite{Sussman_1, Peng, Russo}, coupled particle tracking/level-set \cite{Enright_1, Enright_2, Koh} and coupled level-set/volume-of-fluid \cite{Sussman_2, Wang, Yang} have been proposed to deal with the mass conservation issue, they make the scheme more complex and computationally expensive. To improve the mass conservation property, conservative level-set method was proposed in \cite{Olsson_1, Olsson_2} where the signed distance function was replaced with a hyperbolic tangent function. However, re-initialization was necessary to maintain the width of the hyperbolic tangent profile in this case. One of the recent improvements in the level-set approach is the XFEM which utilizes the enrichment of shape functions of the elements near the interface \cite{Fries_1} to get more accurate solution. Both the level-set and the volume-of-fluid techniques utilize some kind of geometric reconstruction for the modeling of the interface. This could be quite tedious to implement and extend to multi-dimensions for unstructured grids over complex geometries.

Interface capturing for two-phase flows can be further classified under sharp-interface and diffuse-interface methods. In the sharp-interface method, the interface between the two phases is treated as infinitely thin with physical properties such as density and viscosity having a bulk value till the point of the discontinuity at the interface. The equations of motion are solved separately for the two domains and appropriate pressure- and velocity-jump boundary conditions are imposed to have localized interactions at the infinitely thin phase interface.
In the diffuse-interface methods,  however, a gradual and smooth variation of physical properties is assumed across the phase interface of finite thickness. The equations of motion are solved on the entire domain as one-fluid formulation. The physical properties are a function of a conserved order parameter which is solved by minimizing a free energy functional based on thermodynamic arguments \cite{Anderson}. Similar to the VOF and level-set methods, the interface location is defined by the contour levels of the phase-field order parameter. Unlike the sharp-interface description, the phase-field formulation does not require to satisfy the jump conditions at the interface and there is rapid but smooth variations of the order parameter and other physical quantities in the thin interfacial region. The diffuse-interface description has been shown to approach the sharp-interface limit asymptotically \cite{Anderson_1}.

The phase-field models originate from thermodynamically consistent theories of phase transitions via minimization of gradient energy across the diffuse interface. The phases are indicated by an order parameter $(\phi)$ (e.g., $\phi=1$ in one phase and $\phi=-1$ in another phase) which is solved over an Eulerian mesh and the interface is evolved. As they fall under the diffuse-interface description, no conditions at the interface between the two phases are required to be satisfied. The surface tension and capillary effects can be modeled as a function of the phase-field order parameter depending on the minimization of the Ginzburg-Landau energy functional. Therefore, these methods do not require any re-initialization or reconstruction at the interface. Furthermore, the mass conservation property can be imparted in a relatively simple manner. The topology changes in the interface can be handled by the phase-field models easily given that the mesh is refined enough to capture these phenomena.  Due to the mass conservation property and simplicity to handle topology changes, we consider the phase-field technique for physical modeling of multiphase systems.

In the phase-field models, the interface is evolved by solving a transport equation written in the form of a gradient flow of the energy functional, either in the $L^2(\Omega)$ norm (Allen-Cahn (AC) equation \cite{Allen_cahn}) or in the $H^{-1}(\Omega)$ norm (Cahn-Hilliard (CH) equation \cite{Cahn_Hilliard}). The numerical challenges in solving these transport equations are mass conservation, inherent nonlinearities and the high order of the CH equation. While the CH equation conserves mass naturally, the conventional AC equation does not provide the mass conserving property. Method employing Lagrange multiplier which is global and time dependent has been suggested to impart this property to the AC equation in \cite{Rubinstein, Bronsard}. Mass conservation using local and nonlocal Lagrange multiplier was suggested in \cite{Brassel, Kim_2} since it was observed that small geometric features were not maintained by the equation when a global multiplier depending on time was considered \cite{H_G_Lee}. We have implemented the latter strategy to satisfy the mass conservation in the present study. 

A review of the phase-field techniques to handle the nonlinearities in energy stable schemes is discussed in \cite{Tierra}. The treatment of the nonlinear double-well potential function in the energy functional affects the energy stability property of the underlying discretization. The fourth-order of the CH equation is another challenge, which restricts its flexibility for generic unstructured meshes. Under the finite element variational framework, we cannot employ linear finite elements to discretize the equation because the higher order derivatives will be zero. CH equation is thus written as a set of two coupled equations making the system complex and increasing the degrees of freedom. A comparison between the AC and CH equations has been carried out in \cite{Dongsun, Jeong}. A review of the phase-field models for multiphase flows can be found in \cite{Kim}. A second-order time accurate scheme for the CH equation was proposed under mixed-finite element framework in \cite{Gomez}. While the present study deals with the linear finite elements, solving the Allen-Cahn equation is more convenient and less computationally expensive and is thus chosen for the study. Some of the recent works which have used the conservative AC equation and coupled it with Navier-Stokes equations are \cite{Zhang, Yang, Vasconcelos}.
The Allen-Cahn phase-field equation mathematically represents a complex nonlinear convection-diffusion-reaction (CDR) equation. The solution of this equation can exhibit oscillations when the convection or reaction effects are dominant. The basic strategy of the finite difference and finite volume discretizations to deal with such instabilities is classical upwinding technique. Although the algorithm is stable, it is only first-order accurate and leads to smearing of the solution.

%
In the present article, our objective is to solve the Allen-Cahn equation under the variational framework while maintaining boundedness property, mass conservation and energy stability in the underlying discretization. We develop the positivity preserving variational (PPV) technique \citep{PPV} to capture and reduce the oscillations near the interface and maintain boundedness in the solution under a relatively coarser mesh. We add a nonlinear stabilization term to the linear stabilized variational form to achieve the positivity property. The mass of the Allen-Cahn equation is conserved by enforcing a Lagrange multiplier term which has both global and local dependence on the solution. The locally dependent part of the multiplier term is written as a mid-point approximation to achieve the energy stability in the scheme. 
We theoretically prove the mass conservation and energy-stability properties of the proposed scheme. 

The novel features of the present variational scheme are the positivity preserving and energy-stability properties for solving the Allen-Cahn phase-field equation on a general unstructured finite element mesh. The simplicity of the present formulation makes it easier to implement on existing stabilized finite element codes. The PPV-based Allen-Cahn implementation is observed to be second-order accurate in space and more than first-order accurate in time through numerical experiments. We perform the implicit discretizations of the Navier-Stokes  and the Allen-Cahn equations. The time-splitting allows us to decouple the incompressible Navier-Stokes and Allen-Cahn solvers at each time step, which provides an avenue for the staggered solution updates in a straightforward manner. The coupled multiphase solver has been demonstrated to handle high density and viscosity ratios with surface tension effects. We carry out standard benchmarks such as Laplace-Young law and sloshing tank problem to assess the coupled phase-field implementation for incompressible flows. While the Laplace-Young law establishes the ability of the formulation to handle high density ratios, the sloshing tank problem demonstrates the free-surface modeling capabilities of the scheme. Simulations of dam break problem test the ability of the scheme to handle the topological changes of the interface in two- and three-dimensions. These numerical experiments allow to quantify the accuracy of the phase-field scheme in capturing the interface location and to characterize its sensitivity with the model and discretization parameters. 
Finally, a practical problem of 3D wave-structure interaction of a vertical truncated cylinder of circular cross-section is simulated and the wave run-up is compared with the available results in the literature. The presented 3D finite-element based phase-field solver can be easily extended to complex geometries.

The organization of the paper is as follows. Section \ref{proposed method} describes the positivity preserving method dealing with the variational formulation of the Allen-Cahn equation. The variational discretization of the Navier-Stokes equations is presented in Section \ref{section:NS} for the sake of completeness. The details about the coupling and the algorithm employed are discussed in Section \ref{imp_details}. Some numerical tests results assessing the effectiveness of the implementation are shown in Section \ref{tests}, leading to the application to the wave-structure interaction problem in Section \ref{RU_test}. Finally, we conclude the paper with some key results from the study in Section \ref{conclusion}. While Appendix A provides a proof for the discrete 
energy conservation for our Allen-Cahn phase-field formulation, the mass conservation is proven in Appendix B.

\section{Positivity preserving variational scheme for the Allen-Cahn equation}
\label{proposed method}
Before proceeding to the variational formulation based on the positivity preserving scheme \cite{PPV}, we first review the Allen-Cahn phase-field equation for immiscible two-phase flows. The key idea behind the phase-field model is to replace the sharp interface region by thin transition layers via thermodynamic considerations and the conservation laws for the diffuse-interface description of different phases of fluids. The diffuse-interface technique recovers to 
the classical jump discontinuity conditions at the interface asymptotically \citep{Anderson_1}. 

\subsection{The Allen-Cahn equation}
Consider a physical domain $\Omega(\boldsymbol{x},t)$ with spatial and temporal coordinates $\boldsymbol{x}$ and $t$, respectively.
The phases of a binary phase mixture are indicated by an order parameter $\phi(\boldsymbol{x},t)$. Under the diffuse-interface description, the interface has a finite thickness and $\phi(\boldsymbol{x},t)$  varies gradually across the interface. 
The thickness of the diffuse interface layer is represented by $\varepsilon$. Under the thermodynamic arguments, the interface evolution and its dynamics are governed by the Ginzburg-Landau energy functional $\mathcal{E}(\phi)$, which can be expressed as      
\begin{align} \label{energy_functional}
	\mathcal{E}(\phi) = \int_\Omega \bigg( \frac{\varepsilon^2}{2}|\nabla\phi|^2 + F(\phi)\bigg) \mathrm{d}\Omega.
\end{align}
This energy functional consists of two components. The first term in Eq.~(\ref{energy_functional}) represents the interfacial energy which depends on the gradient of the order parameter. On the other hand, the second term is a double-well potential function which represents the free energy of mixing and depends on the local value of $\phi(\boldsymbol{x},t)$. The competition between these two components to minimize the energy  gives rise to the evolution of the phase-field interface. 

It is known that the Allen-Cahn equation is the gradient flow of $\mathcal{E}(\phi)$ in the space $L^2(\phi)$
\begin{align}
	\partial_t\phi = - \bigg( \frac{\delta\mathcal{E}(\phi)}{\delta\phi} \bigg).
\end{align}
Here, $\partial_t\phi$ is the partial temporal derivative of the order parameter $\phi(\boldsymbol{x},t)$ and $F(\phi)$ is the double-well energy potential which is taken as $F(\phi) = \frac{1}{4}(\phi^2-1)^2$ in this study. Therefore, the two stable phases have $\phi=1$ and $\phi=-1$ for the energy potential with the interface over which $\phi$ varies gradually from $-1$ to $1$. To determine the profile of the interface at equilibrium, we minimize the energy functional by taking its variational derivative and equating it to zero. The variational derivative of the energy functional with respect to $\phi$ is called the chemical potential, which is given by
\begin{align} \label{potential}
	\frac{\delta\mathcal{E}}{\delta\phi} = -\varepsilon^2\nabla^2\phi + F'(\phi).
\end{align}
The evolution of $\phi$ with time tends to minimize the energy functional where the diffusive effects of $\nabla^2\phi$ and the reactive effects of $F'(\phi)$ compete with each other. The profile of the interface at equilibrium is obtained by finding the roots of Eq.~(\ref{potential}) as:
\begin{align} \label{eqm_profile}
	\phi(z) = \mathrm{tanh}\bigg( \frac{z}{\sqrt{2}\varepsilon} \bigg),	
\end{align}
where $z$ is the coordinate normal to the interface. The equilibrium interface thickness denoted by $\varepsilon_\mathrm{eqm}$ is the distance over which $\phi$ varies from $-0.9$ to $0.9$ which can be found out as $2\sqrt{2}\varepsilon\mathrm{tanh}^{-1}(0.9) = 4.164\varepsilon$. 

In the present study, we consider the convective form of the Allen-Cahn equation with a Lagrange multiplier to conserve mass. 
For improved capturing of the physical features in the solution, we employ the modified AC equation with both local and global multiplier terms:
\begin{equation}
\left.
\begin{aligned} \label{CAC}
	\partial_t\phi + \boldsymbol{u}\cdot\nabla\phi - \varepsilon^2\nabla^2\phi + F'(\phi) - \beta(t)\sqrt{F(\phi)} = 0,\ \ &\mathrm{on}\ (\boldsymbol{x},t)\in \Omega,\qquad \\
	\boldsymbol{u}|_{\Gamma} = 0,\ \ &\mathrm{on}\ \boldsymbol{x}\in\Gamma, \\
	\frac{\partial \phi}{\partial n}\bigg|_{\Gamma} = \mathbf{n}\cdot\nabla\phi = 0,\ \ &\mathrm{on}\ (\boldsymbol{x},t)\in \Omega,\\
	\phi |_{t=0} = \phi_0,	\ \ &\mathrm{on}\ \boldsymbol{x}\in\Omega, 
\end{aligned}
\right\}
\end{equation}
where $\boldsymbol{u}$ is the convection velocity and $\mathbf{n}$ is the outward normal to the boundary $\Gamma$. $F'(\phi)$ is the derivative of $F(\phi)$ with respect to $\phi$. The third equation in Eq.~(\ref{CAC}) is the homogeneous Neumann boundary condition on $\phi$. The parameter $\beta(t)$ is the time dependent part of the Lagrange multiplier which can be derived using the incompressible flow condition of divergence-free velocity and the given boundary conditions as
\begin{align} \label{eqn:beta}
	\beta(t) = \frac{\int_\Omega F'(\phi)\mathrm{d}\Omega}{\int_\Omega \sqrt{F(\phi)}\mathrm{d}\Omega}.
\end{align}
The Lagrange multiplier is written in such a way that
\begin{align} \label{K_cons}
	\int_\Omega K(\phi) \mathrm{d}\Omega =\ \mathrm{constant},
\end{align}
where $K(\phi) = 0.5(\phi^3/3 - \phi)$.
Our aim here is to solve Eq.~(\ref{CAC}) while preserving positivity and boundedness in the solution. This equation will be later coupled with the incompressible Navier-Stokes equations for two-phase flow modeling. In the next subsection, we present the variational formulation to solve the AC equation using the PPV method.

\subsection{A new positivity preserving variational formulation}
The Allen-Cahn based phase-field equation can be expressed as a convection-diffusion-reaction equation. Variational discretization based on Galerkin method of the equation leads to spurious oscillations in the solution when the convection or reaction effects are dominant  \citep{PPV}. Although linear stabilization methods such as Streamline-Upwind Petrov-Galerkin (SUPG) and Galerkin/least-squares (GLS) can reduce spurious oscillations, the solution has oscillatory behavior near the region of high gradients. The solution thus obtained can affect the physical results and lead to wrong prediction of the underlying phenomenon being studied. With regard to phase-field modeling of two-phase flows, the interface between the two fluids is represented by the order parameter $\phi$ which is solved by the AC equation. The variable $\phi$ is also used to interpolate the physical properties of the fluid phases such as density and viscosity, which should always be positive. Oscillations in $\phi$ can lead to unbounded values which can cause the density or viscosity to be negative, thus producing unstable or unphysical results. The proposed PPV-based formulation \cite{PPV} reduces these oscillations in the solution near the region of high gradients by enforcing the positivity property of the underlying element-level matrix. Before proceeding to the presentation of the spatial PPV-based variational formulation, we first discretize the AC-based phase-field equation in time via variational integration.    

\subsubsection{Temporal discretization}
It is known that variational time integrators have the energy conserving property in comparison to Runge-Kutta time integration \cite{VTI}. To enforce this property, we employ the generalized-$\alpha$ time integration technique \cite{Gen_alpha,Jansen} to discretize the AC equation (Eq.~\ref{CAC}) in time domain. Generalized-$\alpha$ method can be unconditionally stable and second-order accurate for linear problems \cite{Gen_alpha}. The scheme enables user-controlled high frequency damping, which is desirable for a coarser discretization in space and time. This is achieved by specifying a single parameter called the spectral radius $\rho_{\infty}$. This algorithm dampens the spurious high frequency responses, but retains the second-order accuracy. Let $\partial_t{\phi}^{\mathrm{n}+\alpha_\mathrm{m}}$ be the derivative of $\phi$ with respect to time at $t^{\mathrm{n}+\alpha_\mathrm{m}}$. The statement for the time discretized AC-based phase-field equation can be written as:
\begin{align}
	G(\partial_t{\phi}^{\mathrm{n}+\alpha_\mathrm{m}}, \phi^{\mathrm{n}+\alpha}) = \partial_t{\phi}^{\mathrm{n}+\alpha_\mathrm{m}} + \boldsymbol{u}\cdot\nabla\phi^{\mathrm{n}+\alpha} &- \varepsilon^2\nabla^2\phi^{\mathrm{n}+\alpha} + F'(\phi) - \beta(t)\sqrt{F(\phi)} = 0,\\
	\phi^\mathrm{n+1} &= \phi^\mathrm{n} + \Delta t\partial_t{\phi}^\mathrm{n} + \gamma\Delta t(\partial_t{\phi}^\mathrm{n+1} - \partial_t{\phi}^\mathrm{n}), \label{app_1} \\
	\partial_t{\phi}^{\mathrm{n}+\alpha_\mathrm{m}} &= \partial_t{\phi}^\mathrm{n} + \alpha_\mathrm{m} (\partial_t{\phi}^\mathrm{n+1} - \partial_t{\phi}^\mathrm{n}),\label{app_2} \\
	\phi^{\mathrm{n}+\alpha} &= \phi^\mathrm{n} + \alpha (\phi^\mathrm{n+1} - \phi^\mathrm{n}), \label{npalpha}
\end{align}
where $\Delta t$ is the time step size, $\alpha_\mathrm{m}$, $\alpha$ and $\gamma$ are the generalized-$\alpha$ parameters defined as:
\begin{align}
	\alpha_\mathrm{m} = \frac{1}{2}\bigg(\frac{3-\rho_{\infty}}{1+\rho_{\infty}}\bigg),\qquad \alpha = \frac{1}{1+\rho_{\infty}},\qquad \gamma = \frac{1}{2} + \alpha_\mathrm{m} - \alpha.
\end{align}
Here, the energy potential is $F(\phi) = \frac{1}{4}(\phi^2-1)^2$ which is the double-well potential with equal well-depth. For imparting the energy stability property, $F'(\phi)$ is written as \cite{Du}
\begin{align} \label{F_phi}
	F'(\phi) =& \frac{F(\phi^\mathrm{n+1}) - F(\phi^\mathrm{n})}{\phi^\mathrm{n+1} - \phi^\mathrm{n}}.
\end{align}
Let $K'(\phi) = \sqrt{F(\phi)} = 0.5(\phi^2 - 1)$ be written as
\begin{align} \label{K_phi}
	\Aboxed{K'(\phi) &= \sqrt{F(\phi)} = \frac{K(\phi^\mathrm{n+1}) - K(\phi^\mathrm{n})}{\phi^\mathrm{n+1} - \phi^\mathrm{n}}}
\end{align}
The expressions in Eqs.~(\ref{F_phi}) and (\ref{K_phi}) are formulated to provide the energy stability property to the variational scheme. Considering the mid-point approximation of the spatial part of the Lagrange multiplier helps in simplifying the expressions in the derivation of the discrete energy law leading to an energy-stable scheme. A detailed derivation of discrete energy law has been presented in \ref{DEL}.

Using Eq.~(\ref{npalpha}) to replace the expression for $\phi^\mathrm{n+1}$ in $F'(\phi)$ and $\sqrt{F(\phi)}$ and rearranging the terms, the Allen-Cahn equation can be written in the form of convection-diffusion-reaction equation as follows:
\begin{align}
	G(\partial_t{\phi}^{\mathrm{n}+\alpha_\mathrm{m}}, \phi^{\mathrm{n}+\alpha}) = \partial_t{\phi}^{\mathrm{n}+\alpha_\mathrm{m}} + \boldsymbol{u}\cdot\nabla\phi^{\mathrm{n}+\alpha} - k\nabla^2\phi^{\mathrm{n}+\alpha} + s\phi^{\mathrm{n}+\alpha} - f = 0,
\end{align}
where
\begin{align}
	\mathrm{Convection\ velocity} &= \boldsymbol{u},\\
	\mathrm{Diffusion\ coefficient} &= k = \varepsilon^2,\\
	\mathrm{Reaction\ coefficient} &= s = \frac{1}{4}\bigg[ \frac{(\phi^\mathrm{n+\alpha})^2}{\alpha^3} - \bigg(\frac{3}{\alpha^3} - \frac{4}{\alpha^2}\bigg)\phi^\mathrm{n+\alpha}\phi^\mathrm{n} + \bigg( \frac{3}{\alpha^3} - \frac{8}{\alpha^2} + \frac{6}{\alpha} \bigg) (\phi^\mathrm{n})^2 - \frac{2}{\alpha} \bigg]\nonumber \\ &- \frac{\beta(t)}{2}\bigg[ \frac{\phi^\mathrm{n+\alpha}}{3\alpha^2} + \frac{1}{3}\bigg( -\frac{2}{\alpha^2} + \frac{3}{\alpha} \bigg)\phi^\mathrm{n} \bigg],\\
	\mathrm{Source\ term} &= f = -\frac{1}{4}\bigg[ \bigg(-\frac{1}{\alpha^3} +\frac{4}{\alpha^2} - \frac{6}{\alpha} + 4 \bigg)(\phi^\mathrm{n})^3 + \bigg( \frac{2}{\alpha} - 4\bigg)\phi^\mathrm{n} \bigg] \nonumber \\
	&+ \frac{\beta(t)}{2}\bigg[ \frac{1}{3}\bigg( \frac{1}{\alpha^2} - \frac{3}{\alpha} + 3 \bigg)(\phi^\mathrm{n})^2 - 1\bigg].
\end{align}

\subsubsection{Spatial discretization}
Now, we formulate the positivity preserving variational form of the AC-based phase-field equation. Consider the spatial discretization of the computational domain $\Omega$ into $\mathrm{n_{el}}$ number of elements such that $\Omega = \cup_\mathrm{e=1}^\mathrm{n_{el}} \Omega^\mathrm{e}$ and $\emptyset = \cap_\mathrm{e=1}^\mathrm{n_{el}} \Omega^\mathrm{e}$. Let $\mathcal{S}^\mathrm{h}$ be the space of trial solution, the values of which equal the given Dirichlet boundary condition and $\mathcal{V}^\mathrm{h}$ be the space of test functions which vanish on the Dirichlet boundary, $\Gamma_D$. The variational form of the AC equation can be written as: find $\phi_\mathrm{h}(\boldsymbol{x},t^{\mathrm{n}+\alpha}) \in \mathcal{S}^\mathrm{h}$ such that $\forall w_\mathrm{h} \in \mathcal{V}^\mathrm{h}$,
\begin{align}
	\int_\Omega \bigg( w_\mathrm{h}\partial_t{\phi}_\mathrm{h} +  w_\mathrm{h}(\boldsymbol{u}\cdot\nabla\phi_\mathrm{h}) - w_\mathrm{h} k\nabla^2\phi_\mathrm{h} + w_\mathrm{h}s\phi_\mathrm{h} - w_\mathrm{h}f \bigg) \mathrm{d}\Omega = 0.
\end{align}
Using the divergence theorem and the fact that $w_\mathrm{h} = 0$ on $\Gamma_D$,
\begin{align}
	&\int_\Omega \bigg( w_\mathrm{h}\partial_t{\phi}_\mathrm{h} + w_\mathrm{h}(\boldsymbol{u}\cdot\nabla\phi_\mathrm{h}) + \nabla w_\mathrm{h}\cdot(k\nabla\phi_\mathrm{h}) + w_\mathrm{h}s\phi_\mathrm{h} - w_\mathrm{h}f \bigg) \mathrm{d}\Omega - \int_{\Gamma_N} w_\mathrm{h}g \mathrm{d}\Gamma = 0,
\end{align}
where $g$ is the boundary condition on the Neumann boundary $\Gamma_N$ which is zero in this case. The above equation is the standard Galerkin finite element formulation. Next, we apply the positivity preserving stabilization terms to the formulation above:
\begin{align} \label{PPV_AC}
	&\int_\Omega \bigg( w_\mathrm{h}\partial_t{\phi}_\mathrm{h} + w_\mathrm{h}(\boldsymbol{u}\cdot\nabla\phi_\mathrm{h}) + \nabla w_\mathrm{h}\cdot(k\nabla\phi_\mathrm{h} ) + w_\mathrm{h}s\phi_\mathrm{h} - w_\mathrm{h}f \bigg) \mathrm{d}\Omega \nonumber \\
	&+ \displaystyle\sum_\mathrm{e=1}^\mathrm{n_{el}}\int_{\Omega^\mathrm{e}}\bigg( \big(\boldsymbol{u}\cdot\nabla w_\mathrm{h} \big)\tau \big( \partial_t{\phi}_\mathrm{h} + \boldsymbol{u}\cdot\nabla\phi_\mathrm{h} - \nabla\cdot(k\nabla\phi_\mathrm{h}) + s\phi_\mathrm{h} -f \big) \bigg) \mathrm{d}\Omega^\mathrm{e}  \nonumber \\
	&+ \displaystyle\sum_\mathrm{e=1}^\mathrm{n_{el}}\int_{\Omega^\mathrm{e}} \chi \frac{|\mathcal{R}(\phi_\mathrm{h})|}{|\nabla\phi_\mathrm{h}|}k_s^\mathrm{add} \nabla w_\mathrm{h}\cdot \bigg( \frac{\boldsymbol{u}\otimes \boldsymbol{u}}{|\boldsymbol{u}|^2} \bigg) \cdot \nabla\phi_\mathrm{h} \mathrm{d}\Omega^\mathrm{e} + \sum_\mathrm{e=1}^\mathrm{n_{el}} \int_{\Omega^\mathrm{e}}\chi \frac{|\mathcal{R}(\phi_\mathrm{h})|}{|\nabla \phi_\mathrm{h}|} k^\mathrm{add}_{c} \nabla w_\mathrm{h} \cdot \bigg( \mathbf{I} - \frac{\boldsymbol{u}\otimes \boldsymbol{u}}{|\boldsymbol{u}|^2} \bigg) \cdot \nabla\phi_\mathrm{h} \mathrm{d}\Omega^\mathrm{e} \nonumber \\
	&= 0, 	
\end{align}
where $\tau$ is the stabilization parameter given by \cite{Hughes_X}
\begin{align}
	\tau &= \bigg[ \bigg( \frac{2}{\Delta t}\bigg)^2 + \boldsymbol{u}\cdot \boldsymbol{G}\boldsymbol{u} + 9k^2 \boldsymbol{G}:\boldsymbol{G}+ s^2\bigg] ^{-1/2},
\end{align}
where $\boldsymbol{G}$ is the element contravariant metric tensor \cite{Akkerman,Bazilevs_book}, which is defined as
\begin{align}
	\boldsymbol{G} = \frac{\partial \boldsymbol{\xi}^T}{\partial \boldsymbol{x}}\frac{\partial \boldsymbol{\xi}}{\partial \boldsymbol{x}},
\end{align}
where $\boldsymbol{x}$ and $\boldsymbol{\xi}$ are the physical and parametric coordinates respectively
and $\mathcal{R}(\phi_\mathrm{h})$ is the residual of the Allen-Cahn equation given as
\begin{align}
	\mathcal{R}(\phi_\mathrm{h}) = \partial_t\phi_\mathrm{h} + \boldsymbol{u}\cdot\nabla\phi_\mathrm{h} - \nabla\cdot(k\nabla\phi_\mathrm{h}) + s\phi_\mathrm{h} -f. 	
\end{align}
In Eq.~(\ref{PPV_AC}), the first line represents the Galerkin terms. The second line consists of linear stabilization terms in the form of SUPG stabilization with the stabilization parameter $\tau$. 
Note that only the convective term has been taken in the weighting function, in contrast to the combined GLS-subgrid scale methodology of PPV where we also take the reaction term. This new modification has been performed to ensure the mass conservation property in the variational form. The proof of mass conservation property of the modified PPV scheme has been derived in \ref{DMC}. 
Terms in the third line of Eq.~(\ref{PPV_AC}) are the positivity preserving nonlinear stabilization terms which enforce the positivity property to the element-level matrix. 

The idea behind the PPV technique is first to apply the discrete upwind operator to the Galerkin and linear stabilized element-level matrix. This could be seen as addition of diffusion to convert the element matrix to a non-negative matrix in one-dimension. While the discrete upwinding provides the positivity property, it makes the solution first-order accurate. Therefore, the addition of diffusion is restricted to the region near the discontinuity which exhibits oscillations. This is accomplished by the dependence of the nonlinear terms on the residual of the equation. This idea is extended to multi-dimensions by considering the nonlinear terms in both streamline and crosswind directions. The first and second terms in the third line correspond to the streamline and crosswind directions respectively. The details of the derivation of the added diffusions $k_s^\mathrm{add}$, $k_c^\mathrm{add}$ and $\chi$ can be found in \cite{PPV}. For the current context of AC-based phase-field equation,
\begin{align}
	\chi &= \frac{2}{|s|h + 2|\boldsymbol{u}|},\\
	k_s^\mathrm{add} &= \mathrm{max} \bigg\{ \frac{||\boldsymbol{u}| - \tau|\boldsymbol{u}|s|h}{2} - (k + \tau|\boldsymbol{u}|^2) + \frac{sh^2}{6}, 0 \bigg\},\\
	k_c^\mathrm{add} &= \mathrm{max} \bigg\{ \frac{|\boldsymbol{u}|h}{2} - k + \frac{sh^2}{6}, 0 \bigg\},
\end{align} 
where $|\boldsymbol{u}|$ is the magnitude of the convection velocity and $h$ is the characteristic element length defined in \cite{PPV}. Some of the prominent properties of the PPV method applied to the CDR equation are: (i) the method has at least second-order of spatial accuracy in the different convection- and reaction-dominated regions, (ii) a general application of the method to arbitrary topology can be implemented, and (iii) the method captures the high gradient internal and boundary layers in multi-dimensions. The aforementioned properties have been thoroughly investigated and tested for different convection- and reaction-dominated regimes in \cite{PPV}. We next present the Navier-Stokes equations and its variational form of the two-phase solver.

\section{Variational discretization of the Navier-Stokes equations}
\label{section:NS}
For the computation of the multiphase incompressible flows, we next present the one-fluid formulation of the unsteady Navier-Stokes equations in this section.  For the sake of completeness, the governing equations and their variational counterparts are first presented.

\subsection{The incompressible Navier-Stokes equations}
The unsteady Navier-Stokes equations for a viscous incompressible flow on a physical domain $\Omega(\boldsymbol{x},t)$ are
\begin{align} \label{NS}
	\rho(\phi)\frac{\partial {\boldsymbol{u}}}{\partial t} + \rho(\phi){\boldsymbol{u}}\cdot\nabla{\boldsymbol{u}} &= \nabla\cdot {\boldsymbol{\sigma}} + \mathbf{SF}(\phi) + \boldsymbol{b}(\phi),&&\mathrm{on}\ (\boldsymbol{x},t)\in \Omega,\\
	\nabla\cdot{\boldsymbol{u}} &= 0,&&\mathrm{on}\ (\boldsymbol{x},t)\in \Omega,
\end{align}
where ${\boldsymbol{u}} = {\boldsymbol{u}}(\boldsymbol{x},t)$ denotes the fluid velocity defined for each spatial point $\boldsymbol{x} \in \Omega$, $\rho(\phi)$ is the fluid density, $\boldsymbol{b}(\phi)$ is the body force applied on the fluid, $\mathbf{SF}(\phi)$ is the singular force acting at the interface which ensures the pressure jump across the interface and ${\boldsymbol{\sigma}}$ is the Cauchy stress tensor for a Newtonian fluid, given as
\begin{align}
	{\boldsymbol{\sigma}} = -{p}\boldsymbol{I} + \mu(\phi)( \nabla{\boldsymbol{u}}+ (\nabla{\boldsymbol{u}})^T),
\end{align}
where ${p}$ denotes the fluid pressure and $\mu(\phi)$ is the dynamic viscosity of the fluid. The singular force is replaced by a continuum surface force (CSF) \cite{Brackbill}, which depends on the order parameter. Several forms of $\mathbf{SF}(\phi)$ have been used in the literature which are reviewed in \cite{Kim, Kim_3}. In this study, we employ the following definition:
\begin{align}
	\mathbf{SF}(\phi) = \sigma\varepsilon\alpha_\mathrm{sf}\nabla\cdot( |\nabla\phi|^2\mathbf{I} - \nabla\phi \otimes \nabla\phi ), 
\end{align}
where $\sigma$ is the surface tension, $\varepsilon$ is the interface thickness parameter defined in the Allen-Cahn phase-field equation and $\alpha_\mathrm{sf}$ is a constant parameter.
The order parameter $\phi$ needs to be locally in the equilibrium state. Hence, to match the surface tension of the sharp-interface description, $\alpha_\mathrm{sf}$ must satisfy the following condition \cite{Kim_3}
\begin{align}
	\varepsilon\alpha_\mathrm{sf} \int_{-\infty}^{\infty} \bigg( \frac{\mathrm{d}\phi}{\mathrm{d}z} \bigg)^2 dz &= 1,\\
	\alpha_\mathrm{sf} &= \frac{3\sqrt{2}}{4},
\end{align}
where we apply the expression of the equilibrium interface profile $\phi(z)$ given in Eq.~(\ref{eqm_profile}). This gives rise to the continuum surface force
\begin{align}
	\mathbf{SF}(\phi) = \sigma\varepsilon\frac{3\sqrt{2}}{4}\nabla\cdot( |\nabla\phi|^2\mathbf{I} - \nabla\phi \otimes \nabla\phi ).
\end{align}
The physical parameters of the fluid such as $\rho$ and $\mu$ are dependent on the order parameter $\phi$ as
\begin{align}
	\rho(\phi) &= \frac{1+\phi}{2}\rho_1 + \frac{1-\phi}{2}\rho_2, \label{dens}\\
	\mu(\phi) &= \frac{1+\phi}{2}\mu_1 + \frac{1-\phi}{2}\mu_2, \label{visc}
\end{align}
where $(\cdot)_i$ denotes the value of $(\cdot)$ on the phase $i$ of the fluid. The body force $\boldsymbol{b}$ is generally taken as the gravity force dependent on the order parameter $\phi$ as $\boldsymbol{b}(\phi) = \rho(\phi)\boldsymbol{g}$, $\boldsymbol{g}$ being the acceleration due to gravity. Therefore, $\rho=\rho_1$ with $\mu=\mu_1$ on the fluid phase $1$ when $\phi=1$ and $\rho=\rho_2$ with $\mu=\mu_2$ on the fluid phase $2$ when $\phi=-1$. 

\subsection{The variational formulation}
For consistency, the Navier-Stokes equations are discretized using the generalized-$\alpha$ time integration. Let $\partial_t{\boldsymbol{u}}^\mathrm{n+\alpha_m}$ be the temporal derivative of $\boldsymbol{u}$ at time $t^\mathrm{n+\alpha_m}$. The expressions used in the variational formulation are given as
\begin{align}
	{\boldsymbol{u}}_\mathrm{h}^\mathrm{n+1} &= {\boldsymbol{u}}_\mathrm{h}^\mathrm{n} + \Delta t\partial_t{\boldsymbol{u}}_\mathrm{h}^\mathrm{n} + \gamma \Delta t( \partial_t{\boldsymbol{u}}_\mathrm{h}^\mathrm{n+1} - \partial_t{\boldsymbol{u}}_\mathrm{h}^\mathrm{n}), \\
	{\boldsymbol{u}}_\mathrm{h}^\mathrm{n+\alpha} &= {\boldsymbol{u}}_\mathrm{h}^\mathrm{n} + \alpha({\boldsymbol{u}}_\mathrm{h}^\mathrm{n+1} - {\boldsymbol{u}}_\mathrm{h}^\mathrm{n}), \\
	\partial_t\boldsymbol{u}_\mathrm{h}^\mathrm{n+\alpha_m} &= \partial_t{\boldsymbol{u}}_\mathrm{h}^\mathrm{n} + \alpha_\mathrm{m}( \partial_t{\boldsymbol{u}}_\mathrm{h}^\mathrm{n+1} - \partial_t{\boldsymbol{u}}_\mathrm{h}^\mathrm{n}).
\end{align}

Let $\mathcal{S}^\mathrm{h}$ be the space of trial solution, the values of which satisfy the Dirichlet boundary condition and $\mathcal{V}^\mathrm{h}$ be the space of test functions which vanish on the Dirichlet boundary. The variational form of the flow equation can be written as: find $[ {\boldsymbol{u}}_\mathrm{h}^\mathrm{n+\alpha}, {p}_\mathrm{h}^\mathrm{n+1}] \in\mathcal{S}^\mathrm{h}$ such that $\forall [\boldsymbol{\psi}_\mathrm{h}, q_\mathrm{h}] \in \mathcal{V}^\mathrm{h}$,
\begin{align}
	&\int_{\Omega} \rho(\phi) ( \partial_t{\boldsymbol{u}}_\mathrm{h}^\mathrm{n+\alpha_m} + {\boldsymbol{u}}_\mathrm{h}^\mathrm{n+\alpha} \cdot \nabla{\boldsymbol{u}}_\mathrm{h}^\mathrm{n+\alpha})\cdot \boldsymbol{\psi}_\mathrm{h} \mathrm{d\Omega} + \int_{\Omega} {\boldsymbol{\sigma}}_\mathrm{h}^\mathrm{n+\alpha} : \nabla \boldsymbol{\psi}_\mathrm{h} \mathrm{d\Omega} - \int_{\Omega} \mathbf{SF}^\mathrm{n+\alpha}_\mathrm{h}(\phi)\cdot\boldsymbol{\psi}_\mathrm{h} \mathrm{d\Omega}- \int_{\Omega} \nabla q_\mathrm{h}\cdot {\boldsymbol{u}}_\mathrm{h}^\mathrm{n+\alpha} \mathrm{d\Omega}\nonumber \\
	&+ \displaystyle\sum_\mathrm{e=1}^\mathrm{n_{el}}\int_{\Omega^\mathrm{e}} \frac{\tau_\mathrm{m}}{\rho(\phi)} (\rho(\phi){\boldsymbol{u}}_\mathrm{h}^\mathrm{n+\alpha}\cdot \nabla\boldsymbol{\psi}_\mathrm{h}+ \nabla q_\mathrm{h} )\cdot \boldsymbol{\mathcal{R}}_\mathrm{m}({\boldsymbol{u}},{p}) \mathrm{d\Omega^e} + \displaystyle\sum_\mathrm{e=1}^\mathrm{n_{el}}\int_{\Omega^\mathrm{e}} \nabla\cdot \boldsymbol{\psi}_\mathrm{h}\tau_\mathrm{c}\rho(\phi) \boldsymbol{\mathcal{R}}_\mathrm{c}(\boldsymbol{u}) \mathrm{d\Omega^e}\nonumber \\
	& -\displaystyle\sum_\mathrm{e=1}^\mathrm{n_{el}}\int_{\Omega^\mathrm{e}} \tau_\mathrm{m} \boldsymbol{\psi}_\mathrm{h}\cdot (\boldsymbol{\mathcal{R}}_\mathrm{m}({\boldsymbol{u}},{p}) \cdot \nabla {\boldsymbol{u}}_\mathrm{h}^\mathrm{n+\alpha}) \mathrm{d\Omega^e} -\displaystyle\sum_\mathrm{e=1}^\mathrm{n_{el}}\int_{\Omega^\mathrm{e}} \frac{\nabla \boldsymbol{\psi}_\mathrm{h}}{\rho(\phi)}:(\tau_\mathrm{m}\boldsymbol{\mathcal{R}}_\mathrm{m}({\boldsymbol{u}},{p}) \otimes \tau_\mathrm{m}\boldsymbol{\mathcal{R}}_\mathrm{m}({\boldsymbol{u}},{p})) \mathrm{d\Omega^e}\nonumber \\
	&= \int_{\Omega} \boldsymbol{b}(t^\mathrm{n+\alpha})\cdot \boldsymbol{\psi}_\mathrm{h} \mathrm{d\Omega} + \int_{\Gamma_\mathrm{h}} \boldsymbol{h}\cdot \boldsymbol{\psi}_\mathrm{h} \mathrm{d\Gamma},
\end{align}
where the terms in the first line correspond to the Galerkin terms including the transient and convective terms of the momentum equation, the viscous stress and surface tension terms and the variational form of the continuity equation. The second line represents the Galerkin/least-squares stabilization terms for the momentum and continuity equations. With the use of linear finite element spaces, the higher order derivatives of the weighting function related to the viscous stress tensor will be very small and hence, they have been neglected. The two residual terms in the third line are introduced as the approximation of the fine scale velocity on the element interiors and its corresponding convective stabilization based on the multi-scale argument \cite{Hughes_conserve,Hsu,Akkerman}. $\boldsymbol{h}$ is the corresponding Neumann boundary condition for the Navier-Stokes equations. The element-wise residual of the momentum and the continuity equations denoted by $\boldsymbol{\mathcal{R}}_\mathrm{m}$ and $\boldsymbol{\mathcal{R}}_\mathrm{c}$ respectively are given by
\begin{align}
	\boldsymbol{\mathcal{R}}_\mathrm{m}({\boldsymbol{u}},{p}) &= \rho(\phi)\partial_t{\boldsymbol{u}}_\mathrm{h}^\mathrm{n+\alpha_m} + \rho(\phi){\boldsymbol{u}}_\mathrm{h}^\mathrm{n+\alpha} \cdot \nabla{\boldsymbol{u}}_\mathrm{h}^\mathrm{n+\alpha} - \nabla \cdot {\boldsymbol{\sigma}}_\mathrm{h}^\mathrm{n+\alpha} - \mathbf{SF}^\mathrm{n+\alpha}_\mathrm{h}(\phi) - \boldsymbol{b}(t^\mathrm{n+\alpha}), \\
	\boldsymbol{\mathcal{R}}_\mathrm{c}(\boldsymbol{u}) &= \nabla\cdot\boldsymbol{u}^\mathrm{n+\alpha}_\mathrm{h}.
\end{align}
The stabilization parameters $\tau_\mathrm{m}$ and $\tau_\mathrm{c}$ are the least-squares metrics added to the element-level integrals in the stabilized formulation \cite{Brooks,Hughes_X,Tezduyar_1,France_II} and are defined as
\begin{align}
	\tau_\mathrm{m} &= \bigg[ \bigg( \frac{2}{\Delta t}\bigg)^2 + {\boldsymbol{u}}_\mathrm{h}\cdot \boldsymbol{G}{\boldsymbol{u}}_\mathrm{h} + C_I \bigg(\frac{\mu(\phi)}{\rho(\phi)}\bigg)^2 \boldsymbol{G}:\boldsymbol{G}\bigg] ^{-1/2},\\
	\tau_\mathrm{c} &= \frac{1}{\mathrm{tr}(\boldsymbol{G})\tau_\mathrm{m}}, \label{tau_c}
\end{align}
where $C_I$ is a constant derived from the element-wise inverse estimate \cite{Hughes_inv_est}, $\boldsymbol{G}$ is the element contravariant metric tensor and $\mathrm{tr(\boldsymbol{G})}$ in Eq.~(\ref{tau_c}) denotes the trace of the contravariant metric tensor. The stabilization in the variational form provides stability to the velocity field in convection dominated regimes of the fluid domain and circumvents the Babu$\mathrm{\check{s}}$ka-Brezzi condition which is required to be satisfied by any standard mixed Galerkin method \cite{Johnson}. The element metric tensor $\boldsymbol{G}$ deals with different element topology for different mesh discretization and has been greatly studied in the literature \cite{Johnson,Hsu,Tezduyar_1,Hughes_V,Tezduyar_stab}.

\section{Implementation details}
\label{imp_details}
\begin{figure}[!htbp]
\centering
\begin{tikzpicture}[decoration={markings,mark=at position 0.5 with {\arrow[scale=2]{>}}},every node/.style={scale=0.9},scale=0.9]
	\draw[-,black] (0,0) node[anchor=north,black]{} to (0,11);
	\draw[-,black] (6,0) node[anchor=north,black]{} to (6,11);

	\fill[black] (5.85,0.85) rectangle (6.15,1.15) ;
	\fill[black] (5.85,3.85) rectangle (6.15,4.15) ;
	\fill[black] (5.85,6.85) rectangle (6.15,7.15) ;
	\fill[black] (5.85,9.85) rectangle (6.15,10.15) ;

	\draw[black,fill=black] (0,1) circle (0.9ex);
	\draw[black,fill=black] (0,4) circle (0.9ex);
	\draw[black,fill=black] (0,7) circle (0.9ex);
	\draw[black,fill=black] (0,10) circle (0.9ex);
	
	\draw[black,postaction={decorate}] (0,1) to (6,1);
	\draw[black,postaction={decorate}] (4,0) to (6,1);
	\draw[black,postaction={decorate}] (0,10) to (2,11);

	\draw[black,postaction={decorate}] (6,1) to (6,4);
	\draw[black,postaction={decorate}] (6,4) to (0,1);
	\draw[black,postaction={decorate}] (0,1) to (0,4);
	\draw[black,postaction={decorate}] (0,4) to (6,4);

	\draw[black,postaction={decorate}] (6,4) to (6,7);
	\draw[black,postaction={decorate}] (6,7) to (0,4);
	\draw[black,postaction={decorate}] (0,4) to (0,7);
	\draw[black,postaction={decorate}] (0,7) to (6,7);

	\draw[black,postaction={decorate}] (6,7) to (6,10);
	\draw[black,postaction={decorate}] (6,10) to (0,7);
	\draw[black,postaction={decorate}] (0,7) to (0,10);
	\draw[black,postaction={decorate}] (0,10) to (6,10);

	\draw[black,thick] (-0.25,7.6) to (0.25,7.8);
	\draw[black,thick] (-0.25,7.75) to (0.25,7.95);
	\draw[white] (0,7.72) to (0,7.83) ;

	\draw[black,thick] (5.75,7.6) to (6.25,7.8);
	\draw[black,thick] (5.75,7.75) to (6.25,7.95);
	\draw[white] (6,7.72) to (6,7.83) ;

	\draw[black] (-0.2,1) node[anchor=east,black]{$\phi(\boldsymbol{x},t^\mathrm{n})$};
	\draw[black] (-0.2,10) node[anchor=east,black]{$\phi(\boldsymbol{x},t^\mathrm{n+1})$};

	\draw[black] (6.2,1) node[anchor=west,black]{$\boldsymbol{u}(\boldsymbol{x},t^\mathrm{n})$,};
	\draw[black] (6.2,0.5) node[anchor=west,black]{$p(\boldsymbol{x},t^\mathrm{n})$};

	\draw[black] (6.2,10.5) node[anchor=west,black]{$\boldsymbol{u}(\boldsymbol{x},t^\mathrm{n+1})$,};
 	\draw[black] (6.2,10) node[anchor=west,black]{$p(\boldsymbol{x},t^\mathrm{n+1})$};

	\draw (6,4) -- (0,1) node [midway, above, sloped] {$\boldsymbol{u}_\mathrm{k+1}$};
	\draw (3,4.3) node[anchor=west]{$\phi_\mathrm{k+1}$};
	
	\draw (6,2.5) node[anchor=west,black]{[1]};
	\draw (3.4,2.5) node[anchor=west,black]{[2]};
	\draw (0,2.5) node[anchor=east,black]{[3]};
	\draw (3.9,3.7) node[anchor=east,black]{[4]};


	\draw (1.5,2.5) node[anchor=east]{$\mathrm{k}=0$};
	\draw (0.45,2.25) rectangle (1.45,2.7) ;
	\draw (1.5,5.5) node[anchor=east]{$\mathrm{k}=1$};
	\draw (0.45,5.25) rectangle (1.45,5.7) ;
	\draw (0.4,8.5) node[anchor=west]{$\mathrm{k}=\mathrm{nIter}$};
	\draw (0.45,8.25) rectangle (1.95,8.7) ;

	\draw (6.7,4.5) node[rotate=90, anchor=west,black]{Navier-Stokes system};
	\draw (-1.0,7.0) node[rotate=90, anchor=east,black]{Allen-Cahn system};
\end{tikzpicture}
\caption{A schematic of predictor-corrector procedure for Navier-Stokes and Allen-Cahn coupling via nonlinear iterations. Here, $\mathrm{nIter}$ denotes the maximum number of nonlinear iterations to achieve a desired convergence tolerance within a time step  at $ t \in [t^\mathrm{n},t^\mathrm{n+1}]$. Decoupled Navier-Stokes and Allen-Cahn equations are implicitly discretized.}
\label{solver_schematic_1}
\end{figure}
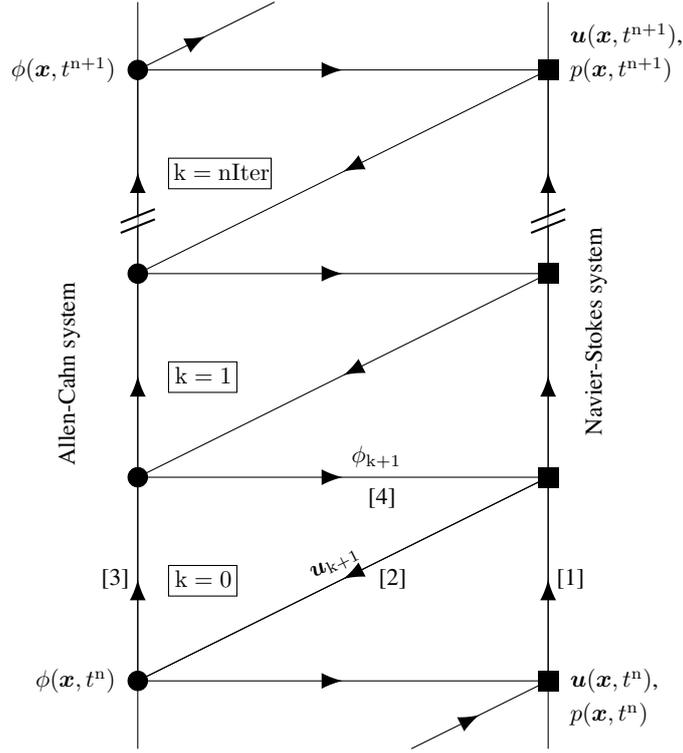

A schematic of the iterative coupling between the Navier-Stokes and the Allen-Cahn equations is shown in Fig. \ref{solver_schematic_1}. 
For efficiency and robustness, we perform the implicit discretizations of the Navier-Stokes  and the Allen-Cahn equations. We employ the time-splitting procedure to decouple the 
incompressible Navier-Stokes and Allen-Cahn updates at the discrete level.
While the Navier-Stokes equations provide a predictor fluid velocity, the Allen-Cahn equation is solved to provide an updated order parameter to interpolate the density, viscosity, capillary and body forces on the Navier-Stokes side. Consider the velocity $\boldsymbol{u}(\boldsymbol{x},t^\mathrm{n})$, pressure $p(\boldsymbol{x},t^\mathrm{n})$  and the order parameter $\phi(\boldsymbol{x},t^\mathrm{n})$ at time $t^\mathrm{n}$. In the first step of the predictor-corrector iteration $\mathrm{k}$, the velocity and pressure are predicted by solving the Navier-Stokes equations. In the second step, the computed fluid velocity is transferred to the Allen-Cahn equation. The convective AC equation is solved using the transferred fluid velocity to obtain an updated order parameter $\phi_\mathrm{k+1}$ in the third step. Finally, in the fourth step, the updated order parameter is utilized in the interpolation of density $\rho(\phi)$, viscosity $\mu(\phi)$, $\mathbf{SF}(\phi)$ and body force $\boldsymbol{b}(\phi)$. These updated quantities are then fed into the Navier-Stokes equations in the next iteration. At the end of the nonlinear iterations when the solver has achieved the convergence criteria, the values at the next time step $t^\mathrm{n+1}$ are updated and the coupled solver is advanced in time.

In this study, the incremental flow velocity, pressure and the order parameter are evaluated by employing Newton-Raphson type iterations at each time step and the generalized-$\alpha$ time integration \cite{Gen_alpha} is applied to update the solver. The linear system for the incompressible Navier-Stokes equations can be written as
\begin{align} \label{LS_NS}
	\begin{bmatrix}
		\boldsymbol{K}_\Omega &  & \boldsymbol{G}_\Omega \\
 & \\
	       -\boldsymbol{G}^T_\Omega &  &\boldsymbol{C}_\Omega
	\end{bmatrix} 
	\begin{Bmatrix}
		\Delta \boldsymbol{u}^\mathrm{n+\alpha_f} \\
\\
		\Delta p^\mathrm{n+1}
	\end{Bmatrix}
	= -\begin{Bmatrix} 
		\boldsymbol{\mathcal{R}}_\mathrm{m} \\
\\
		\boldsymbol{\mathcal{R}}_\mathrm{c}
	  \end{Bmatrix}
\end{align}
where $\boldsymbol{K}_\Omega$ is the stiffness matrix of the momentum equation which consists of inertia, convection, diffusion and stabilization terms, $\boldsymbol{G}_\Omega$ is the discrete gradient operator, $\boldsymbol{G}^T_\Omega$ is the divergence operator and $\boldsymbol{C}_\Omega$ is the pressure-pressure stabilization term. $\Delta \boldsymbol{u}$ and $\Delta p$ are the increments in velocity and pressure respectively. Similar linear system for the Allen-Cahn equation can be expressed as
\begin{align} \label{LS_AC}
	\begin{bmatrix}
		\boldsymbol{K}_{AC}
	\end{bmatrix} 
	\begin{Bmatrix}
		\Delta \phi^\mathrm{n+\alpha}
	\end{Bmatrix}
	= \begin{Bmatrix} 
		\mathcal{R}(\phi)
	  \end{Bmatrix}
\end{align}
where $\boldsymbol{K}_{AC}$ is the stiffness matrix of the Allen-Cahn equation which includes inertia, convection, diffusion, reaction and stabilization terms. $\Delta \phi$ is the increment in the solution for the order parameter. The algorithm is summarized in Algorithm \ref{algorithm_1}. 
In the present investigation, we consider equal-order interpolations for all the quantities ($\boldsymbol{u},p,\phi$) for finite element discretization.
At each time step, the nonlinear errors of fully-coupled implicit systems of the Navier-Stokes and the Allen-Chan equations are minimized by Newton-Raphson iterations.
In our numerical experiments, we have found that about 2-3 nonlinear iterations are enough to obtain a reasonably converged solution. Moreover, we want to emphasize that each nonlinear iteration of the algorithm consists of just one pass through the coupled Navier-Stokes and the Allen-Cahn phase-field system. This single-pass explicit coupling in turn helps in reducing the computational time without compromising with the accuracy and stability of the solution. 
We solve the coupled equations at discrete time steps to capture the transient flow characteristics to converge to a steady state solution, leading to a sequence
of linear system of equations. For the linear system of equations, the matrices are formed and stored using the Harwell-Boeing sparse matrix format.
The linear system  is solved via the Generalized Minimal RESidual (GMRES) algorithm proposed in \cite{saad1986}, which relies on the preconditioned Krylov subspace iteration and the modified Gram-Schmidt orthogonalization.  The number of matrix-vector products determines the dimension of the Krylov space from which a solution is computed.
The phase-field two-phase solver relies on a hybrid parallelism for parallel computing.  It employs a standard master-slave strategy for distributed memory clusters via message passing interface (MPI) based on domain decomposition strategy \cite{mpi}. The parallel implementation for the two-phase solver takes the advantage of state-of-the-art hierarchical memory 
and parallel architectures.
\begin{figure}[H]
\centering
\floatstyle{ruled}
\newfloat{algorithm}{H}{loa}
\floatname{algorithm}{Algorithm}
\begin{algorithm}
\caption{Partitioned staggered coupling of implicit Navier-Stokes and Allen-Cahn solvers}
\label{algorithm_1}
\begin{tabbing}
\     Given $\boldsymbol{u}^0$, $p^0$, $\phi^0$ \\
\qquad    Loop over time steps, $\mathrm{n}=0,1,\cdots$ \\
\qquad\	  Start from known variables $\boldsymbol{u}^\mathrm{n}$, $p^\mathrm{n}$, $\phi^\mathrm{n}$\\
\qquad\   Predict the solution: \\
\qquad\qquad $\boldsymbol{u}^\mathrm{n+1}_{(0)}=\boldsymbol{u}^\mathrm{n}$\\ 
\qquad\qquad $p^\mathrm{n+1}_{(0)} = p^\mathrm{n}$ \\ 
\qquad\qquad $\phi^\mathrm{n+1}_{(0)} = \phi^\mathrm{n}$\\
\qquad\   Loop over the nonlinear iterations, $\mathrm{k}=0,1,\cdots$ until convergence \\
\qquad \begin{tikzpicture}
	\draw (1,0) node(E){[1] \textbf{Navier-Stokes Implicit Solve}};
	\draw (-2,-0.75) node(E1)[anchor=west]{(a) Interpolate solution:};
	\draw (-1.5,-1.25) node(E2)[anchor=west] {$\boldsymbol{u}^\mathrm{n+\alpha_f}_\mathrm{(k+1)} = \boldsymbol{u}^\mathrm{n} + \alpha_\mathrm{f}(\boldsymbol{u}^\mathrm{n+1}_\mathrm{(k)} - \boldsymbol{u}^\mathrm{n})$};
	\draw (-1.5,-1.85) node(E3)[anchor=west] {$p^\mathrm{n+1}_\mathrm{(k+1)} = p^\mathrm{n+1}_\mathrm{(k)}$};
	\draw (-2,-2.75) node(E4)[anchor=west]{(b) Solve for $\Delta \boldsymbol{u}^\mathrm{n+\alpha_f}$ and $\Delta p^\mathrm{n+1}$ in Eq.~({\ref{LS_NS}})};
	\draw (-2,-3.65) node(E5)[anchor=west]{(c) Correct solution:};
	\draw (-1.5,-4.15) node(E6)[anchor=west]{$\boldsymbol{u}^\mathrm{n+\alpha_f}_\mathrm{(k+1)} = \boldsymbol{u}^\mathrm{n+\alpha_f}_\mathrm{(k+1)} + \Delta\boldsymbol{u}^\mathrm{n+\alpha_f}$};	
	\draw (-1.5,-4.75) node(E7)[anchor=west]{$p^\mathrm{n+1}_\mathrm{(k+1)} = p^\mathrm{n+1}_\mathrm{(k+1)} + \Delta p^\mathrm{n+1}$};	
	\draw (-2,-5.65) node(E8)[anchor=west]{(d) Update solution:};
	\draw (-1.5,-6.15) node(E9)[anchor=west]{$\boldsymbol{u}^\mathrm{n+1}_\mathrm{(k+1)} = \boldsymbol{u}^\mathrm{n} + \frac{1}{\alpha_\mathrm{f}}(\boldsymbol{u}^\mathrm{n+\alpha_f}_\mathrm{(k+1)} - \boldsymbol{u}^\mathrm{n})$};	
	\draw (-1.5,-6.75) node(E10)[anchor=west]{$p^\mathrm{n+1}_\mathrm{(k+1)} = p^\mathrm{n+1}_\mathrm{(k+1)}$};	
	\node[fit = (E) (E1) (E2) (E3) (E4) (E5) (E6) (E7) (E8) (E9) (E10),style={block3}](NS){};
\end{tikzpicture}
\hspace{0cm}
\begin{tikzpicture}
	\node (0,0){};
	\draw (0,2) node[anchor=north]{$\xrightarrow{\makebox[2cm]{[2] $\boldsymbol{u}^\mathrm{n+1}_\mathrm{(k+1)}$}}$};
	\draw (0,6) node[anchor=north]{$\xleftarrow{\makebox[2cm]{[4] $\phi^\mathrm{n+1}_\mathrm{(k+1)}$}}$};
	\draw (0,5.25) node[anchor=north]{to interpolate};
	\draw (0,4.75) node[anchor=north]{$\rho(\phi)$, $\mu(\phi)$};
	\draw (0,4.25) node[anchor=north]{$\mathbf{SF}(\phi)$, $\boldsymbol{b}(\phi)$};
\end{tikzpicture}
\begin{tikzpicture}
	\centering
	\draw (0.75,0) node(E){[3] \textbf{Allen-Cahn Implicit Solve}};
	\draw (-2,-0.75) node(E1)[anchor=west]{(a) Interpolate solution:};
	\draw (-1.5,-1.25) node(E2)[anchor=west] {$\phi^\mathrm{n+\alpha_f}_\mathrm{(k+1)} = \phi^\mathrm{n} + \alpha_\mathrm{f}(\phi^\mathrm{n+1}_\mathrm{(k)} - \phi^\mathrm{n})$};
	\draw (-1.5,-1.85) node(E3)[anchor=west] { };
	\draw (-2,-2.75) node(E4)[anchor=west]{(b) Solve for $\Delta \phi^\mathrm{n+\alpha_f}$ in Eq.~(\ref{LS_AC})};
	\draw (-2,-3.65) node(E5)[anchor=west]{(c) Correct solution:};
	\draw (-1.5,-4.15) node(E6)[anchor=west]{$\phi^\mathrm{n+\alpha_f}_\mathrm{(k+1)} = \phi^\mathrm{n+\alpha_f}_\mathrm{(k+1)} + \Delta\phi^\mathrm{n+\alpha_f}$};	
	\draw (-1.5,-4.75) node(E7)[anchor=west]{ };	
	\draw (-2,-5.65) node(E8)[anchor=west]{(d) Update solution:};
	\draw (-1.5,-6.15) node(E9)[anchor=west]{$\phi^\mathrm{n+1}_\mathrm{(k+1)} = \phi^\mathrm{n} + \frac{1}{\alpha_\mathrm{f}}(\phi^\mathrm{n+\alpha_f}_\mathrm{(k+1)} - \phi^\mathrm{n})$};	
	\draw (-1.5,-6.75) node(E10)[anchor=west]{ };	
	\node[fit = (E) (E1) (E2) (E3) (E4) (E5) (E6) (E7) (E8) (E9) (E10),style={block4}](AC){};
\end{tikzpicture}
\end{tabbing}
\end{algorithm}
\vspace{-0.5cm}
\label{solver_schematic_2}
\end{figure}

\section{Numerical tests}
\label{tests}
In this section, we present some numerical tests to assess the scheme for the coupled Allen-Cahn and Navier-Stokes equations. In all the test cases, the location of the interface between the two phases of the fluid is evaluated by linearly interpolating the order parameter field $\phi$ and finding the interface where $\phi=0$.

\subsection{Verification of the Allen-Cahn implementation}
We first validate the Allen-Cahn solver using the volume-conserved motion by curvature in two-dimensions \cite{H_G_Lee}. A square computational domain $[0,1]\times [0,1]$ with varying element sizes is considered. Periodic boundary conditions are imposed on all the boundaries.
\begin{figure}[!htbp]
\centering
	\begin{subfigure}[b]{0.5\textwidth}
\qquad
\begin{tikzpicture}[decoration={markings,mark=at position 1.0 with {\arrow{>}}},scale=5.9]
	\draw (0,0) -- (1,0)-- (1,1) -- (0,1) -- cycle;
	\draw[fill={rgb:black,1;white,2}, fill opacity=0.4] (0.25,0.25) circle (0.1cm);
	\draw[fill={rgb:black,1;white,2}, fill opacity=0.4] (0.57,0.57) circle (0.15cm);
	\draw (0.25,0.25) node(A){$\Omega_1$};
	\draw (0.57,0.57) node(B){$\Omega_1$};
	\node [below=0.3cm of A]{Circle 1};
	\node [below=0.6cm of B]{Circle 2};
	\draw (0.75,0.25) node{$\Omega_2$};
	\draw[thick,postaction={decorate}] (0,0) to (0.2,0);
	\draw[thick,postaction={decorate}] (0,0) to (0,0.2);
	\draw (0.2,0) node[anchor=north]{X};
	\draw (0,0.2) node[anchor=east]{Y};
	\draw[postaction={decorate}] (1.05,0.5) to (1.05,1);
	\draw[postaction={decorate}] (1.05,0.5) to (1.05,0);
	\draw (1.01,0) -- (1.1,0);
	\draw (1.01,1) -- (1.1,1);
	\draw (1.05,0.5) node[anchor=west]{$1$};
	\draw[postaction={decorate}] (0.5,1.05) to (1,1.05);
	\draw[postaction={decorate}] (0.5,1.05) to (0,1.05);
	\draw (0,1.01) -- (0,1.1);
	\draw (1,1.01) -- (1,1.1);
	\draw (0.5,1.05) node[anchor=south]{$1$};
\end{tikzpicture}
	\caption{}
	\end{subfigure}%
	\begin{subfigure}[b]{0.5\textwidth}
		\includegraphics[trim={10cm 0 11.8cm 0cm},clip,width=7cm]{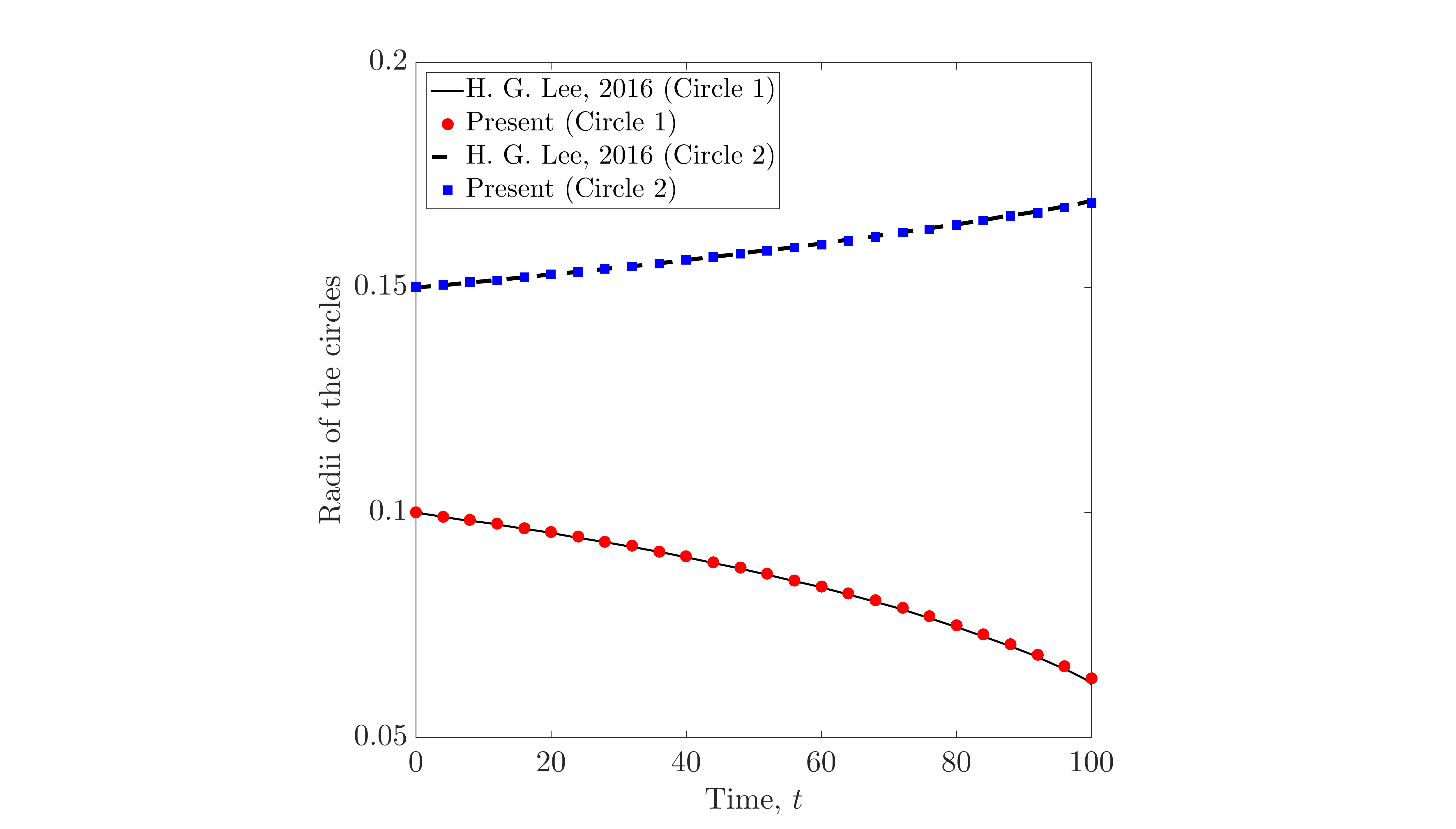}		
	\caption{}
	\end{subfigure}

	\begin{subfigure}[b]{0.5\textwidth}
	\ \ \
		\includegraphics[trim={0.2cm 0.2cm 0.2cm 0.2cm},clip,width=8.5cm]{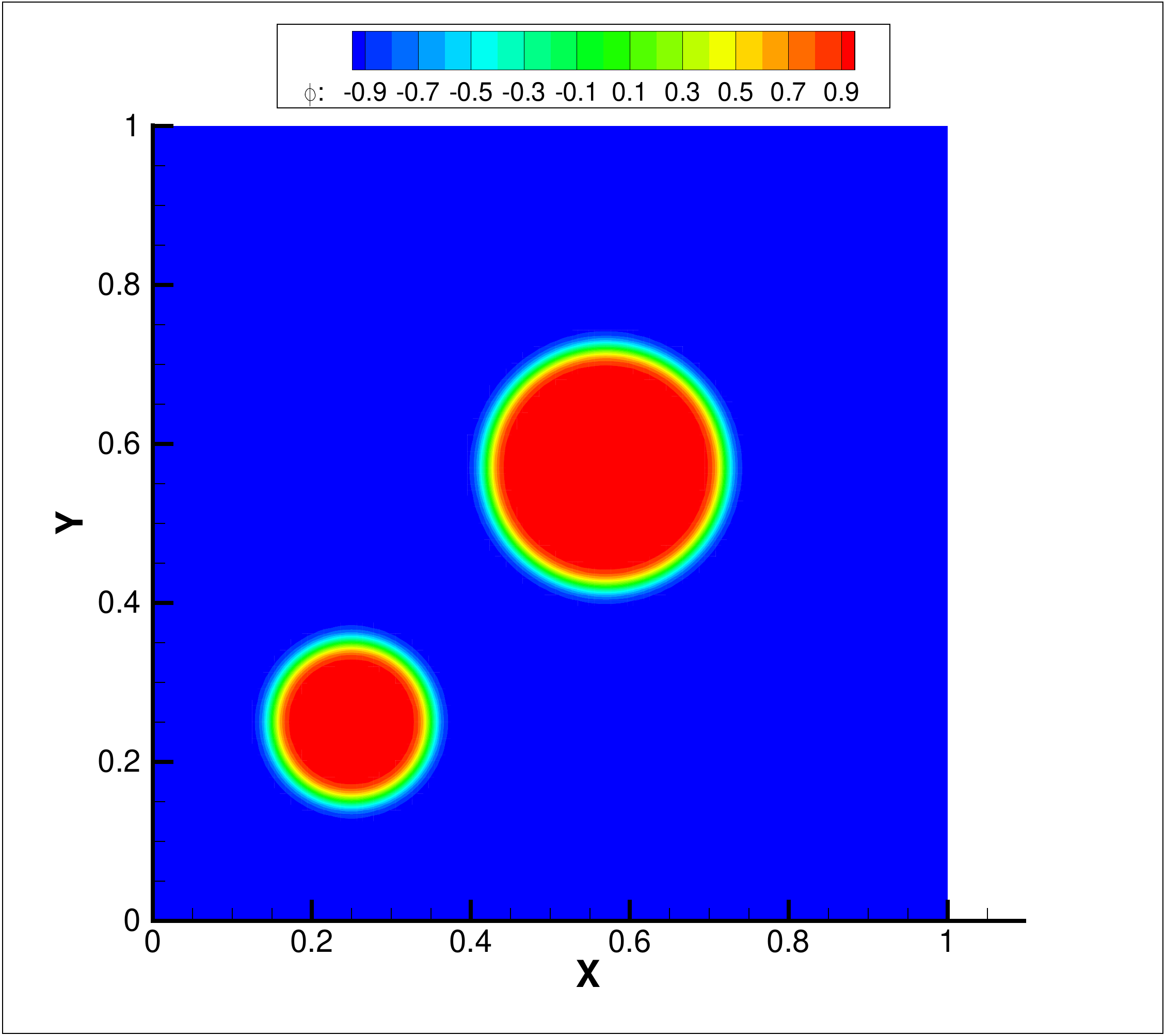}
	\caption{}
	\end{subfigure}%
	\begin{subfigure}[b]{0.5\textwidth}
		\includegraphics[trim={0.2cm 0.2cm 0.2cm 0.2cm},clip,width=8.5cm]{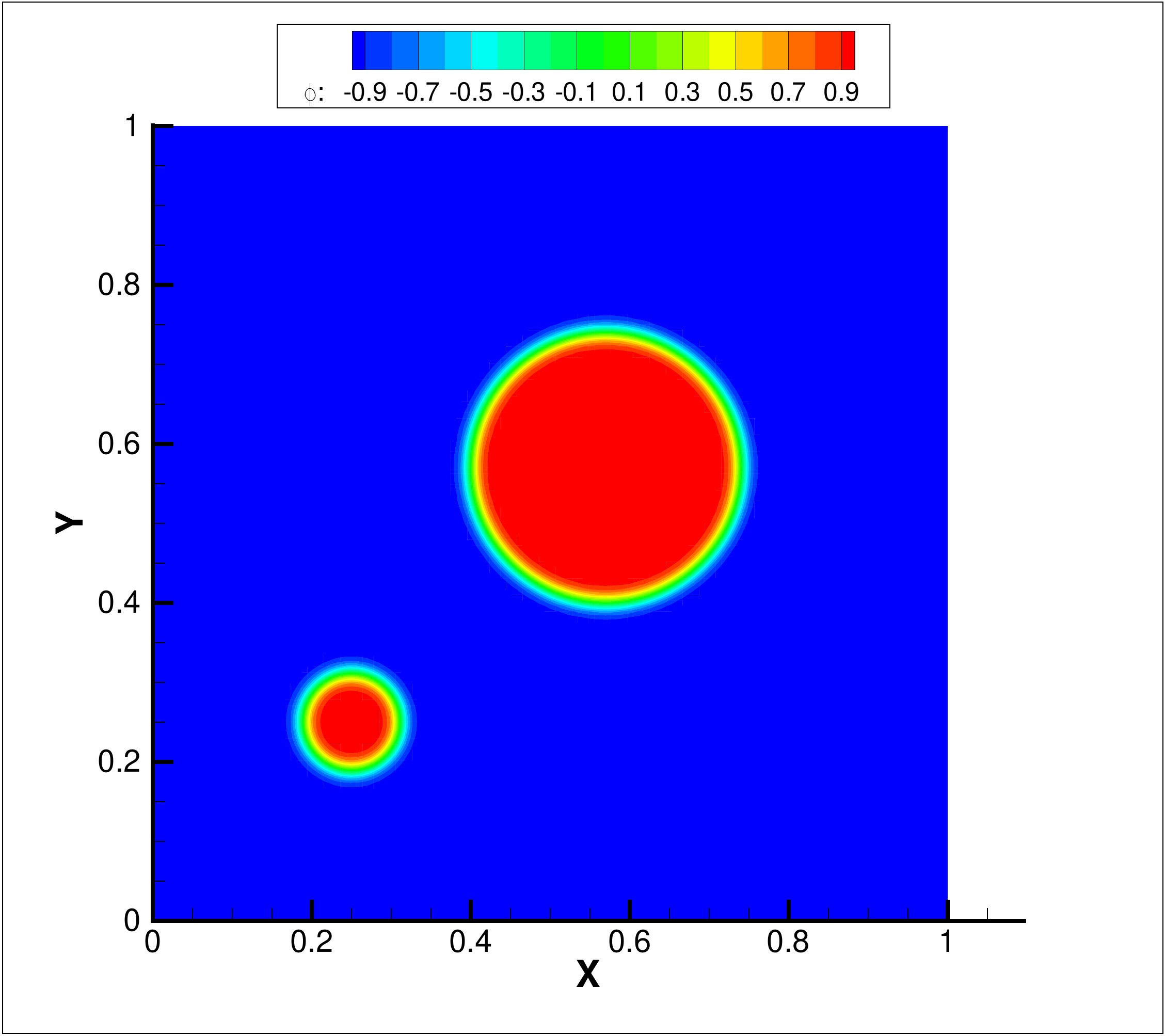}
	\caption{}
	\end{subfigure}
\caption{Evolution of the radii of two-dimensional circles: (a) Schematic diagram showing the computational domain, (b) Validation of the evolution of the radii of the two circles with the literature \cite{H_G_Lee}, and the contour plots of the order parameter $\phi$ at (c) $t=0$, and (d) $t=100$.  In (a), $\Omega_1$ and $\Omega_2$ are the two phases with periodic boundary conditions imposed on all the sides. } 
\label{ac_val}
\end{figure}
The initial condition is given by:
\begin{align}
	\phi(x,y,0) = 1 + \mathrm{tanh}\bigg( \frac{R_1 - \sqrt{(x-0.25)^2 + (y-0.25)^2}}{\sqrt{2}\varepsilon} \bigg) + \mathrm{tanh}\bigg( \frac{R_2 - \sqrt{(x-0.57)^2 + (y-0.57)^2}}{\sqrt{2}\varepsilon} \bigg)
\end{align}
where $R_1=0.1$ and $R_2=0.15$ are the radii of the two circles centered at $(0.25,0.25)$ and $(0.57,0.57)$, respectively. The variation of the change of the radii of the two circles is tracked and validated with the results obtained in \cite{H_G_Lee} for $\varepsilon=0.01$. The time step size in the present simulation is $0.1$ with the final time $t=100$. The problem set-up with the evolution of the radii of the two circles is shown in Fig. \ref{ac_val}. The results are in very close agreement with the reference.

Further analysis is carried out to quantify the appropriate number of elements required across the interface to capture the evolution of the radii. Let $N_\varepsilon$ denote the number of elements across the equilibrium interface thickness, $\varepsilon_\mathrm{eqm}= 4.164\varepsilon$. We simulate the evolution of the radii for $N_\varepsilon \in [3,10]$ and quantify the percentage error ($e_1$) in the radii at the final time $t=100$ defined as
\begin{align}
	e_1 = \frac{|R_i - R_\mathrm{ref}|}{R_\mathrm{ref}} \times 100,
\end{align}
where $R_i$ is the radius corresponding to different resolutions at $t=100$ and $R_\mathrm{ref}$ is the radius obtained for $N_{\varepsilon}=10$ at $t=100$.
The results are summarized in Table \ref{table_ac}. It can be observed that for circle 2 (the large circle), $N_{\varepsilon}>3$ is sufficient for getting the solution within $1\%$ error while for circle 1, $N_{\varepsilon}$ should be $>6$ to obtain results with similar accuracy. The reason is the different curvature of the two circles. With the decreasing radius of circle 1, its curvature increases and more number of elements are needed to sufficiently capture its interface.
\renewcommand{\arraystretch}{0.5}
\begin{table}[!h]
\caption{Percentage error in the final radii of the two circles for different mesh resolutions}
\centering
\begin{tabular}{  M{3cm}  M{1.5cm}  M{1.5cm}  M{1.5cm}  M{1.5cm}  M{1.5cm} M{1.5cm} N }
	\hline
\centering
	\textbf{$N_{\varepsilon}$}$\rightarrow$ & $3$ & $4$ & $5$ & $6$ & $7$ & $8$ &\\[10pt]
	\hline
\centering
	 Circle 1 & 10.39 & 5.29 & 3.26 & 1.54 & 0.91 & 0.58  &\\[10pt]

\centering
	 Circle 2 & 1.28 & 0.60 & 0.36 & 0.20 & 0.15 & 0.02 &\\[10pt]
	\hline
\end{tabular}
\label{table_ac}
\end{table}

To understand the behavior of the solver under spatial and temporal refinements, we evaluate the $L^2$ error of the solution over the whole domain at $t=50$ by decreasing element and time step sizes. The $L^2$ error is calculated as:
\begin{align}
	e_2 = \frac{||\Phi-\Phi_\mathrm{ref}||_2}{||\Phi_\mathrm{ref}||_2},
\end{align}
where $\Phi$ is the vector of the solution of order parameter $\phi$ at the final time $t=50$ over the whole domain for the respective refinement, $\Phi_\mathrm{ref}$ is the vector of the solution of the finest resolution and $||\cdot||_2$ is the $L^2$ norm. The spatial refinement is carried out uniformly by decreasing the element size from $h=1/72$ until $h=1/2304$. The finest grid $h=1/2304$ is selected as the reference solution for evaluating the error norm. We take $\Delta t=0.025$ and decrease it by a factor of $2$ until $\Delta t=3.90625\times 10^{-4}$. In a similar manner, the finest $\Delta t$ is taken as the reference solution. 
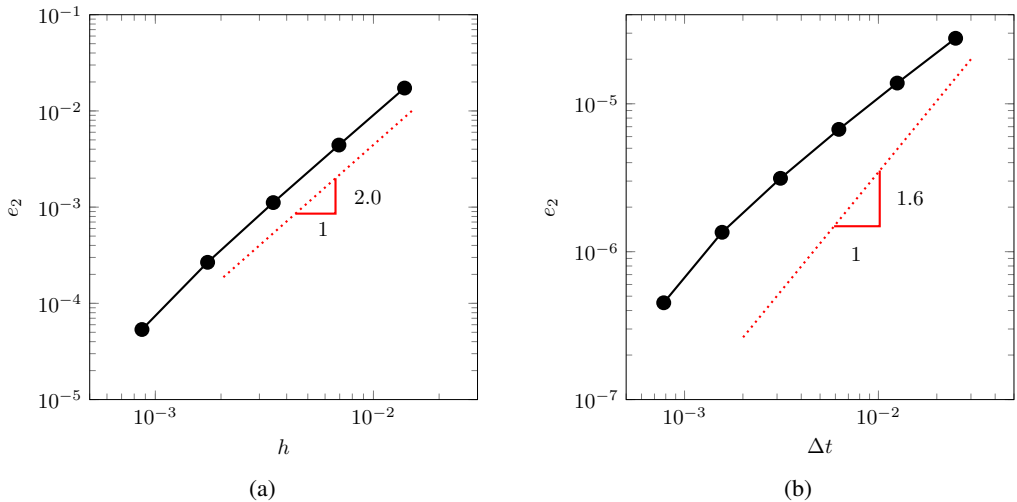
\begin{figure}[H]
\centering
\begin{tikzpicture}[scale=0.8]
\begin{loglogaxis}[
    width=0.5\textwidth,
    height=0.5\textwidth,
    xlabel={$h$},
    ylabel={$e_2$},
    xmin=5e-4, xmax=0.03,
    ymin=1e-5, ymax=0.1,
    legend pos=north west,
]
\addplot[mark=*,black,mark size={3pt},line width={1pt}]
    coordinates {
    (1/72,0.017285213259038)(1/144,0.004418405301747)(1/288,0.001115072427403)(1/576,2.669733237974486e-04)(1/1152,5.346981682268451e-05)
    };
\addplot[dotted,red,mark size={3pt},line width={1pt}]
    coordinates {
    (0.015,0.01)(0.002,1.778e-4)
    }
    coordinate [pos=0.4] (A)
    coordinate [pos=0.6] (B)
    coordinate [pos=0.5] (C)
    coordinate [pos=0.7] (D);
\draw[red,line width={1pt}] (A) |- (B) ;
\draw (C) node[anchor=west,black]{$\ \ \ \ \ \ 2.0$};
\draw (D) node[anchor=west,black]{$\quad\ \ 1$};
\end{loglogaxis}
\draw (3.2,-1.5) node[anchor=east,black,scale=0.9]{(a)};
\end{tikzpicture}
\qquad
\begin{tikzpicture}[scale=0.8]
\begin{loglogaxis}[
    width=0.5\textwidth,
    height=0.5\textwidth,
    xlabel={$\Delta t$},
    ylabel={$e_2$},
    xmin=5e-4, xmax=0.05,
    ymin=1e-7, ymax=4e-5,
    legend pos=north west,
]
\addplot[mark=*,mark options={solid}, black,mark size={3pt},line width={1pt}]
    coordinates {
    (0.025,2.771611524155344e-05)(0.0125,1.380186873448174e-05)(6.25e-3,6.712093119755828e-06)(3.125e-3,3.133188127919196e-06)(1.5625e-3,1.352202843113804e-06)(7.8125e-4,4.513311653281769e-07)
    };
\addplot[dotted,red,mark size={3pt},line width={1pt}]
    coordinates {
    (0.03,2e-5)(2e-3,2.626e-7)
    }
    coordinate [pos=0.4] (A)
    coordinate [pos=0.6] (B)
    coordinate [pos=0.5] (C)
    coordinate [pos=0.7] (D);
\draw[red,line width={1pt}] (A) |- (B) ;
\draw (C) node[anchor=west,black]{$\ \ \ \ \ \ 1.6$};
\draw (D) node[anchor=west,black]{$\quad\ \ 1$};
\end{loglogaxis}
\draw (3.2,-1.5) node[anchor=east,black,scale=0.9]{(b)};
\end{tikzpicture}
\caption{Convergence study for the present method through dependence of non-dimensionalized $L^2$ error $(e_2)$ as a function of: (a) uniform mesh refinement $h$, and (b) uniform temporal refinement $\Delta t$}
\label{error_1}
\end{figure}

The mesh and temporal convergence are plotted in Fig. \ref{error_1}. Some of the peculiar observations from the convergence plots are: (i) the spatial convergence is of second-order with the mesh refinement, and (ii) temporal convergence initially starts with first-order, but increases to a value of $1.6$ as time step is further refined. The order of temporal convergence is not exactly second-order due to the nonlinear energy-stable properties which are imparted by the mid-point approximation in the discretization.

Finally, to analyze the conservation of mass or the order parameter $\phi$, we quantify the percentage change of the mass across the time domain of the experiment compared with the initial mass, which comes out as $5.4577 \times 10^{-5} ~\%$. Note that the mass conservation depends on the tolerance of the linear GMRES solver as well as the nonlinear tolerance. A linear tolerance of $10^{-15}$ with nonlinear tolerance of $10^{-4}$ are chosen for this experiment. Within this limit, it can be concluded that the mass is conserved for the presented scheme.

\subsection{Laplace-Young law}
To verify the coupling between the Allen-Cahn and the incompressible Navier-Stokes equations at high density ratio, we carry out a test to verify the Laplace-Young law. The law states that the pressure difference ($\Delta p$) across the interface of a static bubble in a two-phase fluid system is equivalent to the ratio of the surface tension ($\sigma$) and the radius of curvature ($R$) of the bubble,
\begin{align}
	\Delta p = p_\mathrm{in} - p_\mathrm{out} = \frac{\sigma}{R}.
\end{align} 
In the equilibrium state, the velocity vanishes $\boldsymbol{u}=\mathbf{0}$ and the pressure gradient balances the surface tension force, i.e.,
\begin{align}
	\nabla p = \sigma\varepsilon\frac{3\sqrt{2}}{4}\nabla\cdot( |\nabla\phi|^2\mathbf{I} - \nabla\phi \otimes \nabla\phi ). 
\end{align}
\begin{figure}[H]
\centering
	\begin{subfigure}[b]{0.5\textwidth}
\qquad
\begin{tikzpicture}[decoration={markings,mark=at position 1.0 with {\arrow{>}}},scale=5.9]
	\draw[fill={black!10}] (0,0) -- (1,0)-- (1,1) -- (0,1) -- cycle;
	\draw[fill=white] (0.5,0.5) circle (0.15cm);
	\draw (0.5,0.5) node(A){$\Omega_2$};
	\node [below = -0.2cm of A]{$(\rho_2, \mu_2)$};
	\draw (0.75,0.75) node(B){$\Omega_1$};
	\node [below = -0.2cm of B]{$(\rho_1, \mu_1)$};
	\draw[thick,postaction={decorate}] (0,0) to (0.2,0);
	\draw[thick,postaction={decorate}] (0,0) to (0,0.2);
	\draw (0.2,0) node[anchor=north]{X};
	\draw (0,0.2) node[anchor=east]{Y};
	\draw[postaction={decorate}] (1.05,0.5) to (1.05,1);
	\draw[postaction={decorate}] (1.05,0.5) to (1.05,0);
	\draw (1.01,0) -- (1.1,0);
	\draw (1.01,1) -- (1.1,1);
	\draw (1.05,0.5) node[anchor=west]{$4$};
	\draw[postaction={decorate}] (0.5,1.05) to (1,1.05);
	\draw[postaction={decorate}] (0.5,1.05) to (0,1.05);
	\draw (0,1.01) -- (0,1.1);
	\draw (1,1.01) -- (1,1.1);
	\draw (0.5,1.05) node[anchor=south]{$4$};
	\draw (0.35,0.35) -- (0.35,0.23);
	\draw (0.65,0.35) -- (0.65,0.23);
	\draw[<->] (0.35,0.26) -- (0.65,0.26);
	\draw (0.5,0.26) node[anchor=south] {$2R$};
\end{tikzpicture}
	\caption{}
	\end{subfigure}%
	\begin{subfigure}[b]{0.5\textwidth}
		\includegraphics[trim={10cm 0 11.8cm 0cm},clip,width=7cm]{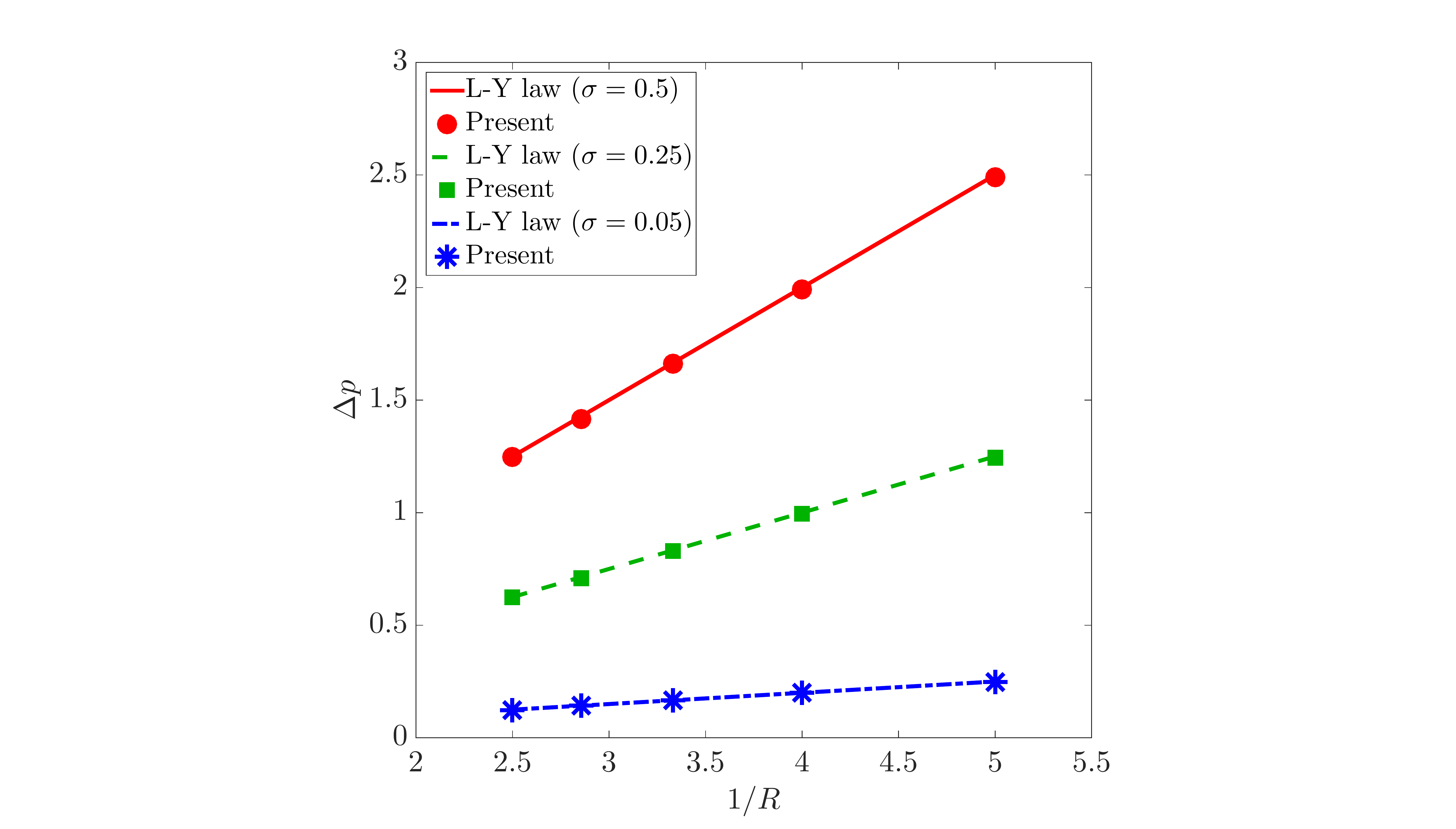}
	\caption{}
	\end{subfigure}
\caption{Laplace-Young law: (a) Schematic diagram showing the computational domain, (b) Comparison of the pressure difference across the interface in a static bubble obtained from the simulation with the Laplace-Young law. In (a), $\Omega_1$ and $\Omega_2$ are the two fluid phases with densities $\rho_1=1000$ and $\rho_2=1$, viscosities $\mu_1=10$, $\mu_2=0.1$ and acceleration due to gravity $\boldsymbol{g}=(0,0,0)$ and all the boundaries have periodic boundary condition.} 
\label{Val_1}
\end{figure}

The density and dynamic viscosity of the two fluids are taken as $\rho_1=1000$, $\rho_2=1$, $\mu_1=10$ and $\mu_2=0.1$. In the numerical tests, we consider a domain size $\Omega = [0,4]\times [0,4]$ with uniform structured mesh of grid size $1/200$ with different radii of the bubble ($0.2$, $0.25$, $0.3$, $0.35$ and $0.4$ units) and three surface tensions ($0.5$, $0.25$ and $0.05$ units). Periodic boundary conditions are applied on all the boundaries. The initial condition is given by:
\begin{align}
	\phi(x,y,0) = -\mathrm{tanh}\bigg( \frac{R - \sqrt{(x-x_c)^2 + (y-y_c)^2}}{\sqrt{2}\varepsilon} \bigg),
\end{align}
where $R$ is the radius of the bubble with its centre at $(x_c,y_c)=(2,2)$. The interface thickness parameter is $\varepsilon=0.01$. The time step size is taken as $\Delta t=0.01$s and the pressure difference is measured after $5000$ time steps. The schematic of the computational domain and the representative results are shown in Fig. \ref{Val_1}. The pressure difference in Fig. \ref{Val_1}(b) shows good agreement with the Laplace-Young law. This verifies the coupling between the Allen-Cahn and the Navier-Stokes equations at high density ratio.

\subsection{Sloshing tank problem}
\begin{figure}[H]
\centering
	\begin{subfigure}[b]{0.5\textwidth}
\qquad
\begin{tikzpicture}[decoration={markings,mark=at position 1.0 with {\arrow{>}}},scale=5]
	\draw[fill=black!10] plot[smooth] coordinates {(0,0.91) (0.1,0.9149) (0.2,0.9291) (0.3,0.9512) (0.4,0.9791) (0.5,1.01) (0.6,1.0409) (0.7,1.0688) (0.8,1.0909) (0.9,1.1051) (1.0,1.11)} -- (1,0) -- (0,0) -- (0,0.91);
	\draw (0,0.91) -- (0,1.5)-- (1,1.5) -- (1,1.11);
	\draw (0.5,0.75) node[anchor=south](A){$\Omega_1$};
	\node [below = 0.0cm of A]{$(\rho_1, \mu_1)$};
	\draw (0.5,1.25) node[anchor=south](B){$\Omega_2$};
	\node [below = 0.0cm of B]{$(\rho_2, \mu_2)$};
	\draw (0.2,0.92) node[anchor=south]{$\Gamma$};
	\draw[thick,postaction={decorate}] (0,0) to (0.2,0);
	\draw[thick,postaction={decorate}] (0,0) to (0,0.2);
	\draw (0.2,0) node[anchor=north]{X};
	\draw (0,0.2) node[anchor=east]{Y};
	\draw[postaction={decorate}] (1.05,0.5) to (1.05,1.5);
	\draw[postaction={decorate}] (1.05,0.5) to (1.05,0);
	\draw (1.01,0) -- (1.1,0);
	\draw (1.01,1.5) -- (1.1,1.5);
	\draw (1.05,0.75) node[anchor=west]{$1.5$};
	\draw[postaction={decorate}] (0.5,1.55) to (1,1.55);
	\draw[postaction={decorate}] (0.5,1.55) to (0,1.55);
	\draw (0,1.51) -- (0,1.6);
	\draw (1,1.51) -- (1,1.6);
	\draw (0.5,1.55) node[anchor=south]{$1$};
	\draw (0.5,0) node{};
\end{tikzpicture}
	\caption{}
	\end{subfigure}%
	\begin{subfigure}[b]{0.5\textwidth}
		\includegraphics[trim={0.2cm 1.0cm 9cm 0.2cm},clip,width=6.5cm]{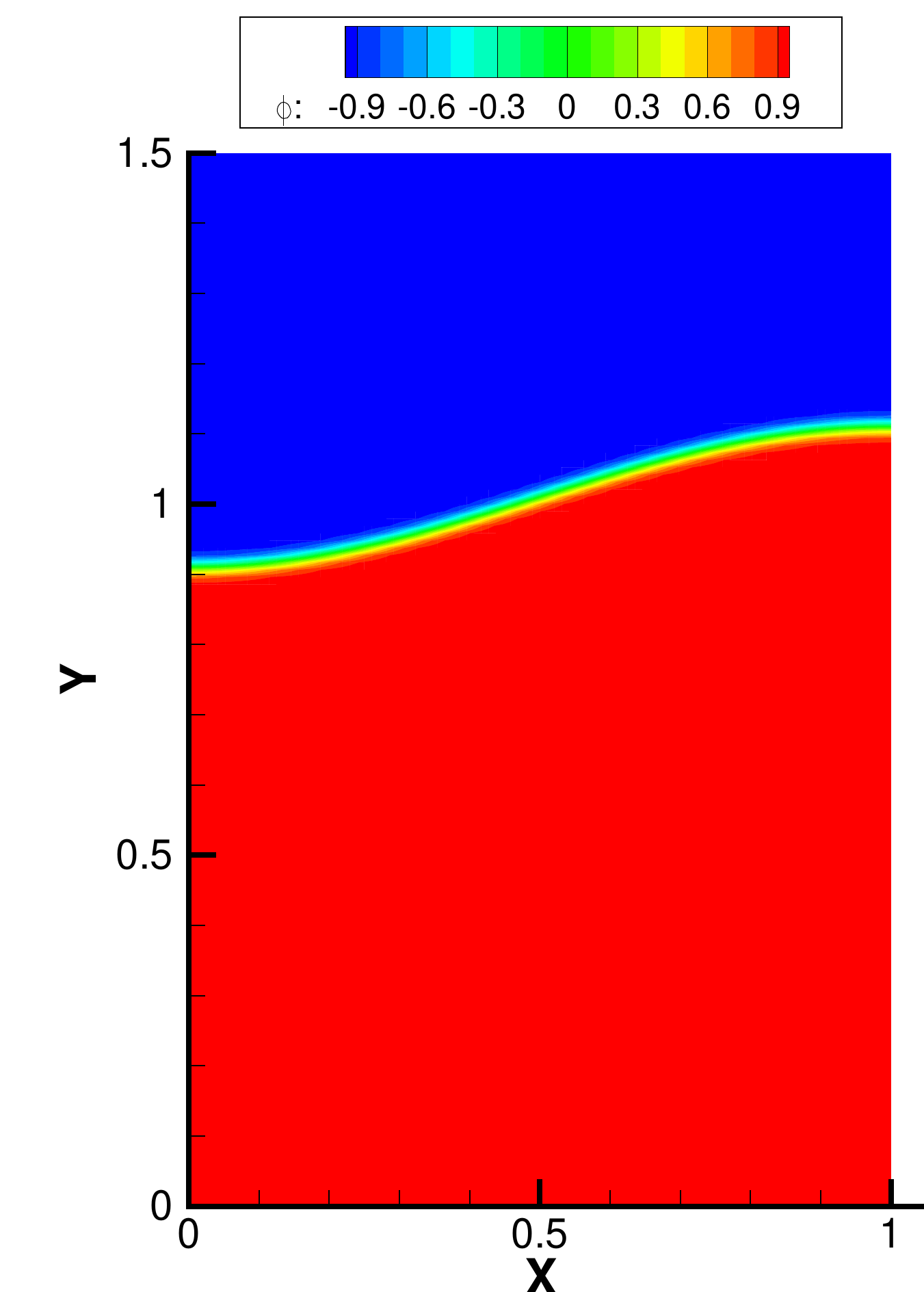}
	\caption{}
	\end{subfigure}
\caption{Sloshing in a rectangular tank: (a) Schematic diagram showing the computational domain,  (b) Contour plot of the order parameter $\phi$ at $t=0$. 
In (a), $\Omega_1$ and $\Omega_2$ are the two fluid phases with densities $\rho_1=1000$ and $\rho_2=1$, viscosities $\mu_1=1$, $\mu_2=0.01$ and acceleration due to gravity $\boldsymbol{g}=(0,-1,0)$ and the walls of the tank have slip boundary condition with $p=0$ at the upper boundary.
} 
\label{ST_1}
\end{figure}

To further validate the coupling, we test a standard sloshing tank problem. A rectangular computational domain $\Omega \in [0,1] \times [0,1.5]$ is considered for the simulation. It is discretized with different resolutions characterized by the number of elements in the equilibrium interface thickness denoted by $N_\varepsilon$. Varying densities and viscosities of the two fluid phases as $\rho_1 = 1000$, $\rho_2 = 1$, $\mu_1 = 1$, $\mu_2 = 0.01$ and $\boldsymbol{g} = (0, -1, 0)$ will also test the ability of the solver to handle high density and viscosity ratios. The capillary effects due to surface tension have been neglected in this test case. The initial condition is defined as:
\begin{align}
	\phi(x,y,0) = - \mathrm{tanh}\bigg( \frac{ y - (1.01 + 0.1\mathrm{sin}((x - 0.5)\pi)) }{\sqrt{2}\varepsilon}\bigg)
\end{align}
Slip boundary condition is set along the walls and a Dirichlet boundary condition is prescribed at the top boundary as $p=0$. The problem set-up is given in Fig. \ref{ST_1}(a) with the contour plor of $\phi$ of the initial condition in Fig. \ref{ST_1}(b).

To have a detailed analysis of the problem, we perform a series of experiments to assess: (a) the effectiveness of the PPV technique, (b) appropriate number of elements ($N_\varepsilon$) required in the equilibrium interface thickness, (c) proper value of $\Delta t$ to obtain results with sufficient accuracy, and (d) the effect of $\varepsilon$ on the two-phase flow solution. 
These convergence and sensitivity studies are hereby presented.
The quantification of error in the following analysis is carried out by evaluating the $L^2$ error $e_3$ defined by
\begin{align}
	e_3 = \frac{|| \Phi - \Phi_\mathrm{ref} ||_2}{|| \Phi_\mathrm{ref} ||_2},
\end{align}
where $\Phi$ is the temporal solution of the interface elevation at the left boundary, $\Phi_\mathrm{ref}$ is the solution of the finest resolution associated with the respective study and $||\cdot||_2$ is the $L^2$ norm.

\subsubsection{Effectiveness of the PPV technique}
First, we emphasize the advantage of using the PPV-based technique for the phase-field equation. In the convection- and reaction-dominated regions, the linear stabilized variational formulation results in oscillations near the regions with high gradients, which can result into unbounded solution. However, the order parameter $\phi$ solved by the Allen-Cahn equation needs to be bounded at all times to ensure that positive values of density and viscosity are transferred to the Navier-Stokes equations. The present variational technique detects the regions of high gradients depending on the residual of the equation and adds diffusion to those regions to eliminate such oscillations.  Further detailed analyses and results in one- and two-dimensions can be found in \cite{PPV}. 

\begin{figure}[H]
		\centering
		\includegraphics[trim={0cm 0.3cm 0cm 0cm},clip,width=10cm]{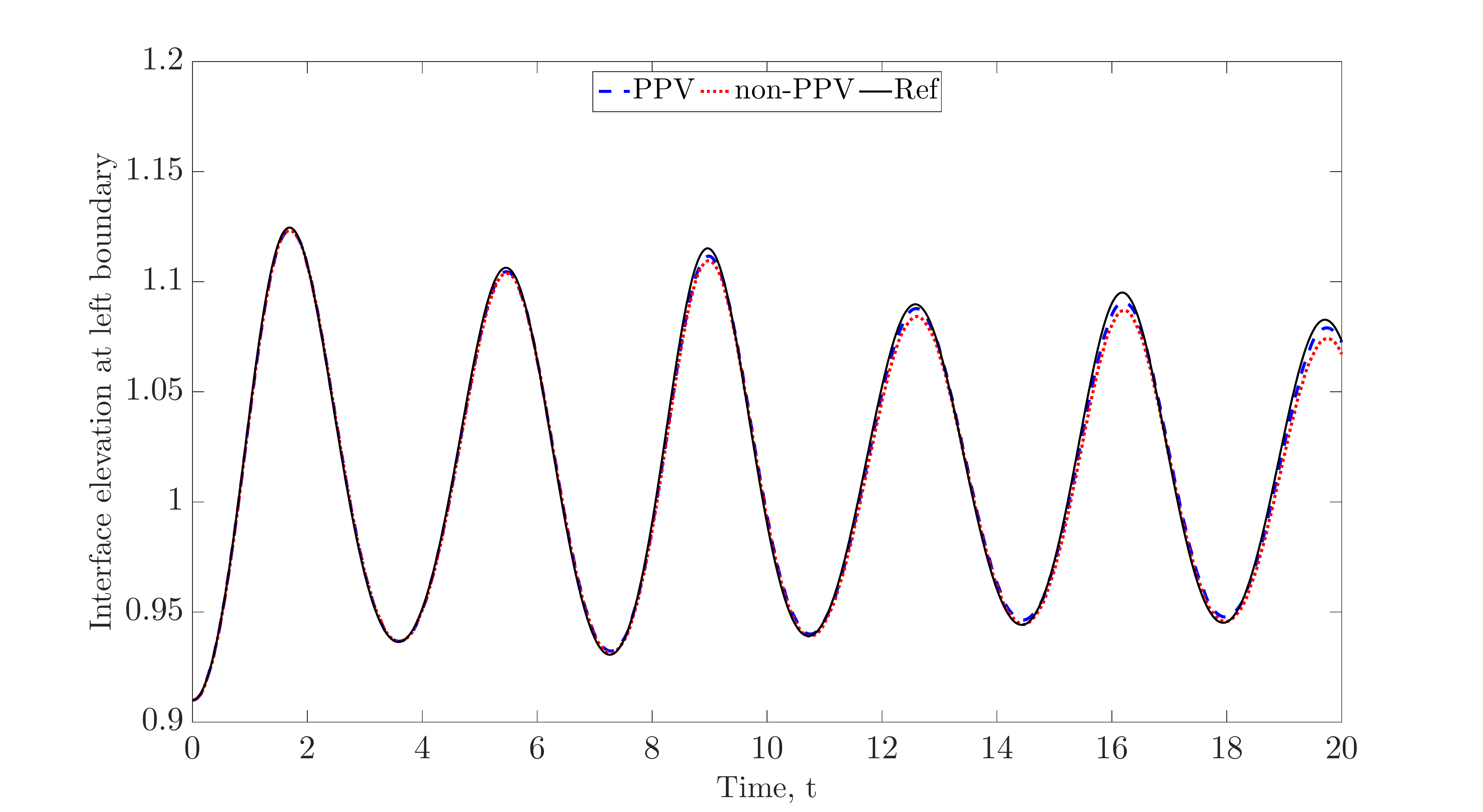}
\caption{Sloshing tank problem: Effect of using the PPV technique on the evolution of the interface. $\varepsilon = 0.01$, $\Delta t=0.001$ and $N_\varepsilon=3$. The reference solution is obtained at $N_\varepsilon=8$.} 
\label{ppv_comp}
\end{figure}

To demonstrate the effect of the PPV technique to the present problem, we simulate two test cases: (a) with nonlinear PPV stabilization term and (b) without nonlinear PPV term. $\Delta t = 0.001$, $\varepsilon=0.01$ and $N_\varepsilon=3$ and $N_\varepsilon=4$ are considered for the tests. The minimum and maximum values of the solution of the interface evolution at the left boundary are evaluated to quantify the bounds of the numerical solution. The values are summarized in Table \ref{table_ppv_comp}. From the table, it is evident that the current formulation reduces the oscillations and preserves the boundedness and positivity in the solution. The solution obtained are also compared in Fig. \ref{ppv_comp} for $N_\varepsilon=3$. The reference solution is the solution for $N_\varepsilon=8$, $\Delta t=0.001$ and $\varepsilon=0.01$. The percentage error for the non-PPV-based solution is $0.3548 \%$ and that for the PPV-based solution is $0.1905\%$. For the case when the nonlinear PPV term is not used, the solution is unbounded. However, for evaluating the density and viscosity values, we chop off the values larger than $1$ and smaller than $-1$ to get physical values for density and viscosity. We observe from Fig. \ref{ppv_comp} that this chopping leads to less accurate solution compared to the reference solution, while the PPV-based solution is more accurate. 
\renewcommand{\arraystretch}{0.5}
\begin{table}[H]
\caption{Bounds in the solution of interface evolution at the left boundary for different techniques}
\centering
\begin{tabular}{  M{3cm}  M{1.5cm}  M{1.5cm}  M{1.5cm}  M{1.5cm}  N }
	\hline
\centering
	\textbf{$N_{\varepsilon}$} & \multicolumn{2}{c}{$\mathrm{min(\phi)}$} & \multicolumn{2}{c}{$\mathrm{max(\phi)}$} &\\[10pt]
	\hline
\centering
	  & non-PPV & PPV & non-PPV & PPV  &\\[10pt]
	\hline
\centering
	3  & -1.0544 & -1.0 & 1.0154 & 0.9999  &\\[10pt]
	\hline
\centering
	4  & -1.0065 & -1.0 & 1.0023 & 0.9999  &\\[10pt]
	\hline
\end{tabular}
\label{table_ppv_comp}
\end{table}
\vspace{-0.5cm}

\subsubsection{Effect of the number of elements in the equilibrium interfacial thickness ($N_\varepsilon$)}
\begin{figure}[H]
		\centering
		\includegraphics[trim={0cm 0.3cm 0cm 0cm},clip,width=10cm]{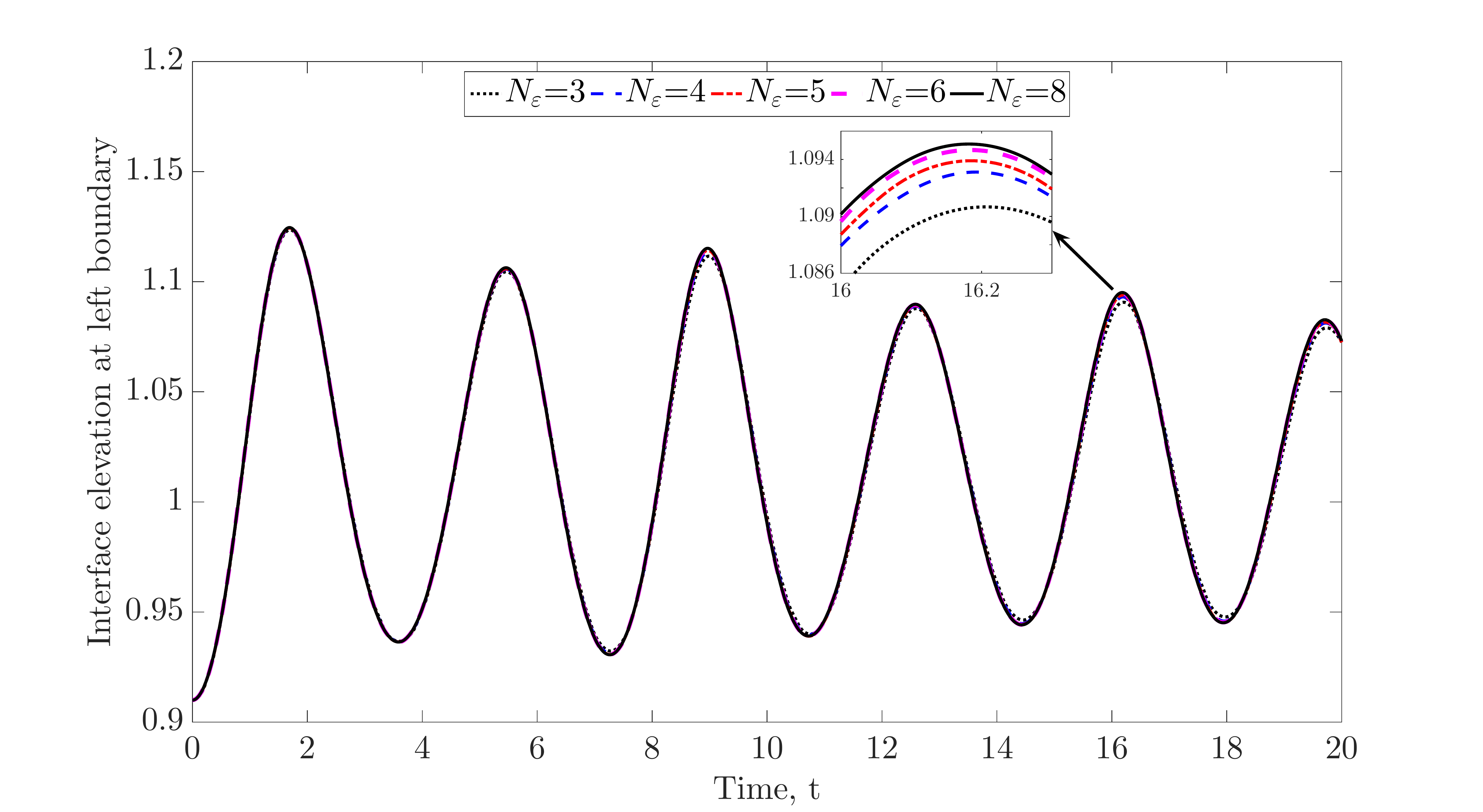}
\caption{Sloshing tank problem: Effect of $N_\varepsilon$ on the evolution of the interface. $\varepsilon = 0.01$ and $\Delta t=0.001$ are kept constant.} 
\label{ST_N_eps}
\end{figure}

We systematically perform mesh convergence studies to quantify the sufficient resolution needed to obtain accurate solution. The mesh resolution is characterized by $N_\varepsilon$. Figure \ref{ST_N_eps} shows the evolution of the interface at the left boundary of the domain with different $N_\varepsilon$ with fixed $\varepsilon=0.01$ and $\Delta t=0.001$. The error is quantified in Table \ref{table_N_eps} with the reference solution taken as that with $N_\varepsilon = 8$.
\vspace{-0.25cm}
\renewcommand{\arraystretch}{0.5}
\begin{table}[H]
\caption{Error ($e_3$) in the solution for different $N_\varepsilon$}
\centering
\begin{tabular}{  M{3cm}  M{1.5cm}  M{1.5cm}  M{1.5cm}  M{1.5cm}  N }
	\hline
\centering
	\textbf{$N_{\varepsilon}$}$\rightarrow$ & $3$ & $4$ & $5$ & $6$ &\\[10pt]
	\hline
\centering
	 $e_3 (\times 10^{-3})$ & 2.491 & 1.094 & 0.564 & 0.208  &\\[10pt]
	\hline
\end{tabular}
\label{table_N_eps}
\end{table}
\vspace{-0.5cm}
We conclude that $N_\varepsilon$ of 4 or 5 is sufficient to capture the interface. Increasing $N_\varepsilon$ gives more accurate solution however the difference is very small. Therefore, $N_\varepsilon$ of 4 or 5 seems to be a good compromise between accuracy and computational cost. 
\vspace{-0.5cm}

\subsubsection{Effect of the time step size ($\Delta t$)}
\begin{figure}[H]
\centering
		\includegraphics[trim={0cm 0.3cm 0cm 0cm},clip,width=10cm]{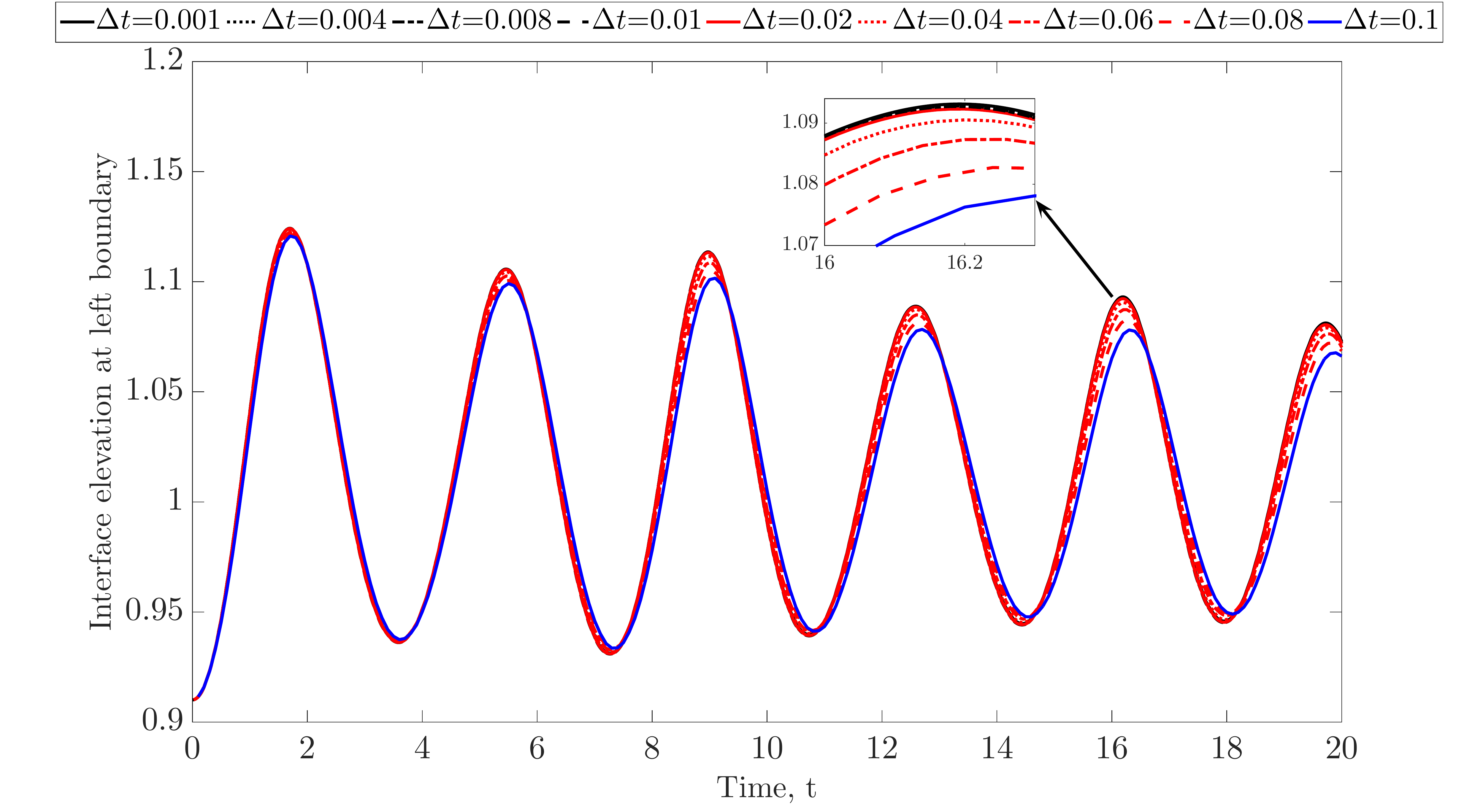}
\caption{Sloshing tank problem: Effect of $\Delta t$ on the evolution of the interface. $\varepsilon = 0.01$ and $N_\varepsilon=4$ are kept constant.} 
\label{ST_dt}
\end{figure}

To observe the effect of $\Delta t$ on the solution, we plot the time convergence plot in Fig. \ref{ST_dt} with $N_\varepsilon = 4$ and $\varepsilon = 0.01$. Varying time steps between $0.1$ and $0.001$ are chosen. The plot shows that the difference in the solution is noticeable till $\Delta t=0.01$, after which further reduction in the time step size has little effect on the accuracy of the solution. The error is shown in Table \ref{table_dt} with solution at $\Delta t=0.001$ as the reference solution.
\vspace{-0.25cm}
\renewcommand{\arraystretch}{0.5}
\begin{table}[H]
\caption{Error ($e_3$) in the solution for different $\Delta t$}
\centering
\begin{tabular}{  M{3cm}  M{1cm}  M{1cm}  M{1cm}  M{1cm}  M{1cm}  M{1cm}  M{1cm}  M{1cm} N }
	\hline
\centering
	\textbf{$\Delta t$}$\rightarrow$ & $0.1$ & $0.08$ & $0.06$ & $0.04$ & $0.02$ & $0.01$ & $0.008$ & $0.004$ &\\[10pt]
	\hline
\centering
	 $e_3 (\times 10^{-3})$ & 10.839 & 6.959 & 3.759 & 1.418 & 0.457 & 0.478 & 0.474 & 0.324  &\\[10pt]
	\hline
\end{tabular}
\label{table_dt}
\end{table}
\vspace{-0.5cm}

\subsubsection{Effect of the interfacial thickness parameter ($\varepsilon$)}
\begin{figure}[H]
\centering
		\includegraphics[trim={0cm 0.3cm 0cm 0cm},clip,width=10cm]{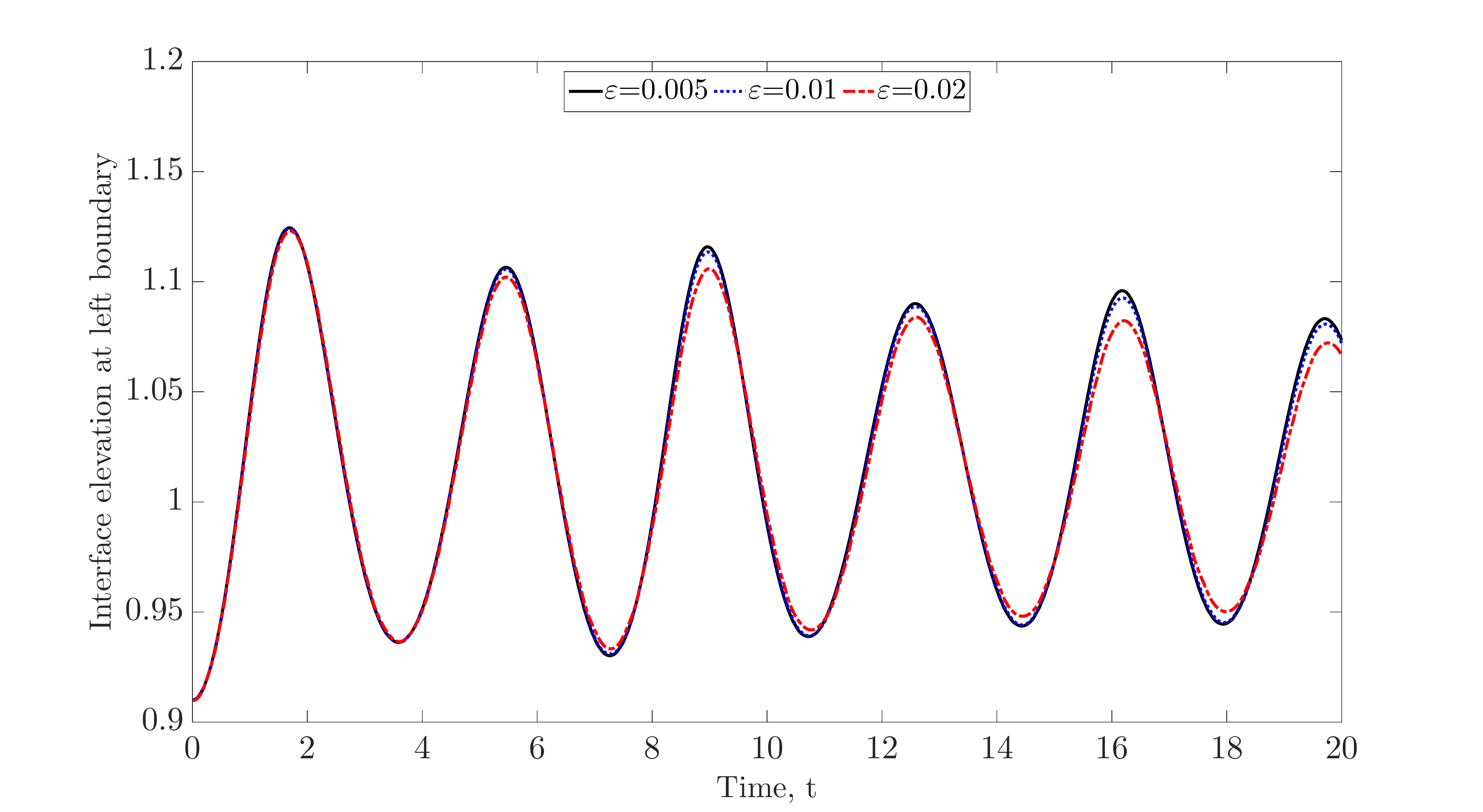}
\caption{Sloshing tank problem: Effect of $\varepsilon$ on the evolution of the interface. $\Delta t=0.01$ and $N_\varepsilon=4$ are kept constant.} 
\label{ST_eps}
\end{figure}

\begin{figure}[H]
\centering
		\includegraphics[trim={0cm 0.3cm 0cm 0cm},clip,width=10cm]{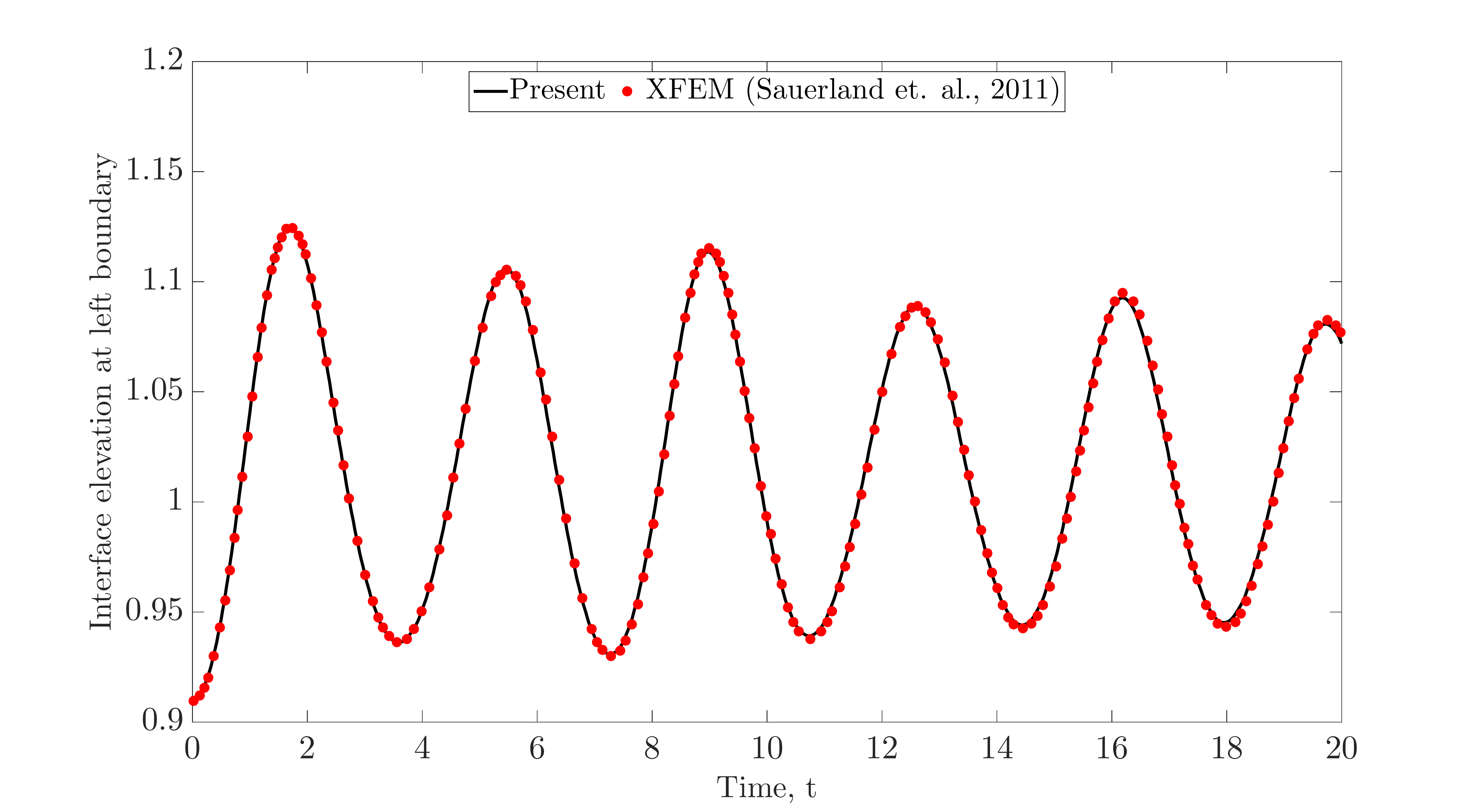}
\caption{Comparison of the solution considering $N_\varepsilon = 4$, $\Delta t=0.01$ and $\varepsilon = 0.01$ with XFEM-based level set approach in \cite{Fries_1}.} 
\label{ST_comp}
\end{figure}

The parameter $\varepsilon$ represents the thickness of the interface. The limit $\varepsilon \to 0$ gives the sharp-interface limit. Although decreasing $\varepsilon$ will indeed give more accurate representation of the interface, it will increase the computational cost due to the requirement of large number of elements in the equilibrium interfacial thickness region. Figure \ref{ST_eps} shows the evolution of the interface for different $\varepsilon$ values with fixed $\Delta t=0.01$ and $N_\varepsilon = 4$. The error is quantified in Table \ref{table_eps} with the solution at $\varepsilon=0.005$ taken as the reference solution. A selection of $\varepsilon=0.01$ is a good compromise between accuracy and cost of computation.
\vspace{-0.25cm}
\renewcommand{\arraystretch}{0.5}
\begin{table}[H]
\caption{Error ($e_3$) in the solution for different $\varepsilon$}
\centering
\begin{tabular}{  M{3cm}  M{2cm}  M{2cm} N }
	\hline
\centering
	\textbf{$\varepsilon$}$\rightarrow$ & $0.02$ & $0.01$ &\\[10pt]
	\hline
\centering
	 $e_3 (\times 10^{-3})$ & 5.469 & 1.269 &\\[10pt]
	\hline
\end{tabular}
\label{table_eps}
\end{table}
\vspace{-0.5cm}
\vspace{-0.5cm}
Finally, we compare and validate the results obtained from the current simulation considering $N_\varepsilon=4$, $\Delta t=0.01$ and $\varepsilon=0.01$ with those obtained using XFEM in \cite{Fries_1}. The method in \cite{Fries_1} employs the local enrichment of the pressure interpolation shape functions and solves the level-set equation to capture the interface with reinitialization procedure. The present method is much simpler to implement without performing any reinitialization or geometric reconstruction. From Fig. \ref{ST_comp}, we observe that sufficiently accurate results can be obtained by applying the diffuse-interface approach. Furthermore, the proposed algorithm consisting of single-pass explicit partitioned staggered coupling discussed in Section \ref{imp_details} helps to reduce the computational time without compromising with the accuracy and stability of the solution. For demonstrating the ability of the solver to handle topological changes of the interface, we simulate the two- and three-dimensional dam break problem for unstructured meshes.
\vspace{-0.25cm}

\subsection{Two-dimensional dam break problem}
\begin{figure}[H]
\centering\hspace{-0.6cm}
	\begin{subfigure}[b]{0.5\textwidth}
\begin{tikzpicture}[decoration={markings,mark=at position 1.0 with {\arrow{>}}},scale=11]
	\draw (0,0) -- (0.584,0)-- (0.584,0.438) -- (0,0.438) -- cycle;
	\draw[fill={rgb:black,1;white,2}, fill opacity=0.4] (0,0)--(0.146,0)--(0.146,0.252) arc (0:90:0.04)--(0,0.292)--(0,0);
	\draw (0.073,0.146) node(A){$\Omega_1$};
	\draw (0.292,0.146) node{$\Omega_2$};
	\draw[thick,postaction={decorate}] (0,0) to (0.05,0);
	\draw[thick,postaction={decorate}] (0,0) to (0,0.05);
	\draw (0.05,0) node[anchor=north]{X};
	\draw (0,0.05) node[anchor=east]{Y};
	\draw[postaction={decorate}] (0.624,0.219) to (0.624,0.438);
	\draw[postaction={decorate}] (0.624,0.219) to (0.624,0);
	\draw (0.594,0) -- (0.654,0);
	\draw (0.594,0.438) -- (0.654,0.438);
	\draw (0.624,0.219) node[anchor=west]{$0.438$};
	\draw[postaction={decorate}] (0.292,0.478) to (0.584,0.478);
	\draw[postaction={decorate}] (0.292,0.478) to (0,0.478);
	\draw (0,0.448) -- (0,0.508);
	\draw (0.584,0.448) -- (0.584,0.508);
	\draw (0.292,0.478) node[anchor=south]{$0.584$};
	\draw[postaction={decorate}] (0.166,0.146) to (0.166,0);
	\draw[postaction={decorate}] (0.166,0.146) to (0.166,0.292);
	\draw (0.156,0.292) -- (0.176,0.292);
	\draw (0.166,0.146) node[anchor=west]{$b$};
	\draw[postaction={decorate}] (0.073,0.312) to (0.146,0.312);
	\draw[postaction={decorate}] (0.073,0.312) to (0,0.312);
	\draw (0.146,0.302) -- (0.146,0.322);
	\draw (0.073,0.312) node[anchor=south]{$a$};
\end{tikzpicture}
	\caption{}
	\end{subfigure}\hspace{0.5cm}%
	\begin{subfigure}[b]{0.5\textwidth}
	\includegraphics[trim={1cm 0.5cm 2cm 2cm},clip,width=7.25cm]{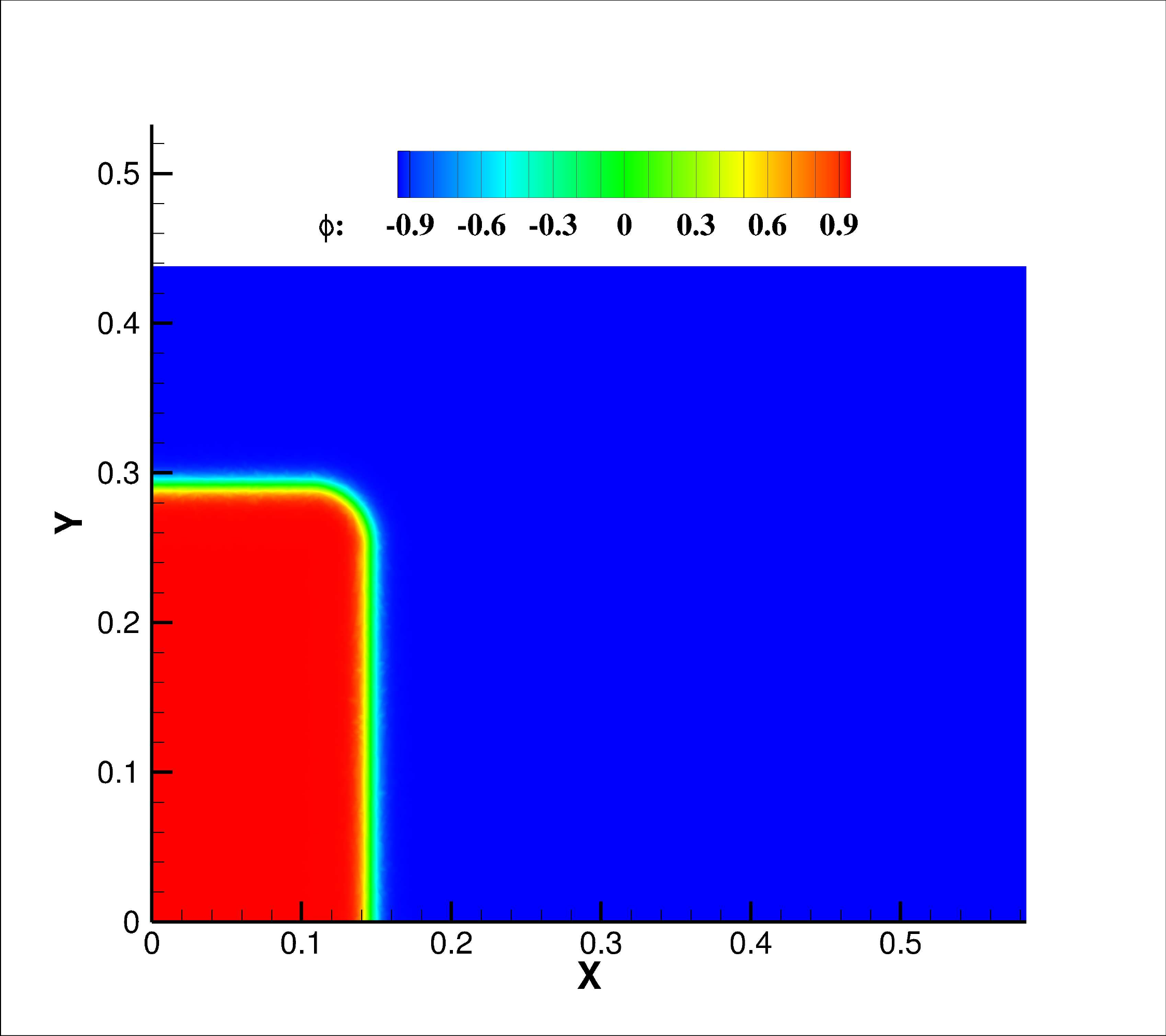}		
	\caption{}
	\end{subfigure}
\caption{Two-dimensional dam break problem: (a) schematic diagram showing the computational domain, and (b) the contour plot of the order parameter $\phi$ at $t=0$. In (a), $\Omega_1$ and $\Omega_2$ are the two phases with $\rho_1=1000$, $\rho_2=1$, $\mu_1=10^{-3}$ and $\mu_2=10^{-5}$, acceleration due to gravity is taken as $\mathbf{g}=(0,-9.81,0)$ and slip boundary condition is imposed on all the boundaries.} 
\label{db_2D}
\end{figure}

\begin{figure}[H]
\centering
	\begin{subfigure}[b]{0.5\textwidth}
		\includegraphics[trim={10cm 0 11.8cm 0cm},clip,width=7cm]{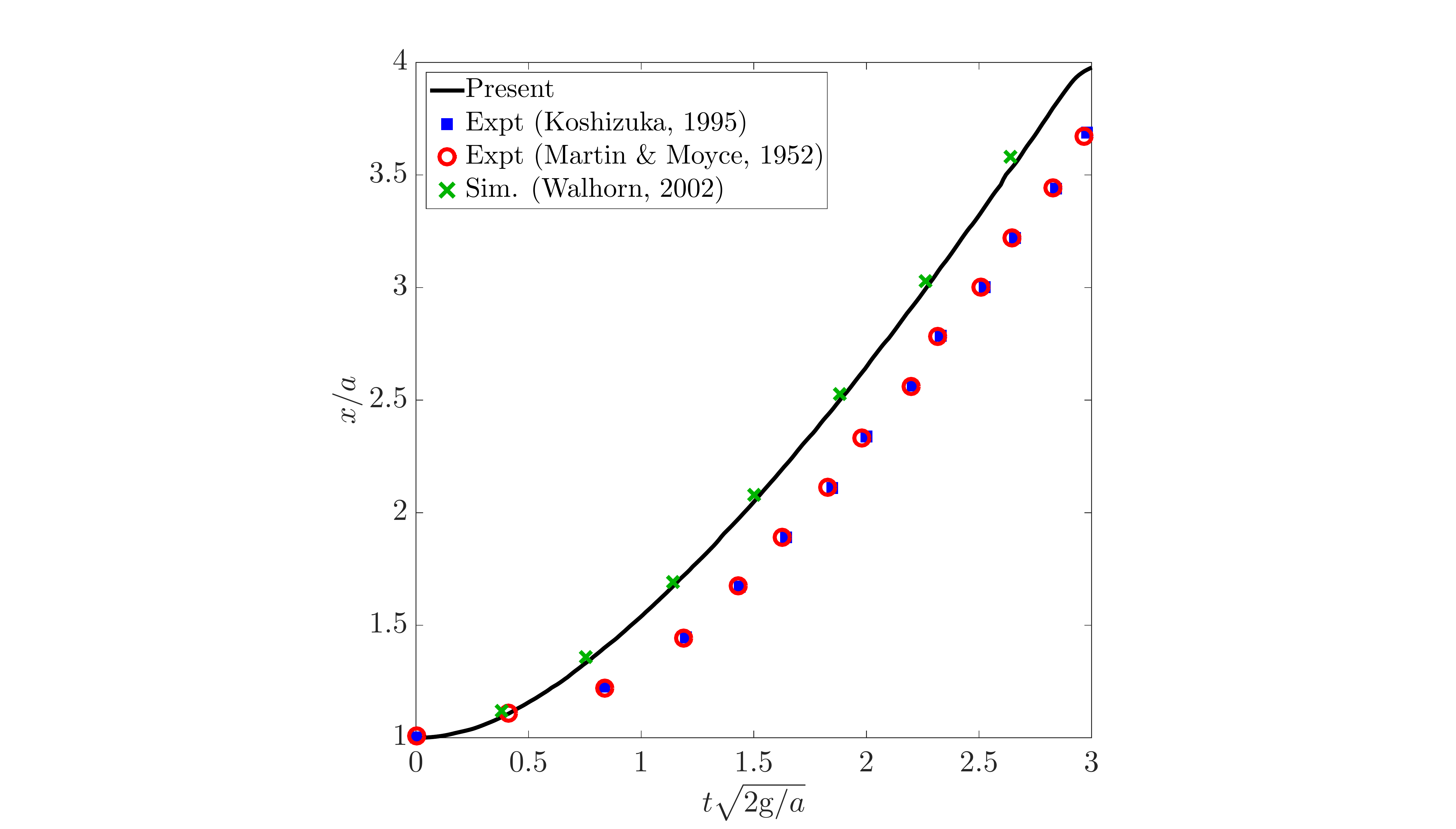}
	\caption{}
	\end{subfigure}%
	\begin{subfigure}[b]{0.5\textwidth}
		\includegraphics[trim={10cm 0 11.8cm 0cm},clip,width=7cm]{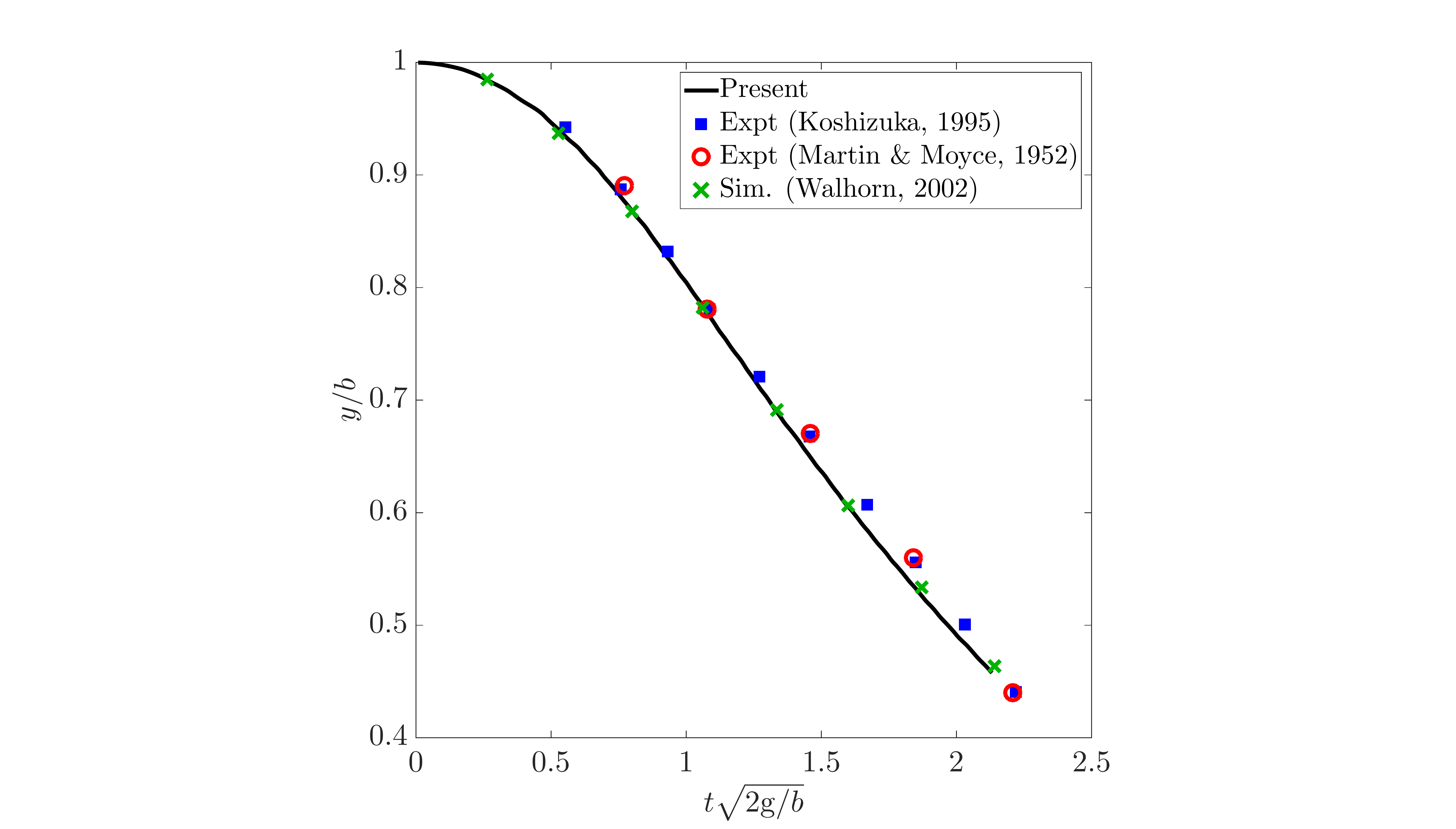}
	\caption{}
	\end{subfigure}		
\caption{Two-dimensional dam break problem: temporal evolution of non-dimensional water column (a) width and (b) height. The results are in good agreement with the literature.} 
\label{db_2D_val}
\end{figure}

In this subsection, we test the two-dimensional dam break problem to assess the ability of the solver to handle breaking and merging of the interface for the two-phase flow problem. A rectangular computational domain $[0,0.584]\times [0,0.438]$ shown in Fig. \ref{db_2D}(a) is considered for the present study. A water column of size $0.146 \times 0.292$ units is placed at the left boundary of the domain at time $t=0$. 
The initial condition is given by:
\begin{align}
	\phi(x,y,0) = \begin{cases}
		-\mathrm{tanh}\big( \frac{y-b}{\sqrt{2}\varepsilon} \big),\ \mathrm{for}\ x \leq (a - r), y \geq (b-r) \\
		 \\
		-\mathrm{tanh}\big( \frac{x-a}{\sqrt{2}\varepsilon} \big),\ \mathrm{for}\ x > (a-r), y < (b-r) \\
		\\
		\mathrm{tanh}\big( \frac{r - \sqrt{(x-(a-r))^2 + (y-(b-r))^2} }{\sqrt{2}\varepsilon} \big),\ \mathrm{for}\ x \geq (a-r), y \geq (b-r) \\
		 \\
		 1,\ \mathrm{elsewhere}
		\end{cases}
\end{align}
where $a=0.146$, $b=0.292$ and $r=0.04$ are the width, height and the radius of the curve of the water column respectively, $\phi=1$ and $\phi=-1$ correspond to the order parameter on $\Omega_1$ and $\Omega_2$ respectively. A non-uniform mesh consisting of about 13,500 nodes and 13,300 four node quadrilaterals is employed for the study. The interfacial thickness parameter $\varepsilon$ was selected as $0.005$. The total computational time for simulating $1000$ time steps with step size of $\Delta t=0.001$ was $0.453$ hour on $4$ CPUs. 
The interface location at the left and bottom boundaries are tracked with time for validation with experiments \cite{Martin_Moyce, Ubbink_thesis} and interface tracking simulation \cite{Walhorn_thesis}. The temporal variation of the interface location based on the non-dimensional water column width/height is compared with the results from the literature in Fig. \ref{db_2D_val} where a good agreement is found. The evolution of the height of the water column is quantified very well. However, the expansion of the water column (in width) in the experiment is slower than what is predicted from the simulation. This delay has been pointed out in \cite{Fries_1} to be due to the time required to remove the partition which holds the water column to its initial profile in the experiment. 

\begin{figure}[H]
\centering	
	\begin{subfigure}[b]{0.33\textwidth}
	\includegraphics[trim={1cm 0.5cm 2cm 2cm},clip,width=5cm]{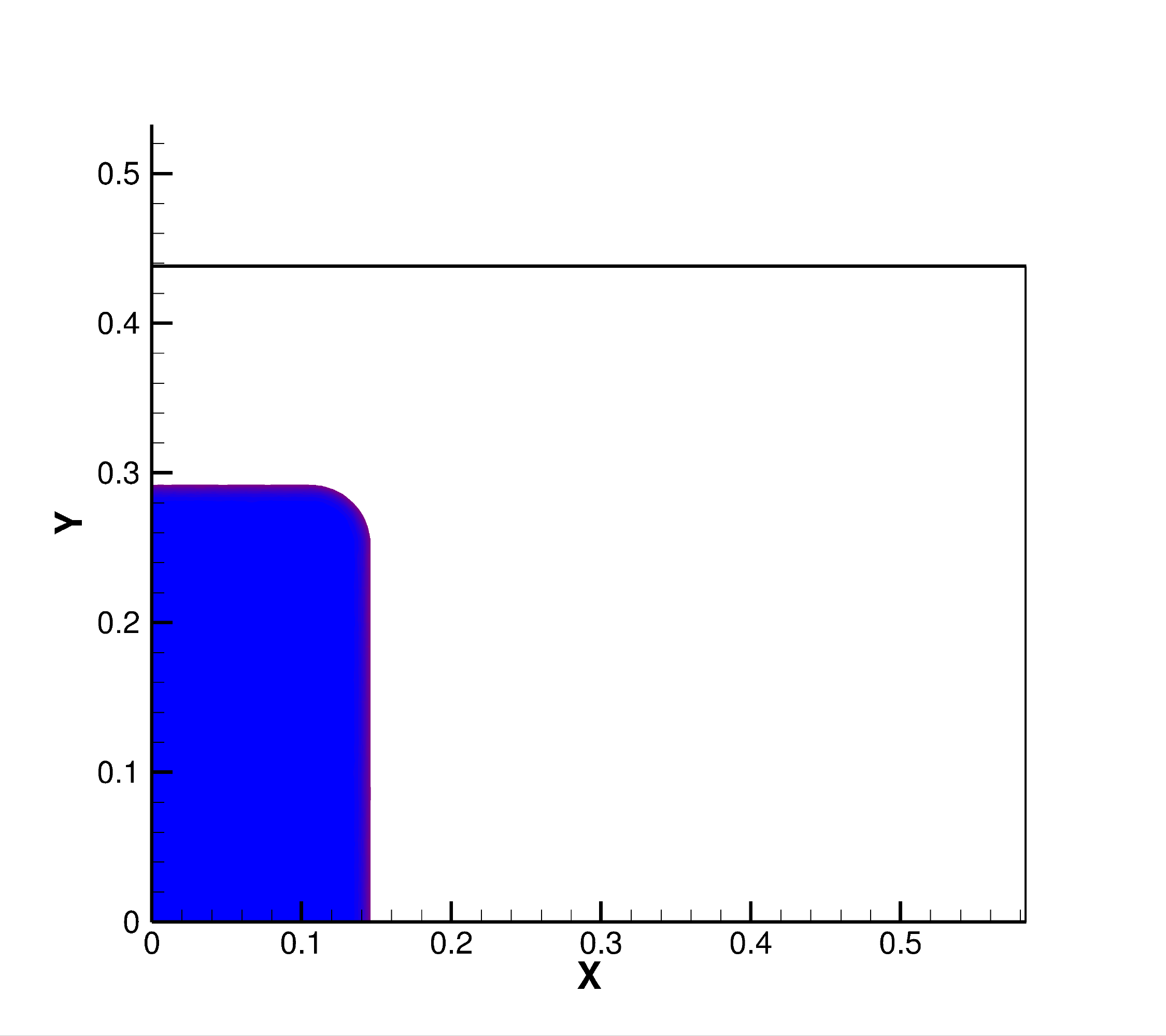}		
	\caption{}
	\end{subfigure}%
	\begin{subfigure}[b]{0.33\textwidth}
	\includegraphics[trim={1cm 0.5cm 2cm 2cm},clip,width=5cm]{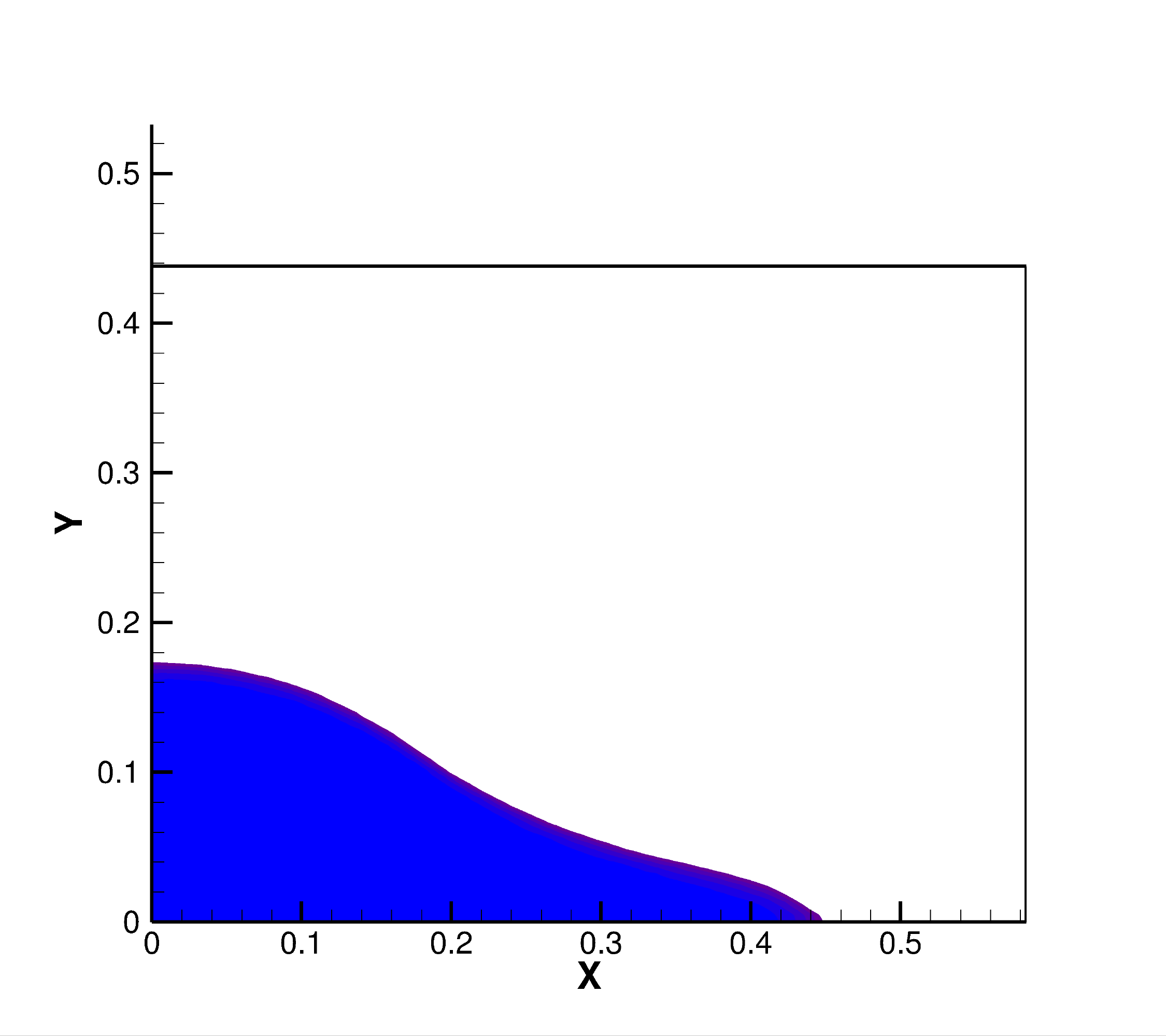}		
	\caption{}
	\end{subfigure}%
	\begin{subfigure}[b]{0.33\textwidth}
	\includegraphics[trim={1cm 0.5cm 2cm 2cm},clip,width=5cm]{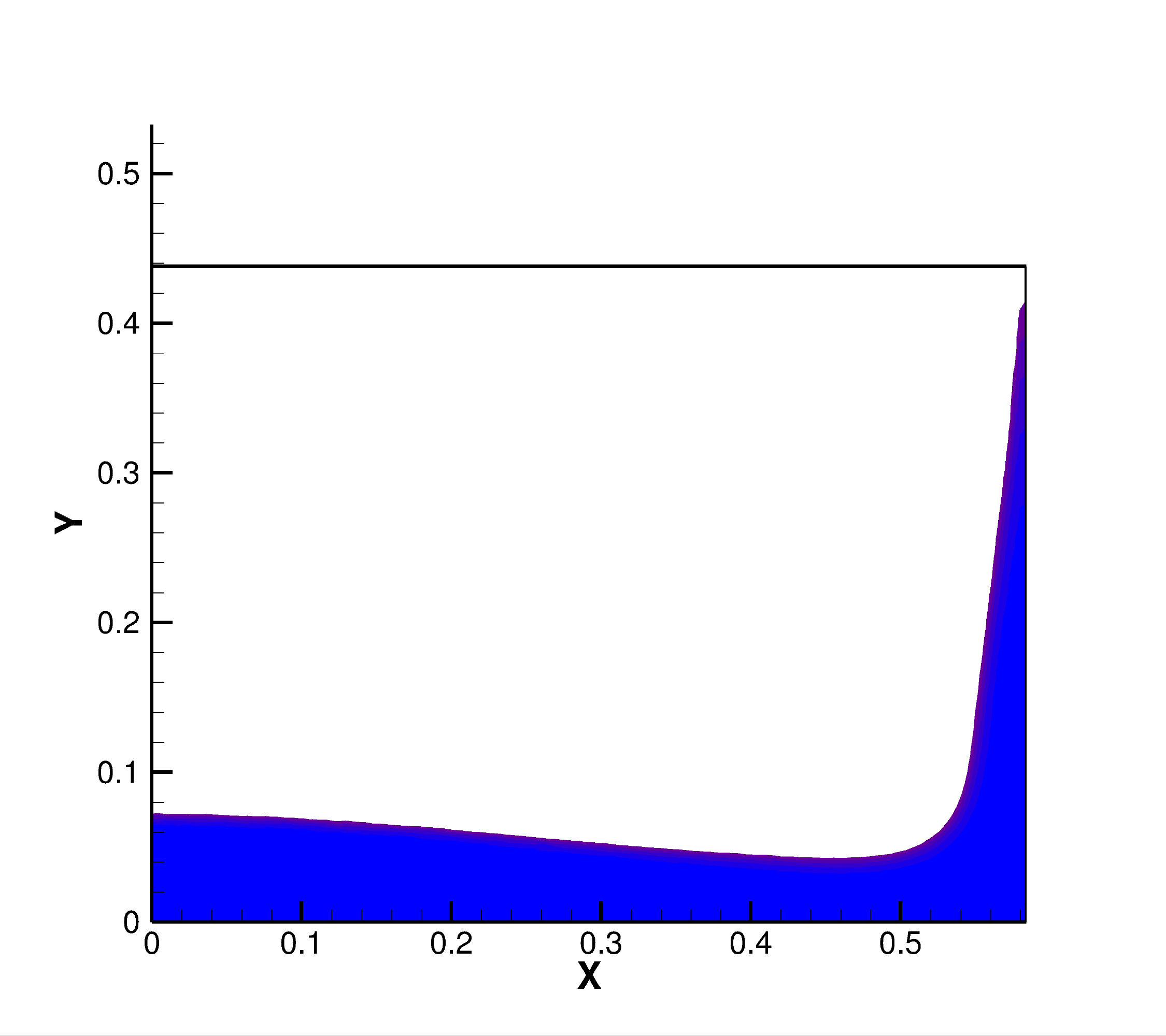}		
	\caption{}
	\end{subfigure}
	
	\begin{subfigure}[b]{0.33\textwidth}
	\includegraphics[trim={1cm 0.5cm 2cm 2cm},clip,width=5cm]{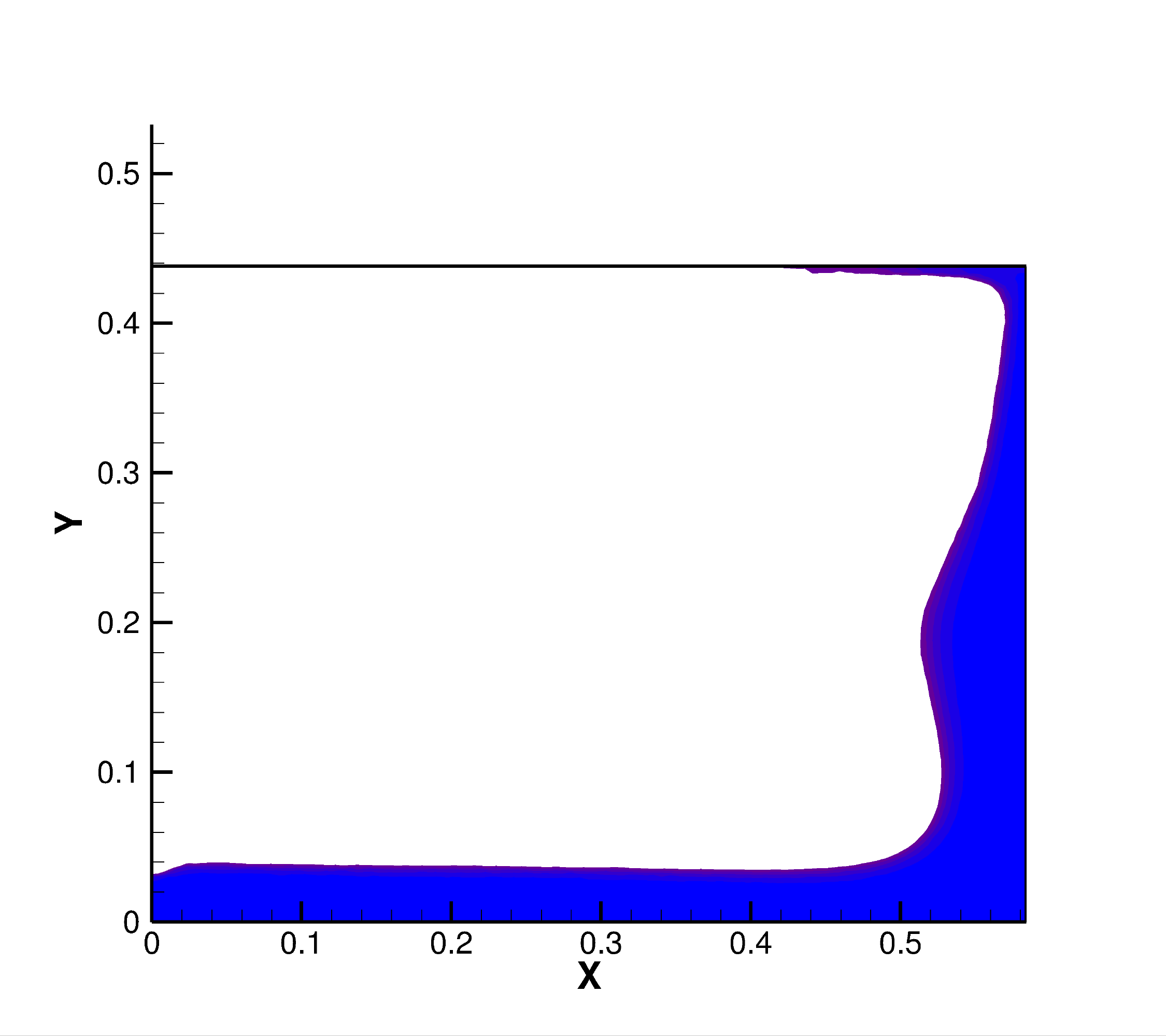}		
	\caption{}
	\end{subfigure}%
	\begin{subfigure}[b]{0.33\textwidth}
	\includegraphics[trim={1cm 0.5cm 2cm 2cm},clip,width=5cm]{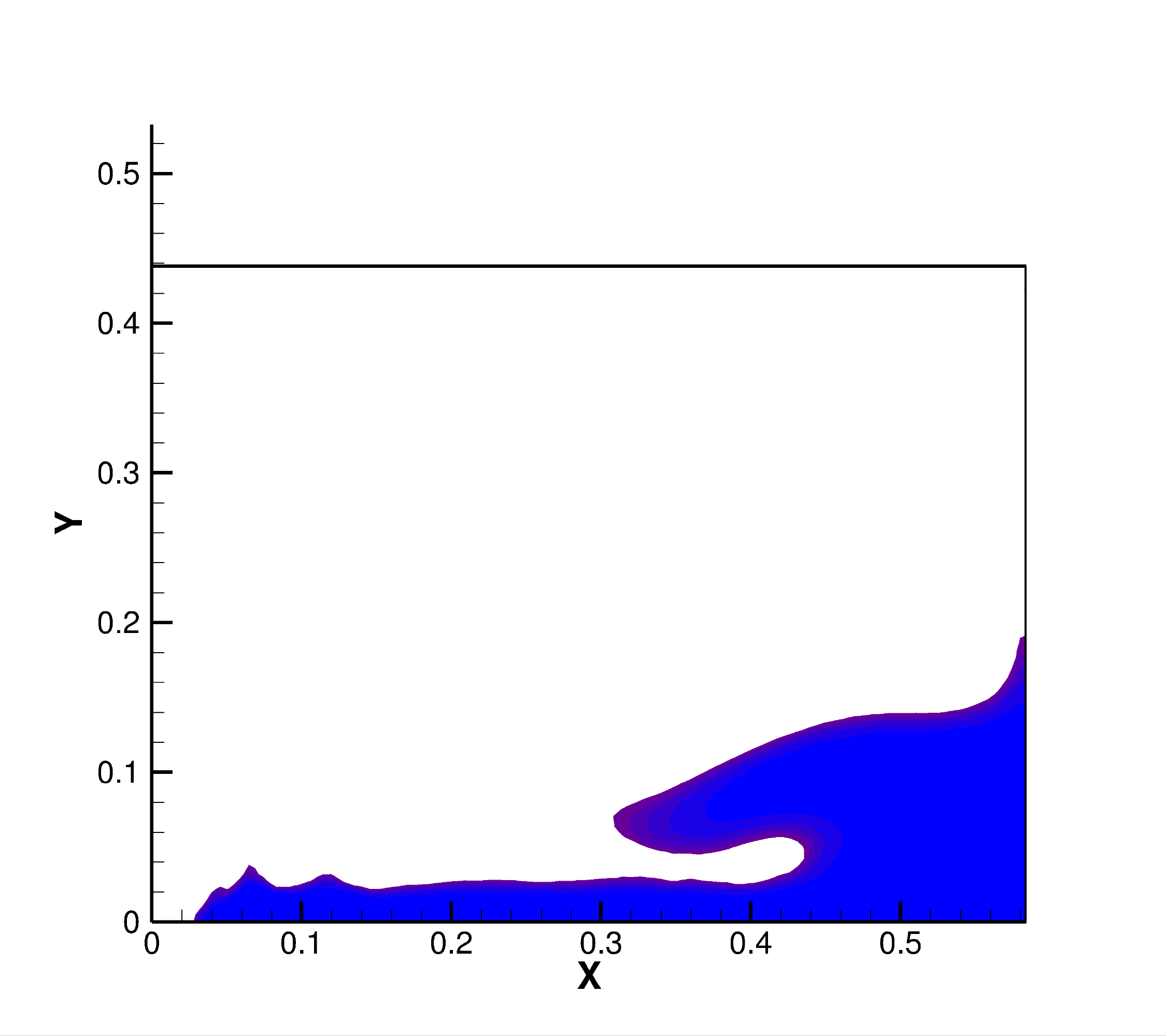}		
	\caption{}
	\end{subfigure}%
	\begin{subfigure}[b]{0.33\textwidth}
	\includegraphics[trim={1cm 0.5cm 2cm 2cm},clip,width=5cm]{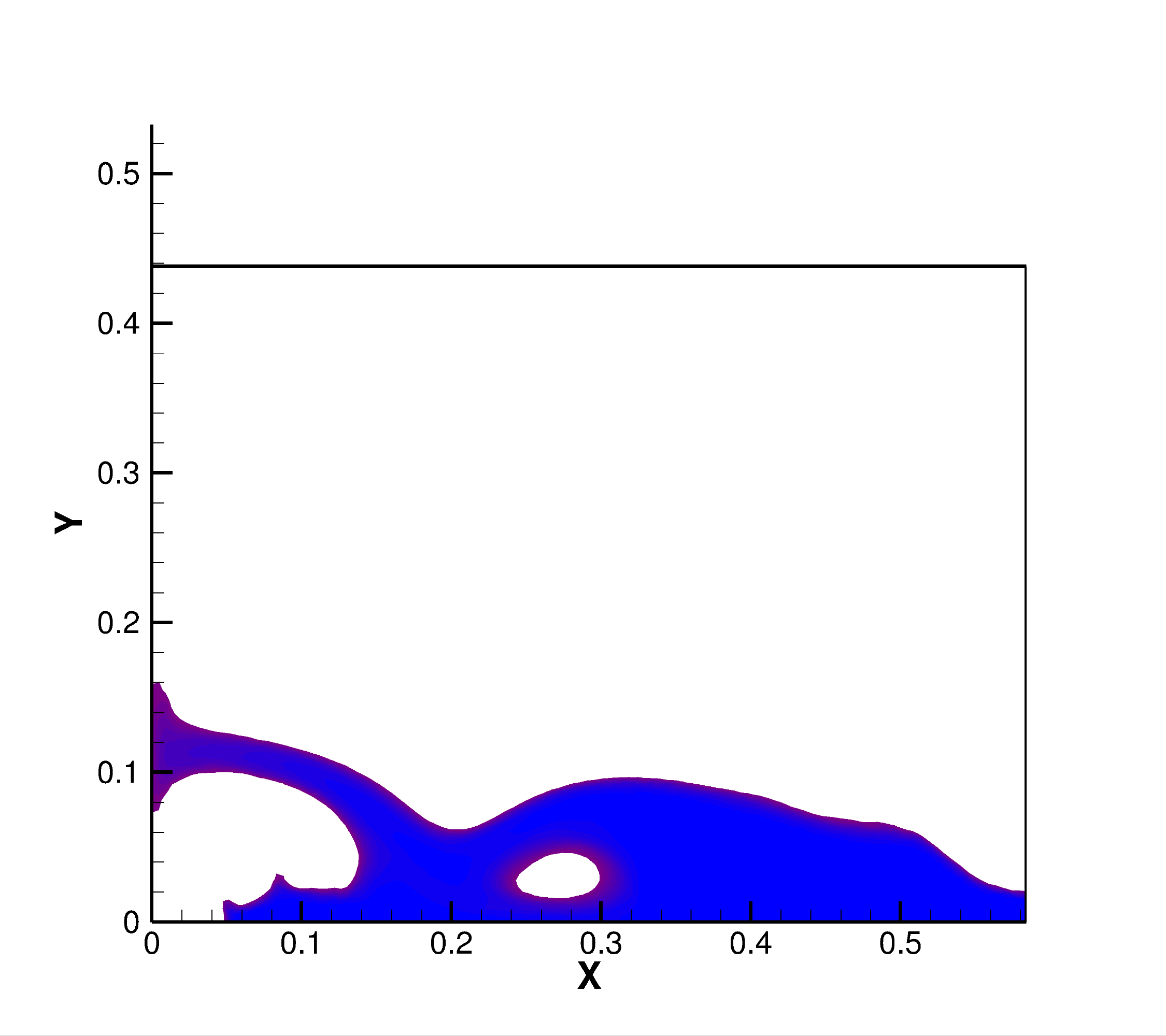}		
	\caption{}
	\end{subfigure}
\caption{Interface profiles for two-dimensional dam break problem: (a) $t=0.0$, (b) $t=0.2$, (c) $t=0.4$, (d) $t=0.6$, (e) $t=0.8$ and (f) $t=1.0$. The profile shown depicts the contour of the order parameter cutoff below $0$, i.e., it only shows the domain consisting water.} 
\label{db_2D_profile}
\end{figure}

The profiles of the interface evolution is shown in Fig. \ref{db_2D_profile}. It is in good agreement with the profiles obtained in the literature. Some variations can be found which may be due to the closed domain condition in the numerical study.

\subsection{Three-dimensional dam break with obstacle}
\begin{figure}[!htbp]
\centering
	\begin{subfigure}[b]{\textwidth}
\qquad
\begin{tikzpicture}[decoration={markings,mark=at position 1.0 with {\arrow{>}}},scale=2.5,every node/.style={scale=0.833}]
	\draw (0,0) -- (0,1)-- (4,2) -- (4,1) -- cycle;
	\draw (0,0) -- (-1,0.5) -- (-1,1.5) -- (0,1);
	\draw (-1,1.5) -- (3,2.5) -- (4,2);
	\draw[dashed] (3,2.5) -- (3,1.5)--(4,1);
	\draw[dashed] (3,1.5)--(-1,0.5);
	\draw[fill={rgb:black,1;white,2}, fill opacity=0.4] (0,0)--(-1,0.5)--(-1,1)--(0,0.5)--(0,0);
	\draw[fill={rgb:black,1;white,2}, fill opacity=0.4] (0,0)--(0,0.5)--(1.2,0.8)--(1.2,0.3)--(0,0);
	\draw[fill={rgb:black,1;white,2}, fill opacity=0.4] (0,0.5)--(-1,1)--(0.2,1.3)--(1.2,0.8)--(0,0.5);
	\draw (2.4,0.8)--(2.4,1)--(2.6,1.05)--(2.6,0.85)--(2.4,0.8);
	\draw (2.4,0.8)--(2,1)--(2,1.2)--(2.4,1);
	\draw (2,1.2)--(2.2,1.25)--(2.6,1.05);
	
	\draw[dotted] (-1,0.5)--(-1.3,0.425);
	\draw[dotted] (0,0)--(-0.3,-0.075);
	\draw[<->] (-0.25,-0.0625) -- (-1.25,0.4375)node[midway,sloped,above]{$1.0$};
	\draw[dotted] (0,0)--(0.5,-0.25);
	\draw[dotted] (4,1)--(4.5,0.75);
	\draw[<->] (0.5,-0.25) -- (4.5,0.75)node[midway,sloped,above]{$3.22$};
	\draw[dotted] (4,2)--(4.2,2.05);
	\draw[dotted] (4,1)--(4.2,1.05);
	\draw[<->] (4.1,2.025) -- (4.1,1.025)node[midway,sloped,above]{$1.0$};
	\draw[dotted] (1.2,0.8)--(1.2,1.0);
	\draw[<->] (0,0.6) -- (1.2,0.9)node[midway,sloped,above]{$1.228$};
	\draw[dotted] (1.2,0.8)--(1.35,0.8375);
	\draw[<->] (1.3,0.825) -- (1.3,0.325)node[midway,sloped,above]{$0.55$};
	
	\draw[dotted] (2.4,0.8)--(2.6,0.7);
	\draw[<->] (2.5,0.75) -- (2.7,0.8) node[midway,sloped,below]{$0.161$};
	\draw[dotted] (2.6,1.05)--(2.75,1.0875);
	\draw[dotted] (2.6,0.85)--(2.75,0.88);
	\draw[<->] (2.7,1.075) -- (2.7,0.875)node[midway,sloped,above]{$0.161$};
	\draw[dotted] (2.4,0.8)--(2.25,0.7625);
	\draw[dotted] (2,1)--(1.85,0.9625);
	\draw[<->] (1.9,0.975) -- (2.3,0.775)node[midway,sloped,below]{$0.403$};
	
	\draw[dotted] (2.6,0.85) -- (3.2,0.55);
	\draw[<->] (0.25,-0.125) -- (3.1,0.6)node[midway,sloped,above]{$2.5565$};;
	\draw (0.6,0.34) node{$\Omega_1$};
	\draw (1,1.5) node{$\Omega_2$};
	\draw[thick,postaction={decorate}] (0,0) to (0.2,0.05);
	\draw[thick,postaction={decorate}] (0,0) to (0,0.2);
	\draw[thick,postaction={decorate}] (0,0) to (0.2,-0.1);
	\draw (0.2,0.2) node[anchor=north]{X};
	\draw (0.15,0.2) node[anchor=east]{Y};
	\draw (0.3,-0.04) node[anchor=east]{Z};
\end{tikzpicture}
	\end{subfigure}
\caption{Schematic diagram showing the computational domain for three-dimensional dam break with obstacle. $\Omega_1$ and $\Omega_2$ are the two phases with $\rho_1=1000$, $\rho_2=1.225$, $\mu_1=1.002\times 10^{-3}$ and $\mu_2=1.983\times 10^{-5}$, acceleration due to gravity is taken as $\mathbf{g}=(0,-9.81,0)$ and slip boundary condition is imposed on all the boundaries.} 
\label{db_3D}
\end{figure}
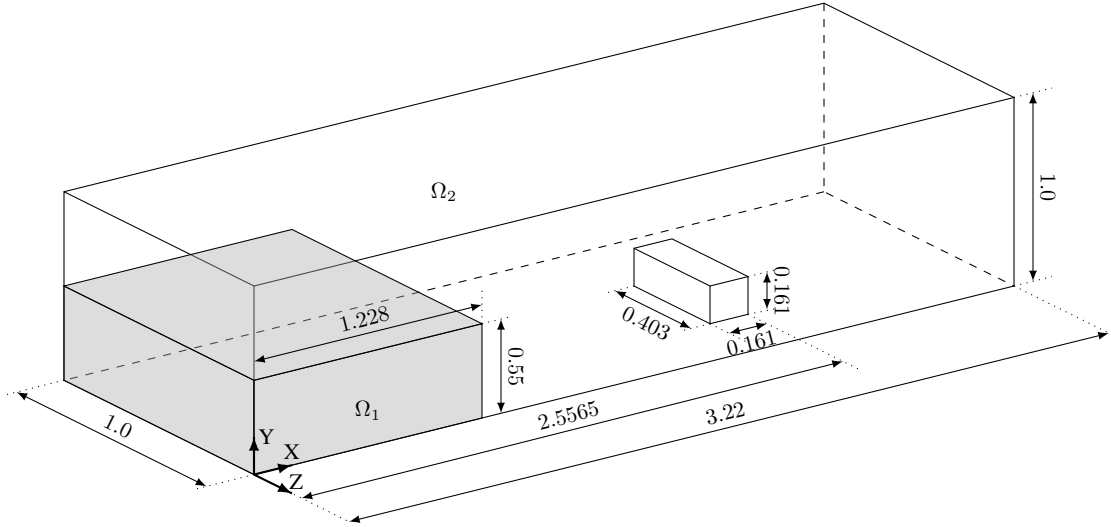

\begin{figure}[H]
\centering
	\begin{subfigure}[b]{0.5\textwidth}
		\includegraphics[trim={10cm 0 11.8cm 0cm},clip,width=7cm]{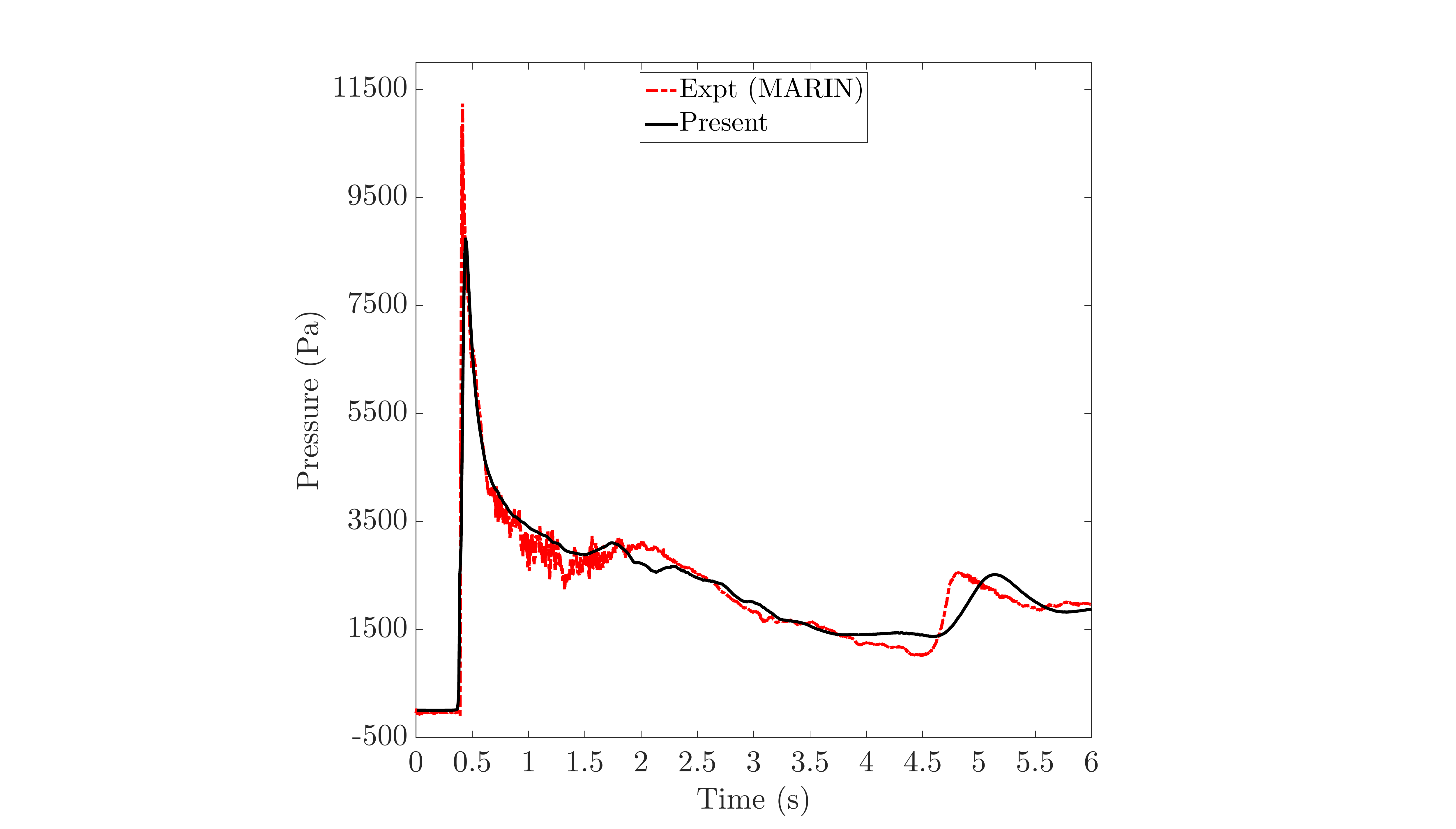}
	\caption{}
	\end{subfigure}%
	\begin{subfigure}[b]{0.5\textwidth}
		\includegraphics[trim={10cm 0 11.8cm 0cm},clip,width=7cm]{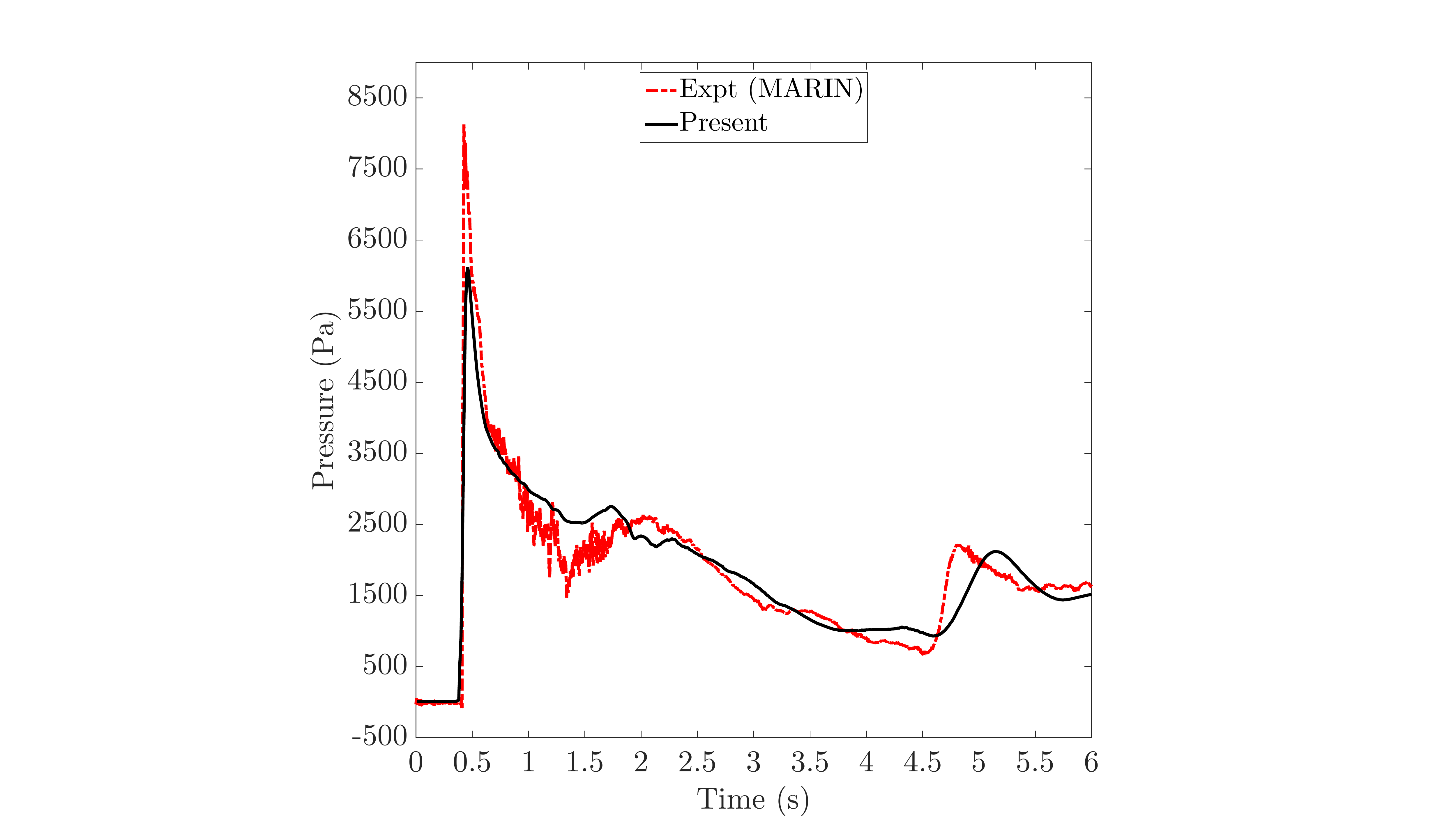}
	\caption{}
	\end{subfigure}	
	
\caption{Three-dimensional dam break problem: temporal evolution of the pressure at the probe points: (a) P1 and (b) P2. The results are in good agreement with the literature.} 
\label{db_3D_val_p}
\end{figure}

\begin{figure}[H]
\centering
	\begin{subfigure}[b]{0.5\textwidth}
		\includegraphics[trim={10cm 0 11.8cm 0cm},clip,width=7cm]{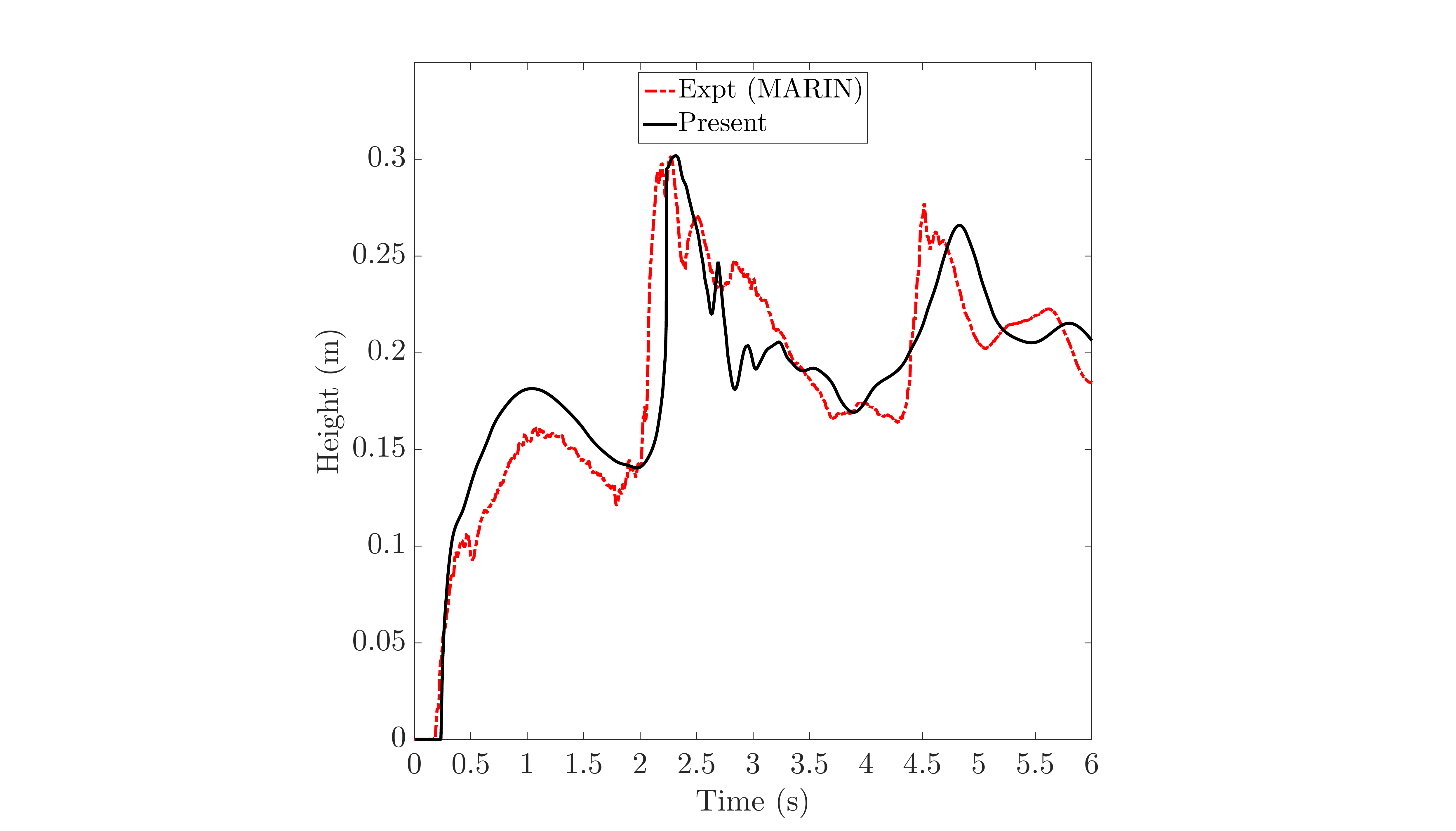}
	\caption{}
	\end{subfigure}%
	\begin{subfigure}[b]{0.5\textwidth}
		\includegraphics[trim={10cm 0 11.8cm 0cm},clip,width=7cm]{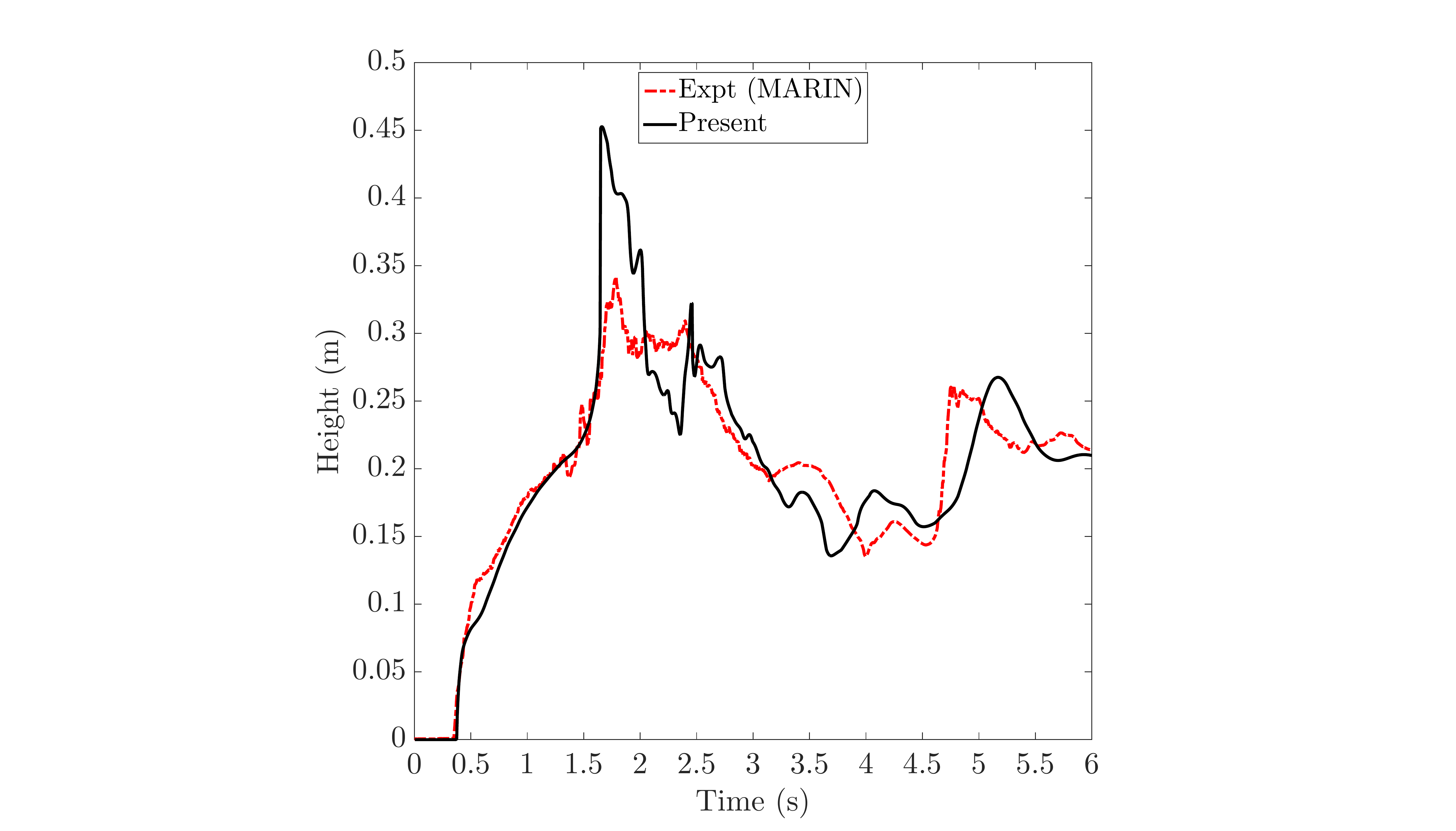}
	\caption{}
	\end{subfigure}			
\caption{Three-dimensional dam break problem: temporal evolution of the height at the probe points: (a) H1 and (b) H2. The results are in good agreement with the literature.} 
\label{db_3D_val_h}
\end{figure}

\begin{figure}[H]
\centering
	\begin{subfigure}[b]{0.5\textwidth}
		\includegraphics[trim={0.2cm 2.5cm 0.2cm 0.2cm},clip,width=7cm]{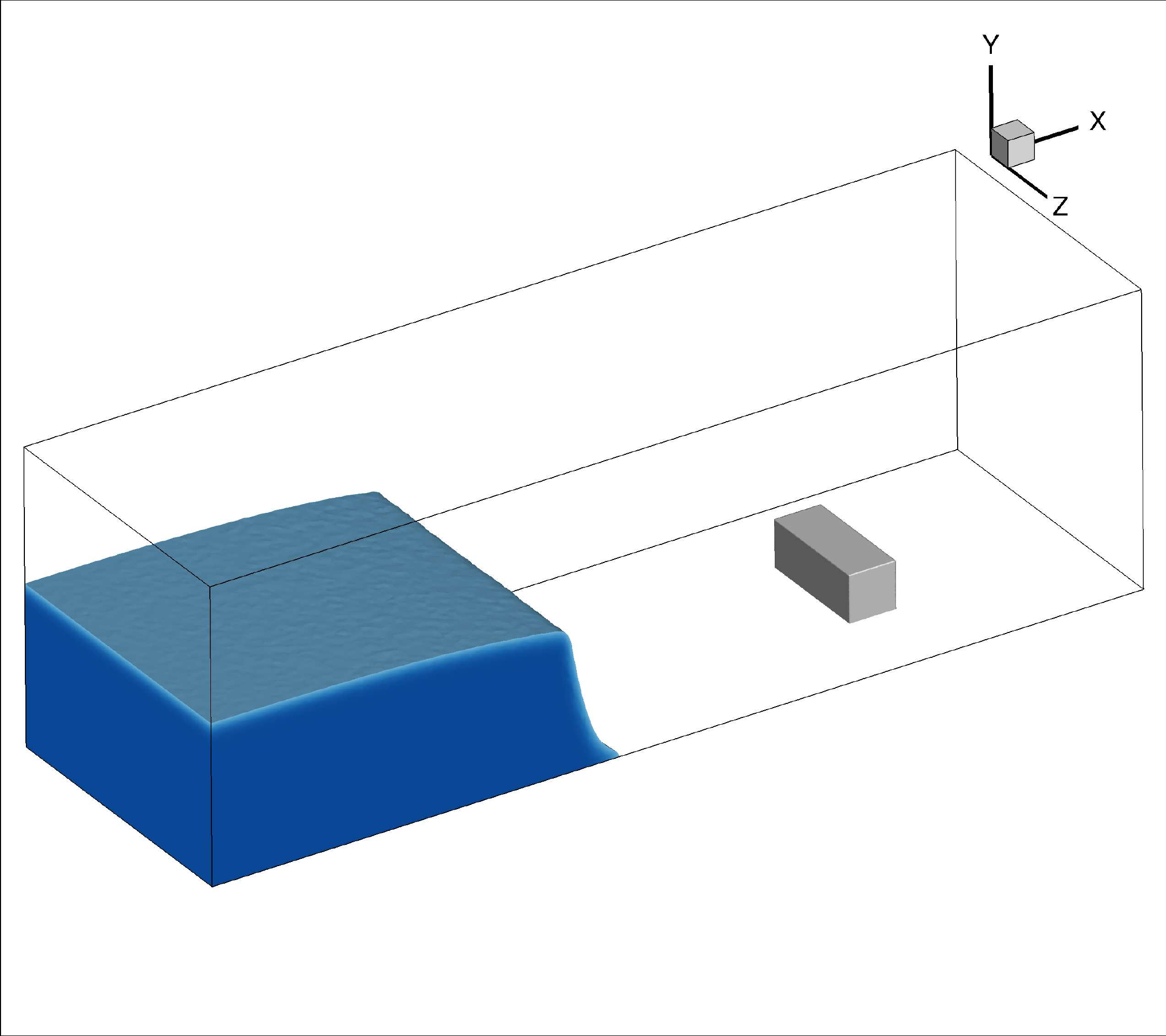}
	\caption{}
	\end{subfigure}%
	\begin{subfigure}[b]{0.5\textwidth}
		\includegraphics[trim={0.2cm 2.5cm 0.2cm 0.2cm},clip,width=7cm]{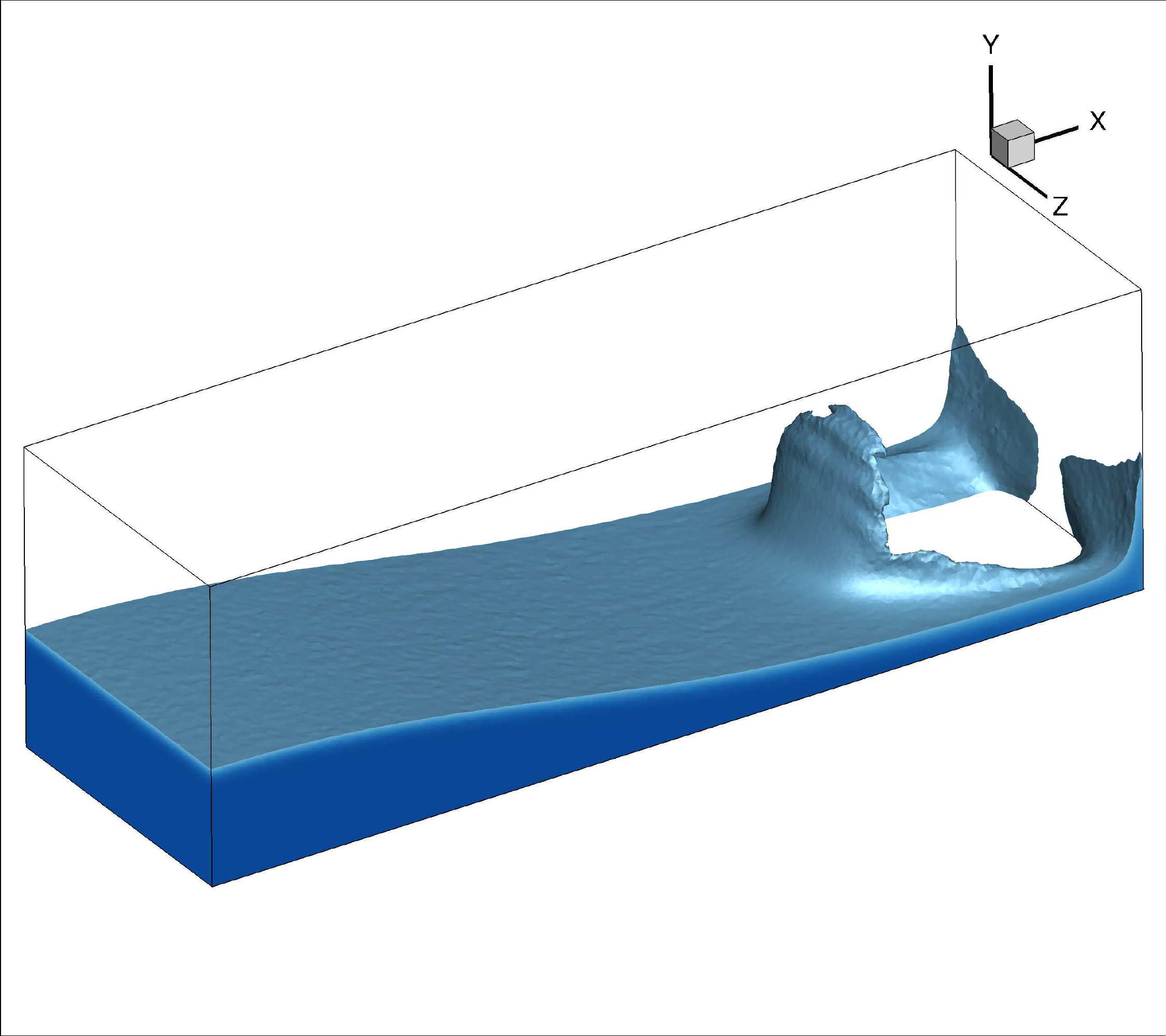}
	\caption{}
	\end{subfigure}
	
	\begin{subfigure}[b]{0.5\textwidth}
		\includegraphics[trim={0.2cm 2.5cm 0.2cm 0.2cm},clip,width=7cm]{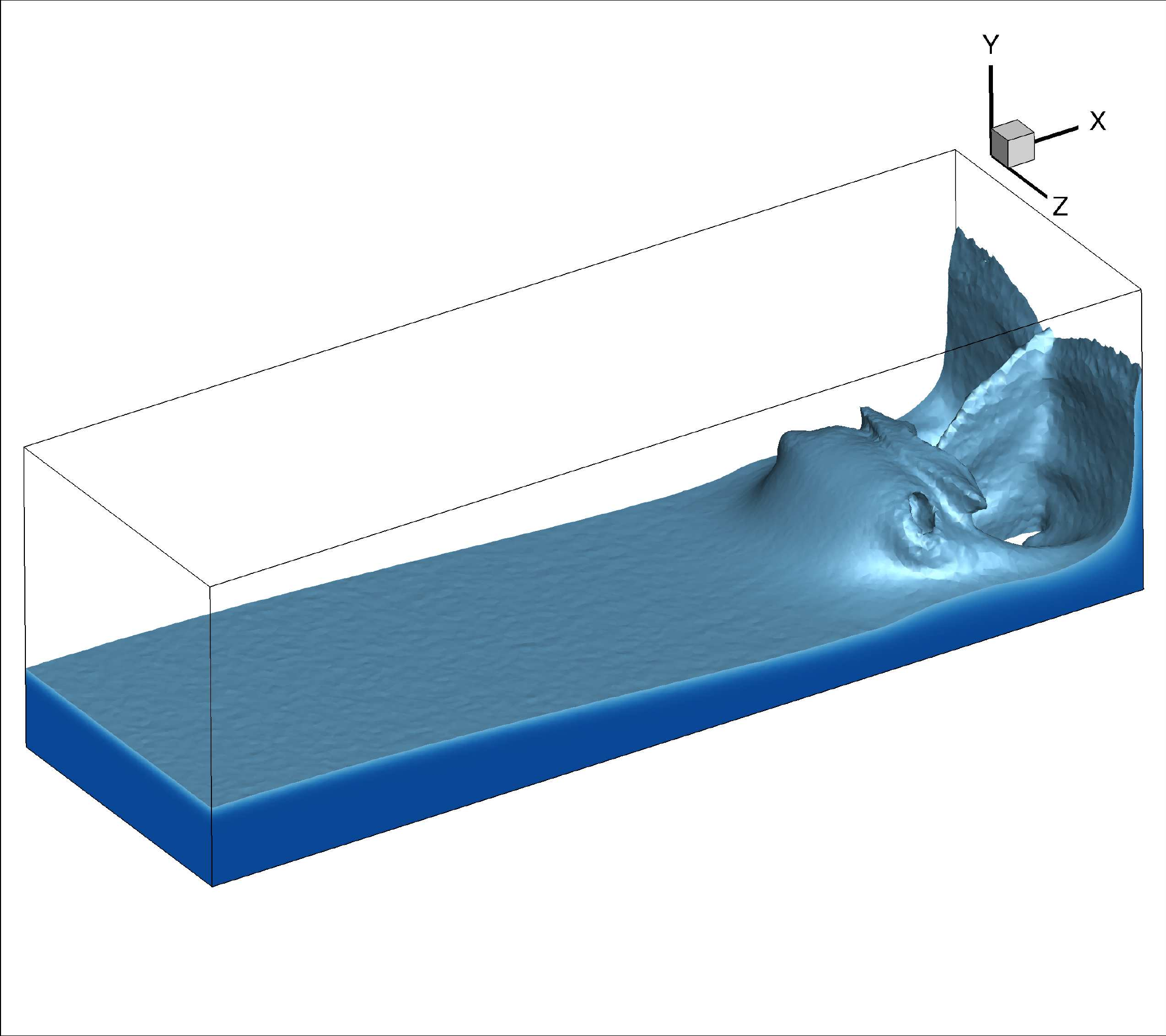}
	\caption{}
	\end{subfigure}%
	\begin{subfigure}[b]{0.5\textwidth}
		\includegraphics[trim={0.2cm 2.5cm 0.2cm 0.2cm},clip,width=7cm]{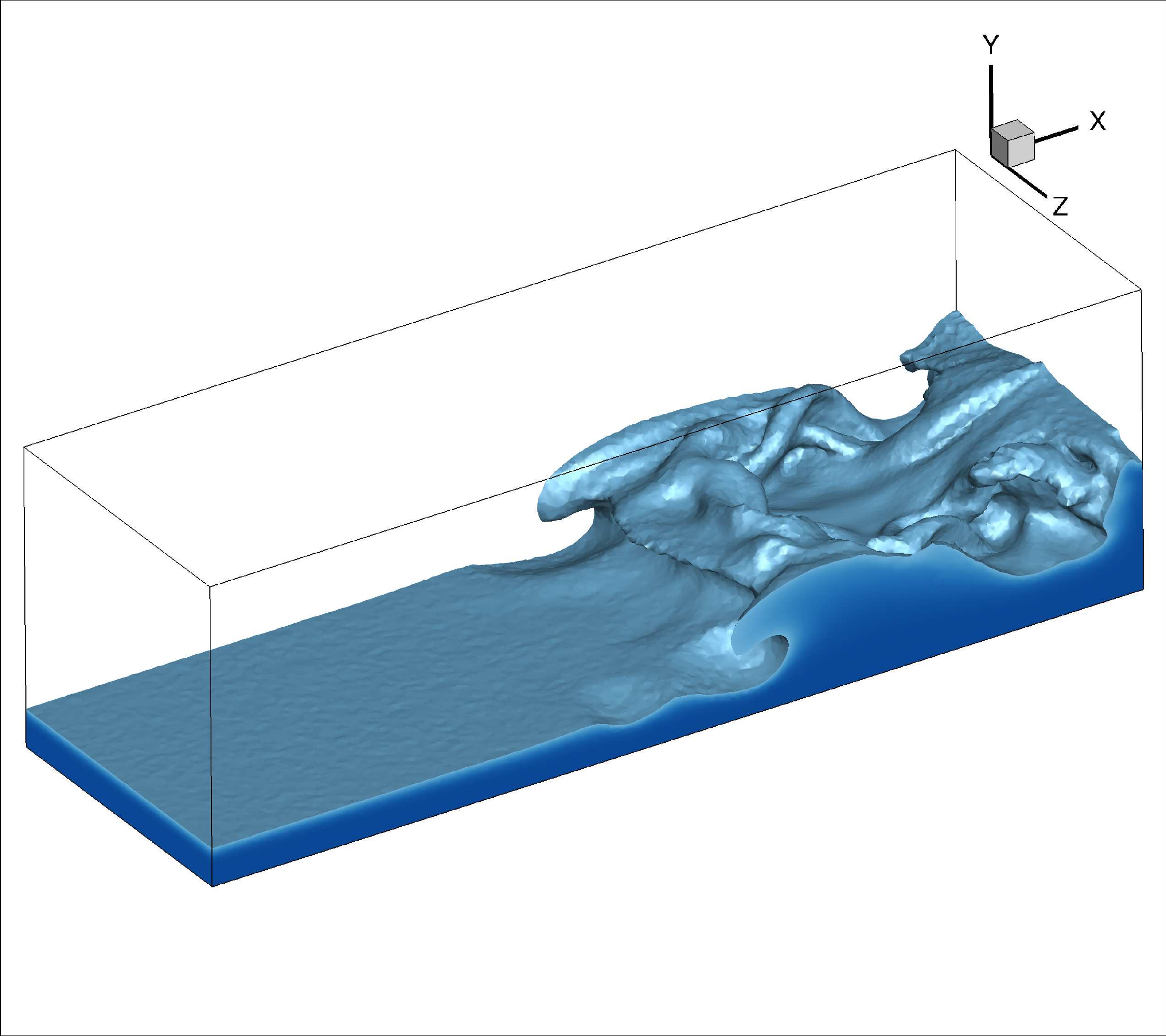}
	\caption{}
	\end{subfigure}	
	
	\begin{subfigure}[b]{0.5\textwidth}
		\includegraphics[trim={0.2cm 2.5cm 0.2cm 0.2cm},clip,width=7cm]{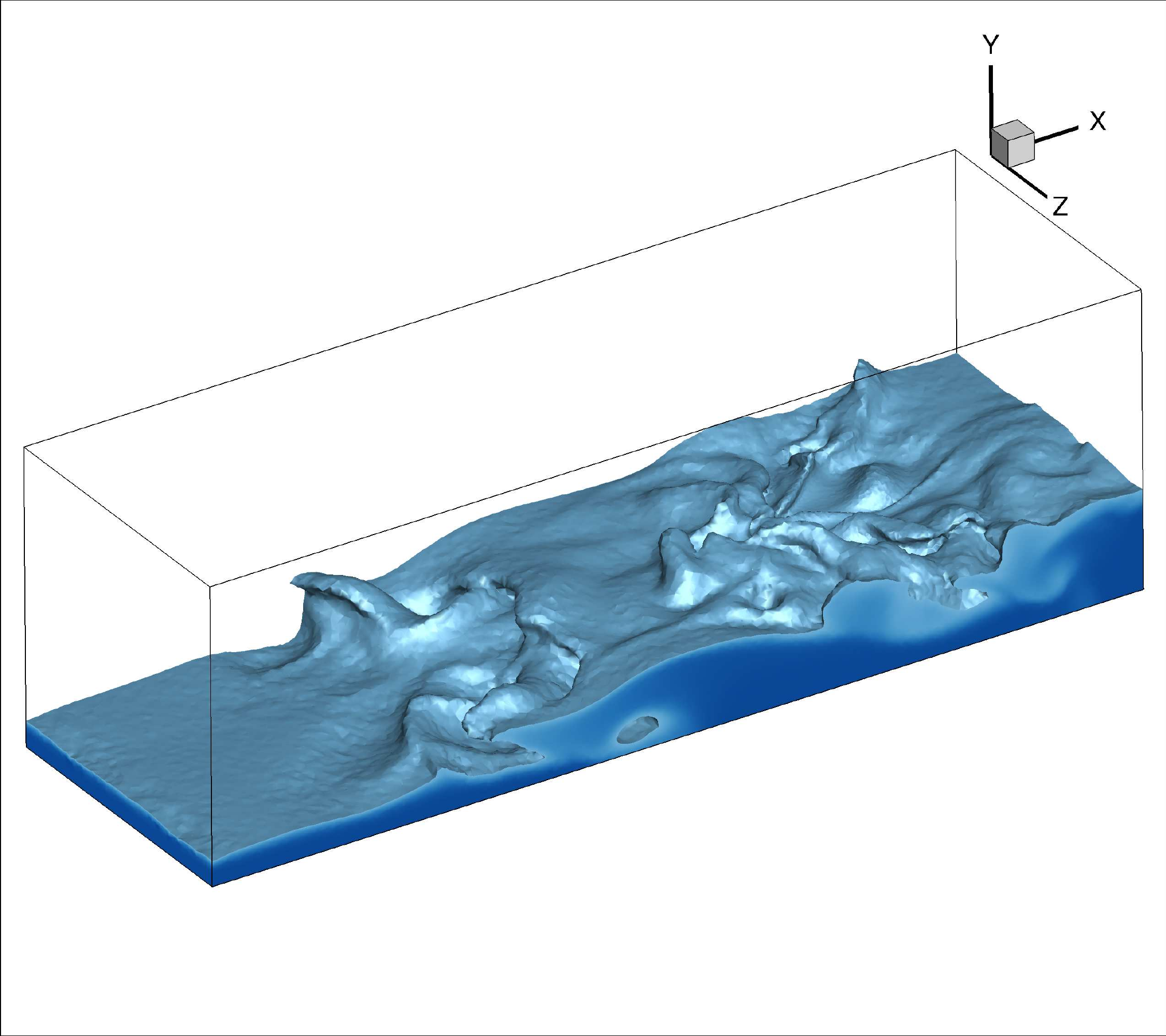}
	\caption{}
	\end{subfigure}%
	\begin{subfigure}[b]{0.5\textwidth}
		\includegraphics[trim={0.2cm 2.5cm 0.2cm 0.2cm},clip,width=7cm]{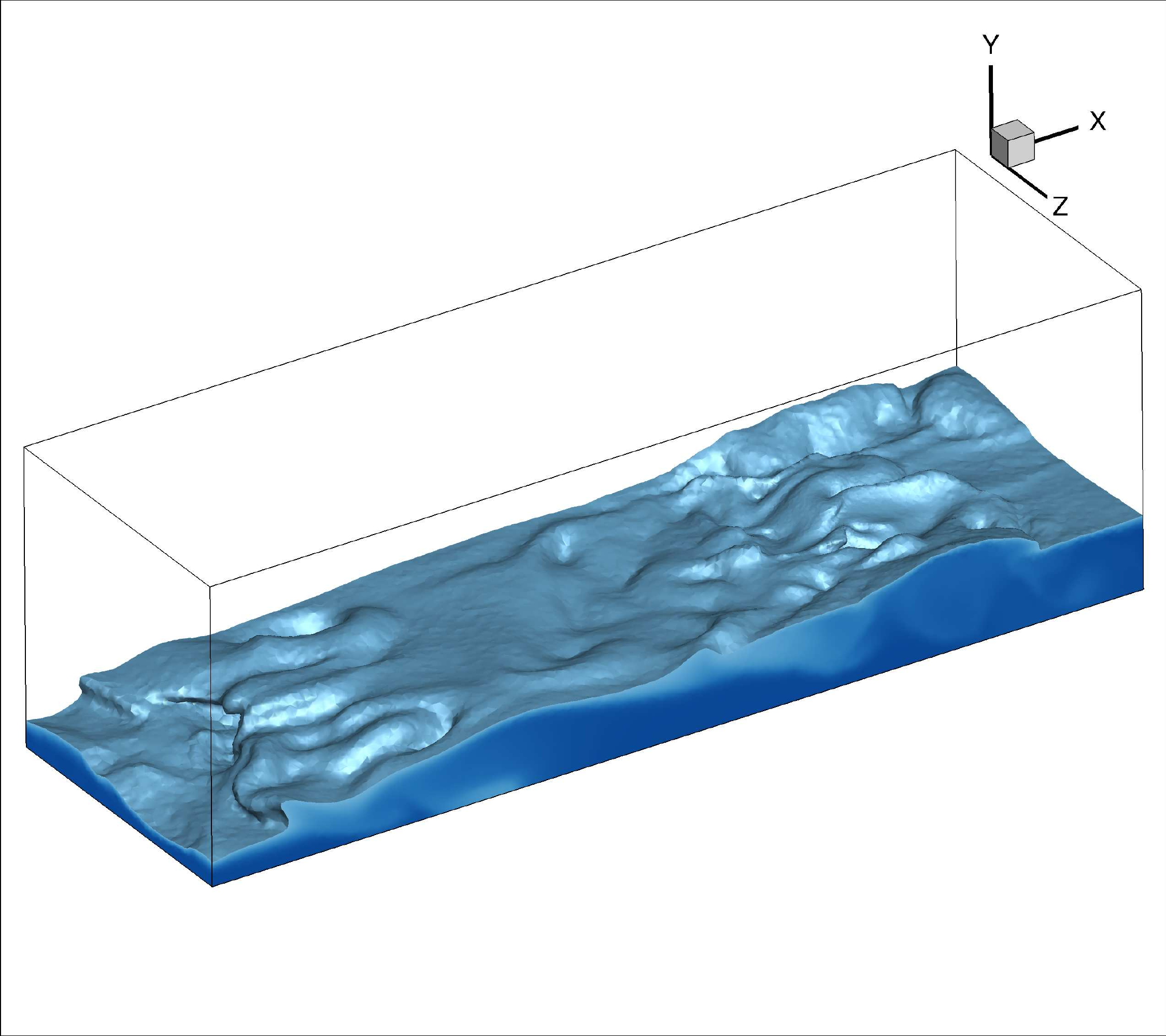}
	\caption{}
	\end{subfigure}			
\caption{Three-dimensional dam break problem: temporal evolution of the iso-contours of water $(\phi>0)$ at time: (a) $0.125$ s, (b) $0.75$ s, (c) $1.125$ s, (d) $2$ s, (e) $2.5$ s and (f) $3$ s.} 
\label{db_3D_visual}
\end{figure}

To further assess the solver, we simulate the three-dimensional dam break problem with a rectangular obstacle. The computational domain is a cuboid of size $[0,3.22] \times [0,1] \times [0,1]$ with a water column of size $1.228 \times 0.55 \times 1$ units. The rectangular cuboid obstacle is of size $0.161 \times 0.161 \times 0.403$. The dimensions of the domain are given in Fig. \ref{db_3D}. To extract the pressure for comparison with the experiment, we had two probes at the following locations (with respect to the axis orientation given in the Fig. \ref{db_3D}):
\begin{itemize}
	\item $(2.3955, 0.021, -0.5255)$: Pressure value (P1)
	\item $(2.3955, 0.061, -0.5255)$: Pressure value (P2)
\end{itemize}
The notation in the parentheses is the name of the probe in the experiment conducted by MARIN. To obtain the variation in the water level (or interface), the height of the water is tracked at two locations, viz., $(X,Z)=(2.724,-0.5)$ and $(X,Z)=(2.228,-0.5)$ corresponding to H1 and H2 probes of the experiment respectively. 
The data of the experiment is taken from \cite{Issa}. The computational domain is discretized into $430,000$ nodes with $2.5$ million tetrahedral elements. The interfacial thickness parameter is chosen as $\varepsilon=0.02$. The total computational time for the simulation of $1200$ time steps with $\Delta t=0.005$ was $2.2$ hours on $24$ CPUs. The comparison between the results obtained from the present simulation and the experiment is depicted in Fig. \ref{db_3D_val_p} and \ref{db_3D_val_h}. The change in the iso-contours of water $(\phi>0)$ have been plotted in Fig. \ref{db_3D_visual} which shows a good qualitative agreement with the results from the literature \cite{Elias,Akkerman_LS,Kleefsman}. 

These tests confirm the ability of the solver to capture the air-water interface with topological changes.
After assessing the present formulation for the Allen-Cahn equation and its coupling with the Navier-Stokes equations, we next demonstrate its application to the wave-structure interaction problem.

\section{Application to wave-structure interaction}
\label{RU_test}
\begin{figure}[H]
\centering
\begin{tikzpicture}[decoration={markings,mark=at position 1.0 with {\arrow[scale=1]{>}}},every node/.style={scale=0.9},scale=0.9]
	\draw[-,black] (-2,0) node[anchor=north,black]{} to (12,0);
	\draw[-,line width=2pt,black] (-2,-4) node[anchor=north,black]{} to (12,-4);
	\draw (1,-1) cos (2,0) sin (3,1) cos (4,0) sin (5,-1) cos (6,0) sin(7,1) cos(8,0) sin(9,-1);
	\draw (11,0) node(A){};
	\node[above=0.35cm of A,draw,minimum size=0.1cm, regular polygon, regular polygon sides=3, rotate=180] (11,0.5){};
	\draw[line width=1.5pt,postaction={decorate}] (-1,0) to (0.5,0);
	\draw[line width=1.5pt,postaction={decorate}] (-1,0) to (-1,1.5);
	\draw (0.5,0) node[anchor=north]{X};
	\draw (-1,1.5) node[anchor=east]{Z};

	\draw[->] (11.5,-2) -- (11.5,0);
	\draw[->] (11.5,-2) -- (11.5,-4);
	\draw (11.5,-2) node[anchor=west] {$d$};

	\draw[->] (7,0.5) -- (7,1);
	\draw[->] (7,0.5) -- (7,0);
	\draw (7,0.5) node[anchor=west]{$A$};

	\draw[->] (10,0) -- (10,1);
	\draw[->] (10,0) -- (10,-1);
	\draw[-,dotted] (5,-1) -- (10,-1);
	\draw[-,dotted] (7,1) -- (10,1);
	\draw (10.25,0.1) node[anchor=south]{$H$};

	\draw[->] (4,-2) -- (2,-2);
	\draw[->] (4,-2) -- (6,-2);
	\draw[-,dotted] (2,0) -- (2,-2);
	\draw[-,dotted] (6,0) -- (6,-2);
	\draw (4,-2) node[anchor=north]{$\lambda$};

	\draw (3,1) node[anchor=south]{$\eta(x,t)$};
	\node[right=1cm of A] {Mean water level};
	\draw (5,-4) node[anchor=south]{Wall};
	\draw[->] (3,2) -- (4,2) ;
	\draw (3.5,2) node[anchor=south]{Direction of wave motion};
\end{tikzpicture}
\caption{Notations and coordinate system in the description of the second-order Stokes wave theory: $\eta(x,t)$ is the profile of the free-surface, $d$ is the water depth, $A$ is the amplitude of the wave, $H=2A$ is the height of the wave and $\lambda$ is the wavelength.}
\label{notations_WT}
\end{figure}
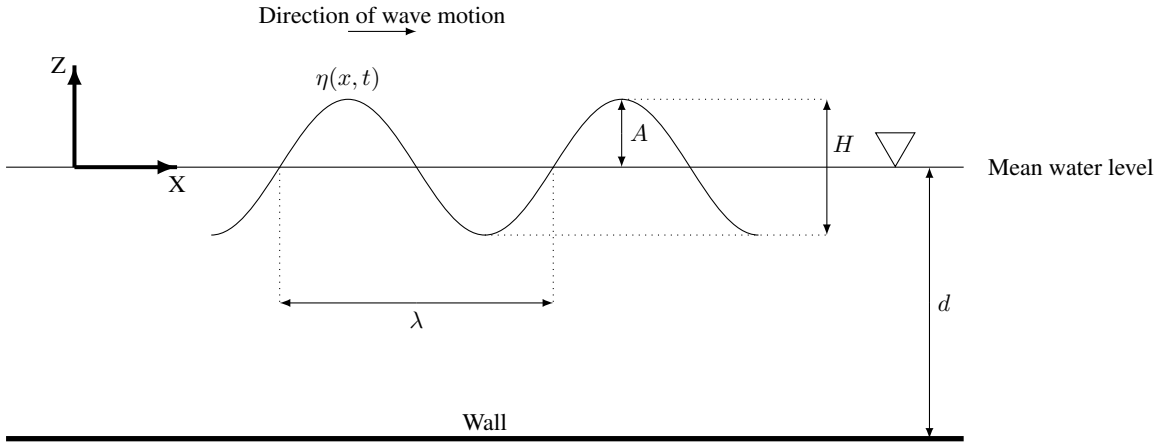

Before we consider the phase-field two-phase solver to wave run-up problem, we first briefly present a background theory of ocean waves. Propagation of free-surface waves is a very complex phenomenon due to the irregular nature of the waves and nonlinear effects. It is very challenging to develop an extensive mathematical formulation to predict this phenomenon. The simplest model is the linear or first-order wave theory where it is assumed that the fluid is incompressible, inviscid and irrotational, density is uniform throughout the fluid, waves are planar and monochromatic. Nonlinear boundary conditions at the free-surface are linearized to further simplify the model \cite{Sorensen, Chakrabarti}. A more accurate model involving the nonlinearities of the wave phenomenon is the second-order Stokes wave theory developed in \cite{Stokes}. The notations used in the description of the wave theory are shown in Fig. \ref{notations_WT}. Apart from the notations shown, $T$ is the time period of the wave, $\omega = 2\pi/T$ is the angular frequency and $k=2\pi/\lambda$ is the wave number. We summarize the results obtained by the second-order Stokes wave theory. The free-surface profile of the wave is given as
\begin{align} \label{interface_profile}
	\eta(x,t) = \frac{H}{2}\mathrm{cos}(kx-\omega t) + \frac{\pi H^2}{8\lambda}\frac{(\mathrm{cosh}(kd))(2+\mathrm{cosh}(2kd))}{\mathrm{sinh}^3(kd)} \mathrm{cos}[2(kx-\omega t)].
\end{align}
The horizontal and vertical components of the velocity to generate the above profile are given as
\begin{align}
	u(x,z,t) &= \frac{Hgk}{2\omega}\frac{\mathrm{cosh}[k(d+z)]}{\mathrm{cosh}(kd)}\mathrm{cos}(kx-\omega t) + \frac{3H^2\omega k}{16}\frac{\mathrm{cosh}[2k(d+z)]}{\mathrm{sinh}^4(kd)}\mathrm{cos}[2(kx-\omega t)], \label{u_eqn}\\
	w(x,z,t) &= \frac{Hgk}{2\omega}\frac{\mathrm{sinh}[k(d+z)]}{\mathrm{cosh}(kd)}\mathrm{sin}(kx-\omega t) + \frac{3H^2\omega k}{16}\frac{\mathrm{sinh}[2k(d+z)]}{\mathrm{sinh}^4(kd)}\mathrm{sin}[2(kx-\omega t)]. \label{w_eqn}
\end{align}

With the description of the second-order Stokes waves, we set up a numerical wave tank to demonstrate the wave run-up problem across a vertical truncated cylinder. The vertical truncated cylinder is one 
of the most common structural members in many offshore structures, e.g., gravity-based 
structures, semi-submersibles and tension-leg platforms. 
The wave run-up on a submerged structure is measured by the run-up ratio, $R/A$ defined as the ratio of the highest free-surface elevation of the fluid (water in this case) at the front face of the cylinder to the amplitude of the incident incoming wave. The run-up depends on two crucial non-dimensional parameters: the wave steepness, $kA$ and the wave scattering parameter, $ka$, which are given by
\begin{align}
	kA = \frac{2\pi A}{\lambda},\\
	ka = \frac{2\pi a}{\lambda},
\end{align} 
where $A$ is the amplitude of the incident wave, $a$ is half of the cross-sectional width of the submerged structure (radius of the cylinder in this case) and $\lambda$ is the wavelength of the incoming wave. If the steepness of the incoming wave is increased ($kA$ is increased) or if the cross-sectional width of the structure is increased ($ka$ is increased) leading to more resistance to the flow, the run-up ratio increases. The quantification of the run-up ratio with $kA$ and $ka$ is of a particular interest in ocean engineering. Some of the experimental and numerical works dealing with the run-up ratio of different cross-sectional structures are discussed in \cite{Morris_thesis, Thiagarajan}.

\begin{figure}[H]
\centering
\begin{tikzpicture}[very thick,decoration={markings,mark=at position 0.5 with {\arrow{>}}},,every node/.style={scale=1.0},scale=1.0]
	\draw[fill=black!10] (0,3.5) -- (1,4.5) -- (13,4.5) -- (12,3.5) -- (0,3.5);
	\draw[fill=black!10] (0,0) -- (0,3.5) -- (12,3.5) -- (13,4.5) -- (13,1) -- (12,0);
	\draw (0,0) node[left]{} -- (0,6) node[right]{} -- (12,6) node[above]{} -- (12,0) node[above]{} -- cycle;
	\draw[black,dotted] (1,1) to (1,7);
	\draw[black] (1,7) to (13,7);
	\draw[black] (13,7) to (13,1);
	\draw[black,dotted] (13,1) to (1,1);

	\draw[black] (0,6) to (1,7);
	\draw[black] (12,6) to (13,7);
	\draw[black] (12,0) to (13,1);
	\draw[black,dotted] (0,0) to (1,1);

	\draw[->](-2.5,4) to (-0.5,4);
	\draw (-1.5,4) node[anchor=north]{$\mathrm{u}=u(x,z,t)$};
	\draw (-1.5,3.5) node[anchor=north]{$\mathrm{v}=0$};
	\draw (-1.5,3) node[anchor=north]{$\mathrm{w}=w(x,z,t)$};

	\draw (4,6) node[anchor=south]{$\sigma_\mathrm{xx}=0,$};
	\draw (5.7,6) node[anchor=south]{$\sigma_\mathrm{zx}=0,$};
	\draw (7.4,6) node[anchor=south]{$\sigma_\mathrm{yx}=0,$};
	\draw (8.8,6) node[anchor=south]{$p=0$};

	\draw (13.2,1) to (13.6,1);
	\draw (13.2,4.5) to (13.6,4.5);
	\draw[<->] (13.4,1) to (13.4,4.5);
	\draw (13.4,2.75) node[anchor=west]{$d$};

	\draw (3.1,5) to (3.3,5);
	\draw (3.1,3) to (3.3,3);
	\draw[<->] (3.2,5) to (3.2,3);
	\draw (3.2,4) node[anchor=west]{$h_c$};

	\draw (2.5,5) -- (2.5,4);
	\draw[dotted] (2.5,4) -- (2.5,3);
	\draw[dotted] (3,3) -- (3,4);
	\draw (3,4) -- (3,5);
	\draw (2.75,5) ellipse (0.26cm and 0.13cm);
	\draw[dotted] (2.75,3) ellipse (0.26cm and 0.13cm);
	\draw (2.75,4) ellipse (0.26cm and 0.13cm);
	\draw[->] (2,2.7) to (2.5,2.7);
	\draw[->] (3.5,2.7) to (3,2.7);
	\draw (2.5,2.8) -- (2.5,2.5);
	\draw (3,2.8) -- (3,2.5);
	\draw (2.75,2.6) node[]{$2a$};

	\draw (6,3) node[anchor=north]{$\Omega_1$};
	\draw (6,2) node[anchor=south]{$(\rho_1, \mu_1)$};

	\draw (6,5.7) node[anchor=north]{$\Omega_2$};
	\draw (6,4.7) node[anchor=south]{$(\rho_2, \mu_2)$};

	\draw (0,-0.2) to (0,-1.2);;
	\draw (12,-0.2) to (12,-1.2);
	\draw[<->] (0,-1) to (12,-1);
	\draw (6,-1) node[anchor=north]{$L$};
	\draw[<->] (0,-0.4) to (2.75,-0.4);
	\draw (2.75,-0.2) to (2.75,-0.6);
	\draw (1.375,-0.4) node[anchor=north]{$L_c$};
	\draw (12.2,0) to (12.6,0);
	\draw[<->] (12.4,0) to (13.4,1);
	\draw (12.9,0.4) node[anchor=west]{$W$};

	\draw[->,red] (0.5,4) to (1.5,4);
	\draw (1.5,4) node[anchor=west,red]{X};
	\draw[->,red] (0.5,4) to (0.5,5);
	\draw (0.5,5) node[anchor=west,red]{Z};
	\draw[->,red] (0.5,4) to (1.2,4.7);
	\draw (1.2,4.7) node[anchor=west,red]{Y};
\end{tikzpicture}
\caption{A schematic of the wave-structure problem. The computational setup and boundary conditions are shown for the Navier-Stokes equations. Here, $\boldsymbol{u} = (\mathrm{u},\mathrm{v},\mathrm{w})$ denotes the components of the fluid velocity which are given by Eqs.~(\ref{u_eqn}) and (\ref{w_eqn}) with $\mathrm{v}=0$. Stress-free boundary condition with zero pressure is given at the top surface of the wave tank. Slip boundary condition is imposed at the X-Z plane at $y=-1$ and $y=1$. All other boundaries have the no-slip condition. Moreover, zero Neumann boundary condition is imposed for the order parameter $\phi$ on all the boundaries.}
\label{RU}
\end{figure}
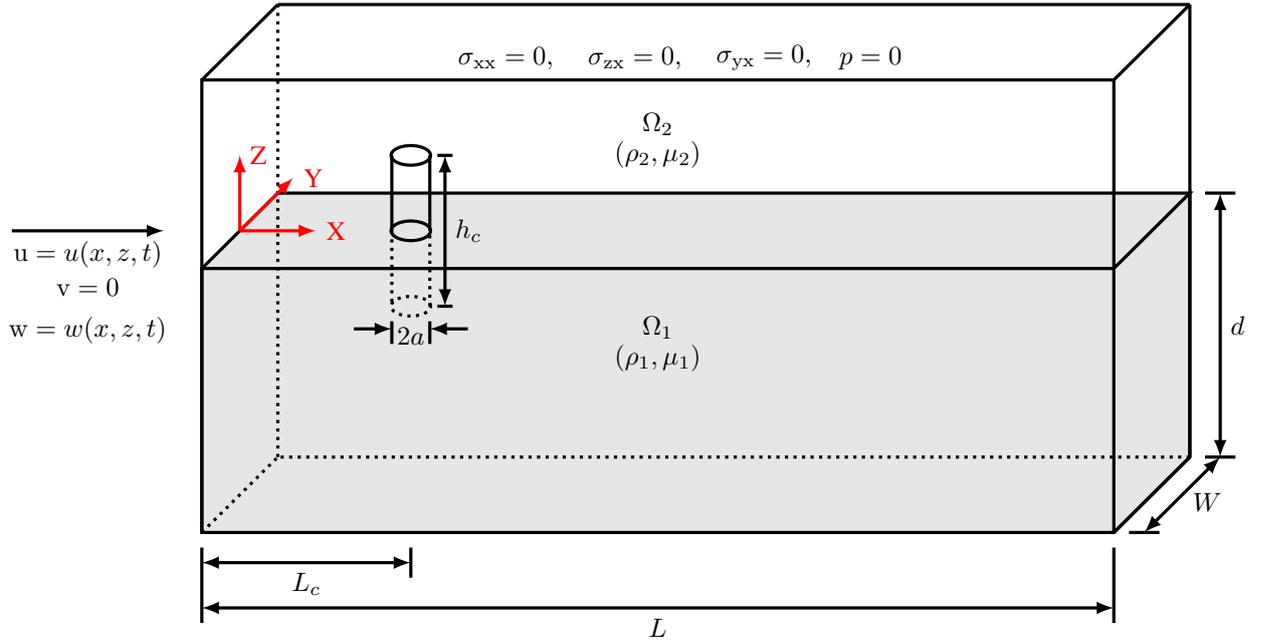

For the present demonstration, we consider a truncated circular cylinder as the structure being impinged by the incoming waves. We keep the parameter $ka$ constant and quantify the run-up on the cylinder by varying $kA$ values. The computational setup is similar to that employed in \cite{Thiagarajan} and is depicted as a schematic in Fig. \ref{RU}. The size of the computational domain is $24$m$\times$$2$m$\times$$2$m, i.e., $L=24$m and $W=2$m. The depth of the water is $d=1.2$m and the draft of the submerged cylinder is $0.4$m. The diameter of the cylinder is $2a=0.2$m and its total height is $h_c = 0.8$m. The centre of the cylinder is at a distance of $L_c=3.6$m from the left boundary. The boundary conditions employed in the simulation are as follows: velocity inlet according to Eqs.~(\ref{u_eqn}) and (\ref{w_eqn}) to simulate Stokes waves at the left boundary, stress-free boundary condition with zero pressure at the top boundary and slip boundary condition at the two sides at $y=-1$m and $y=1$m. All other boundaries (bottom and right) have no-slip boundary condition. The order parameter $\phi$ is initialized with the interface at $z=0$m with $\Omega_1$ phase depicting water ($\rho_1=1000, \mu_1=1.002\times 10^{-3}$) and $\Omega_2$ representing air ($\rho_2=1.225, \mu_2=1.983\times 10^{-5}$). The interface thickness parameter for the Allen-Cahn solver is taken as $\varepsilon=0.02$. Initial numerical tests suggested that employing a much sharper interface by taking smaller $\varepsilon$ does not affect the solution for this large scale problem. The time-step size is taken as $\Delta t=0.0025$ with the total number of time steps as $5000$.  The capillary effects due to surface tension have been neglected for the modeling of the free-surface waves.

Based on the experimental campaign in \cite{Morris_thesis}, we select the $ka$ value of $0.208$ which gives a wavelength of $\lambda=3.0208$m. For the Stokes waves, the time period is a function of wavelength as $\lambda = \frac{gT^2}{2\pi}\mathrm{tanh}\big(\frac{2\pi d}{\lambda}\big)$, which gives $T = 1.40045$s. We vary the $kA$ values by changing the amplitude of the wave and quantify the run-up ratio $R/A$ for the truncated cylinder.

\begin{figure}[H]
\centering
	\begin{subfigure}[b]{0.5\textwidth}
		\includegraphics[trim={0.2cm 0.2cm 0.2cm 0.2cm},clip,width=7.5cm]{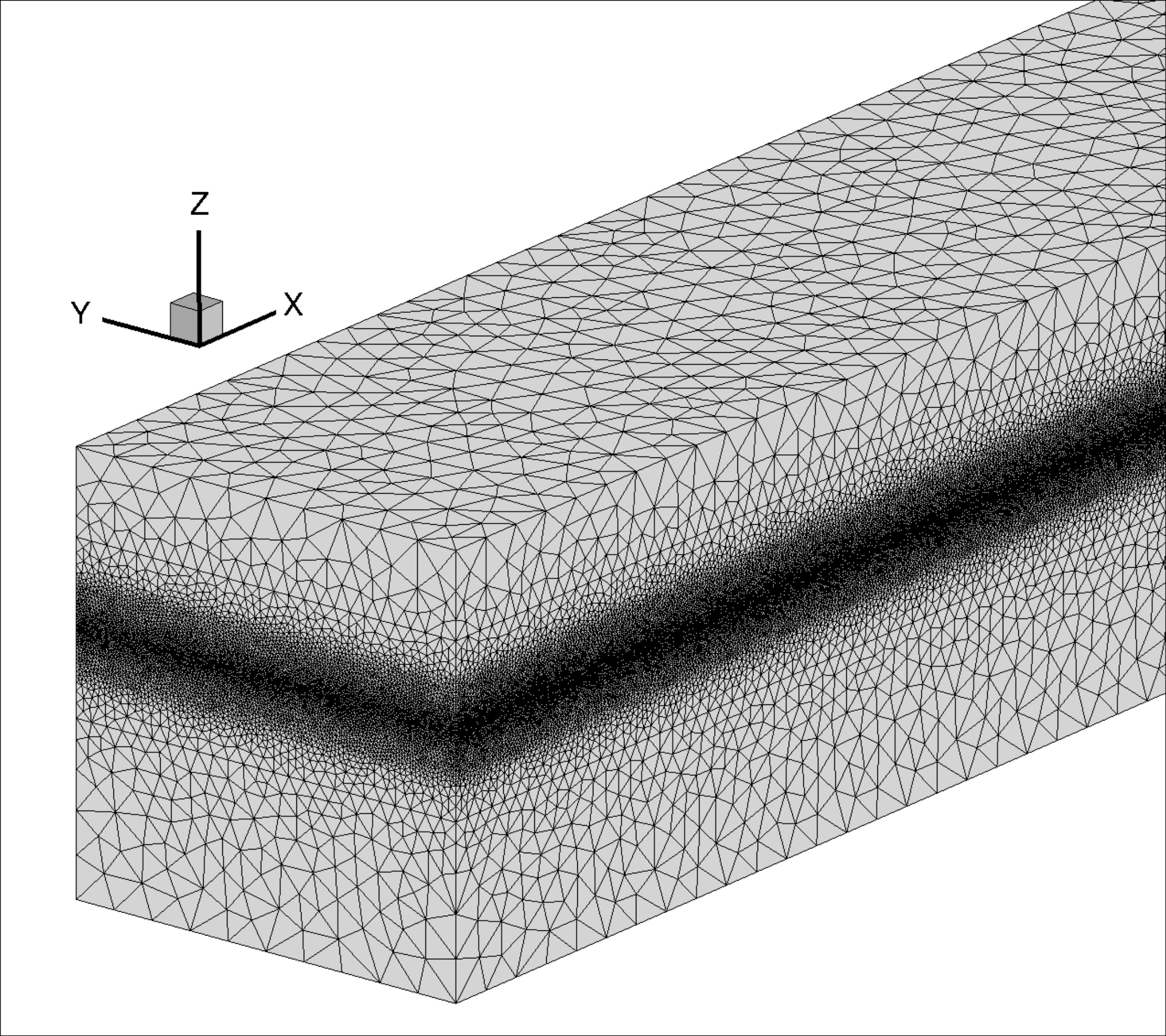}
	\caption{}
	\end{subfigure}%
	\begin{subfigure}[b]{0.5\textwidth}
		\includegraphics[trim={0.2cm 0.2cm 0.2cm 0.2cm},clip,width=8cm]{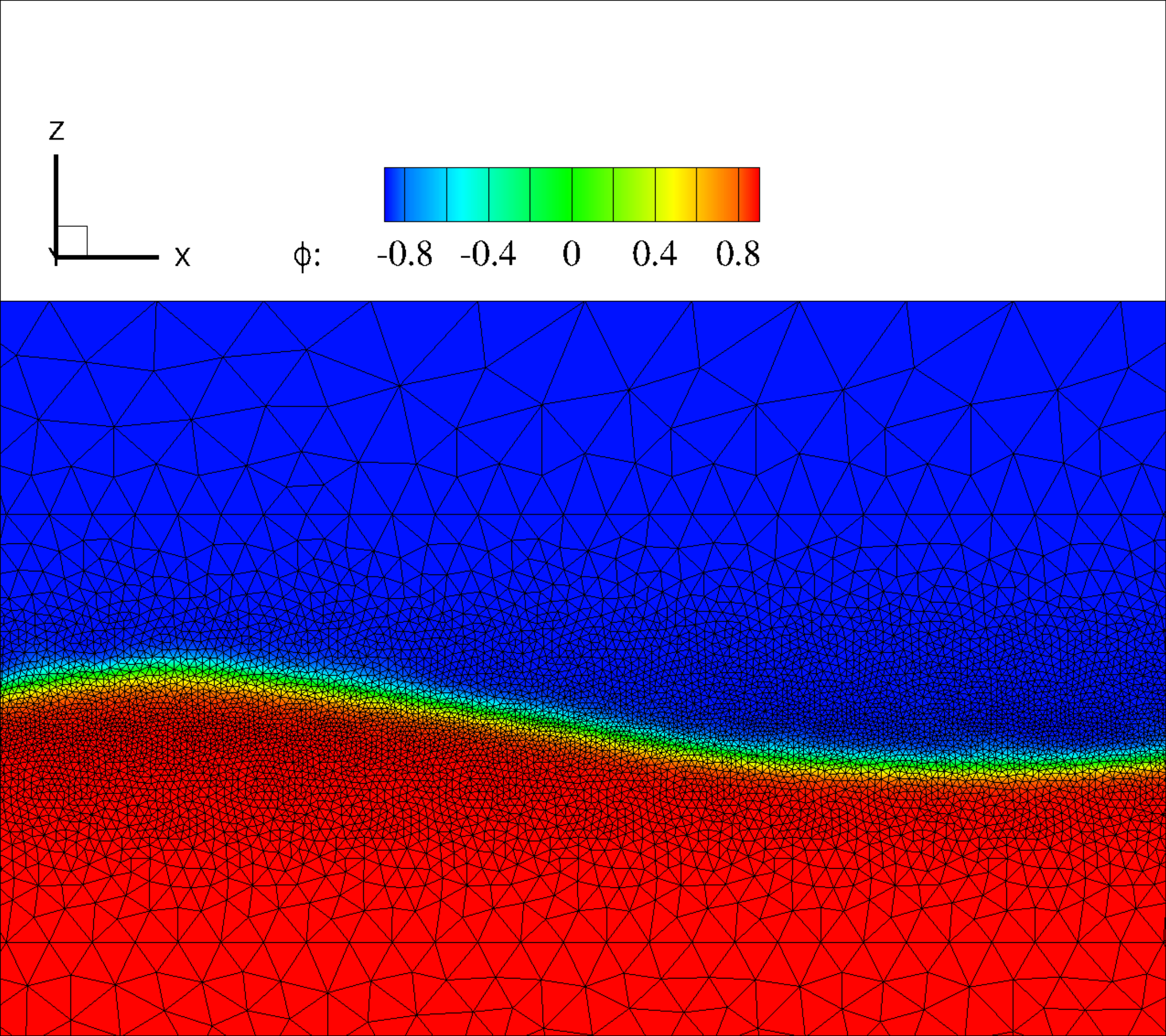}
	\caption{}
	\end{subfigure}
\caption{Three-dimensional computational mesh and contours of order parameter for the run-up problem: (a) refined interfacial region with unstructured finite element mesh, (b) contour plot of order parameter $\phi$ superimposed with the unstructured mesh.} 
\label{Mesh_1}
\end{figure}

An unstructured finite element mesh is constructed for the numerical wave tank 
via open-source mesh generator \cite{gmsh}. A gradually coarsening mesh from mesh size $\delta=0.01$ at $z=0$ to $\delta =0.02$ is used to capture the interface region from $z=-0.2$ to $z=0.2$. This resolution also ensures that at least $4$ number of elements fall under the equilibrium interfacial region to capture the interface properties accurately. Furthermore, the resolution of the mesh is increased to a value of $\delta=0.01$ enveloping the cylinder. The total number of grid nodes in the mesh are around $5.6$ million with approximately $34$ million tetrahedral unstructured elements. The mesh is depicted in Fig. \ref{Mesh_1}. The simulations are performed using 600 CPU cores with MPI parallelism and the total time taken to complete the $5000$ time steps is approximately $6.5$ hours with approximately $5$ seconds to complete a minimum of $3$ nonlinear iterations in a time step.

\begin{figure}[H]
\centering
	\begin{subfigure}[b]{0.5\textwidth}
		\centering
		\includegraphics[trim={0.2cm 4cm 0.2cm 0.2cm},clip,width=7cm]{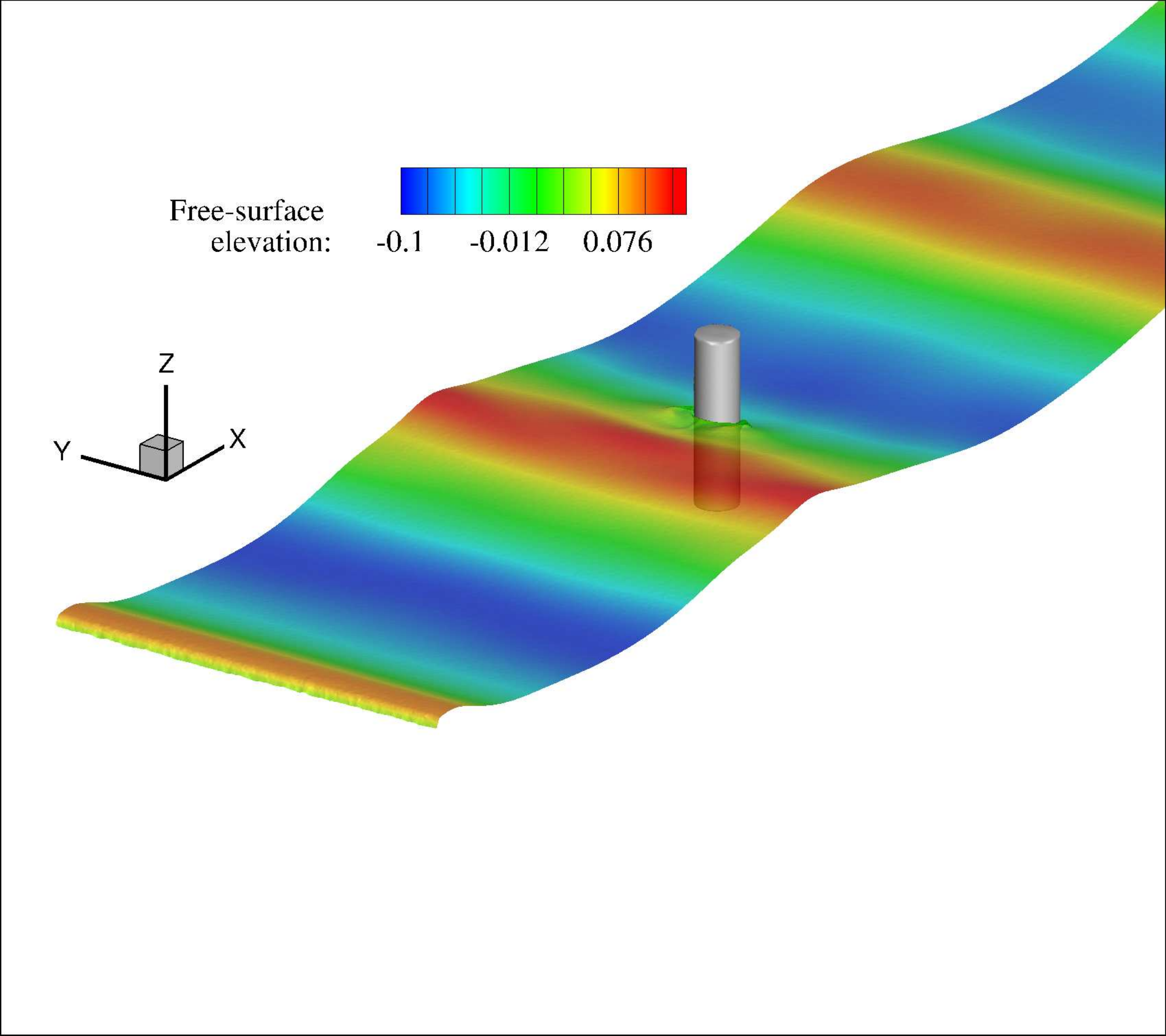}
	\caption{}
	\end{subfigure}%
	\begin{subfigure}[b]{0.5\textwidth}
		\centering
		\includegraphics[trim={0.2cm 4cm 0.2cm 0.2cm},clip,width=7cm]{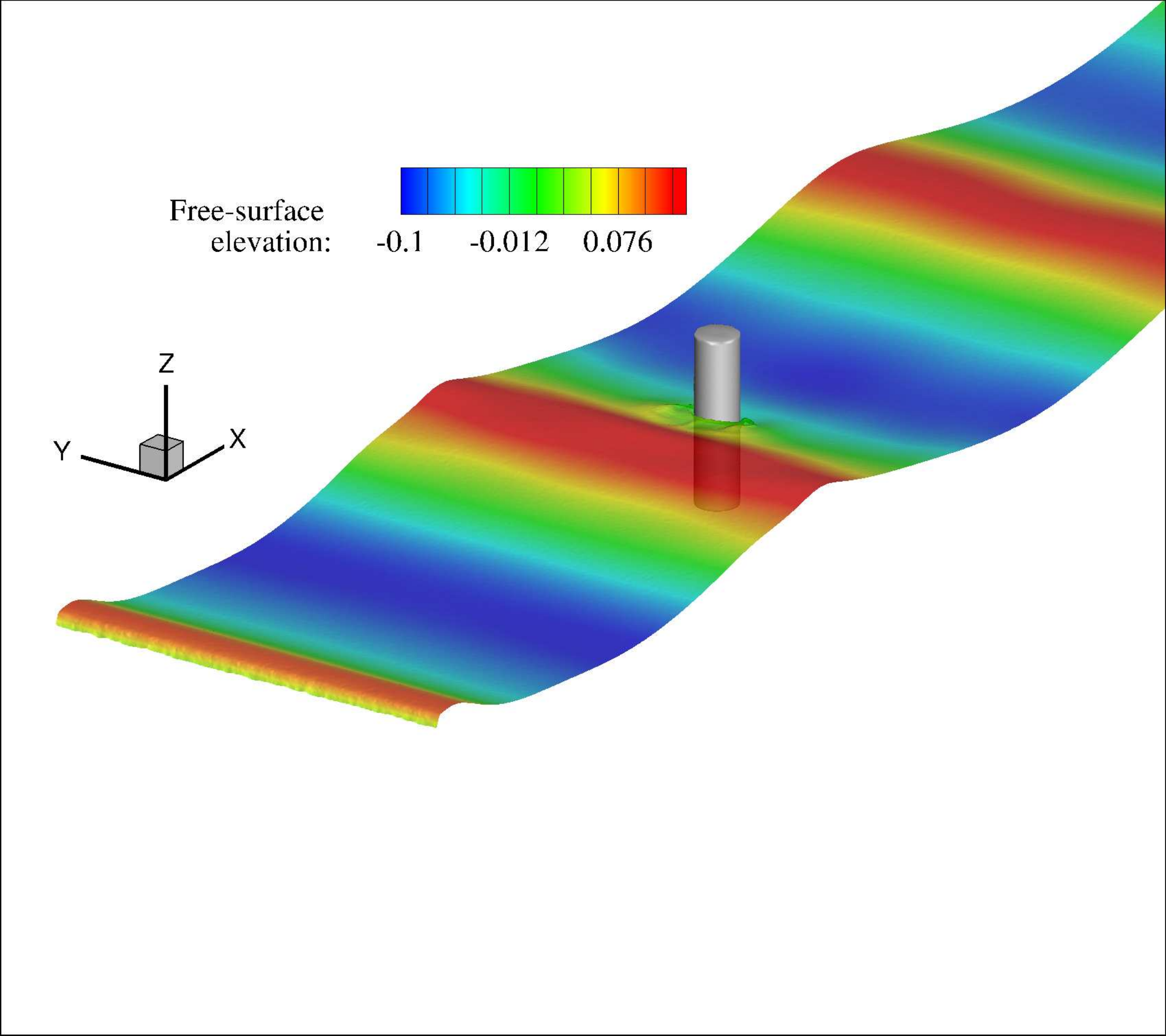}
	\caption{}
	\end{subfigure}
\caption{Water wave run-up on a truncated cylinder: Snapshots of iso-contours $\phi=0$ colored by the free-surface elevation for two representative wave steepness parameters: 
(a) $kA=0.1837$ and (b) $0.2146$ at $t=12.5$.} 
\label{Plt_2}
\end{figure}

\begin{figure}[H]
\centering
	\begin{subfigure}[b]{0.5\textwidth}
		\centering
		\includegraphics[trim={10cm 0.2cm 10cm 0.2cm},clip,width=7cm]{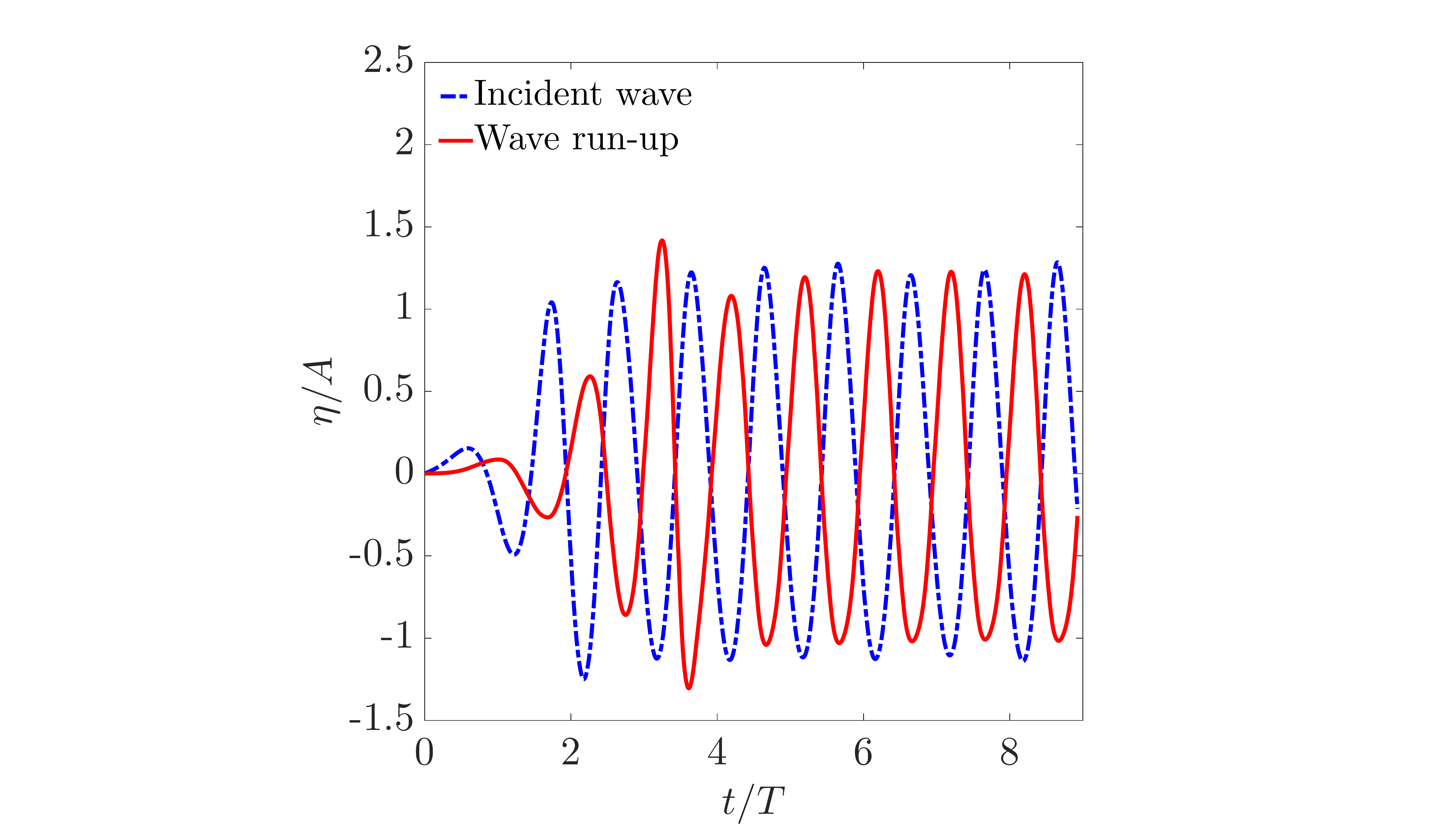}
	\caption{}
	\end{subfigure}%
	\begin{subfigure}[b]{0.5\textwidth}
		\centering
		\includegraphics[trim={10cm 0.2cm 10cm 0.2cm},clip,width=7cm]{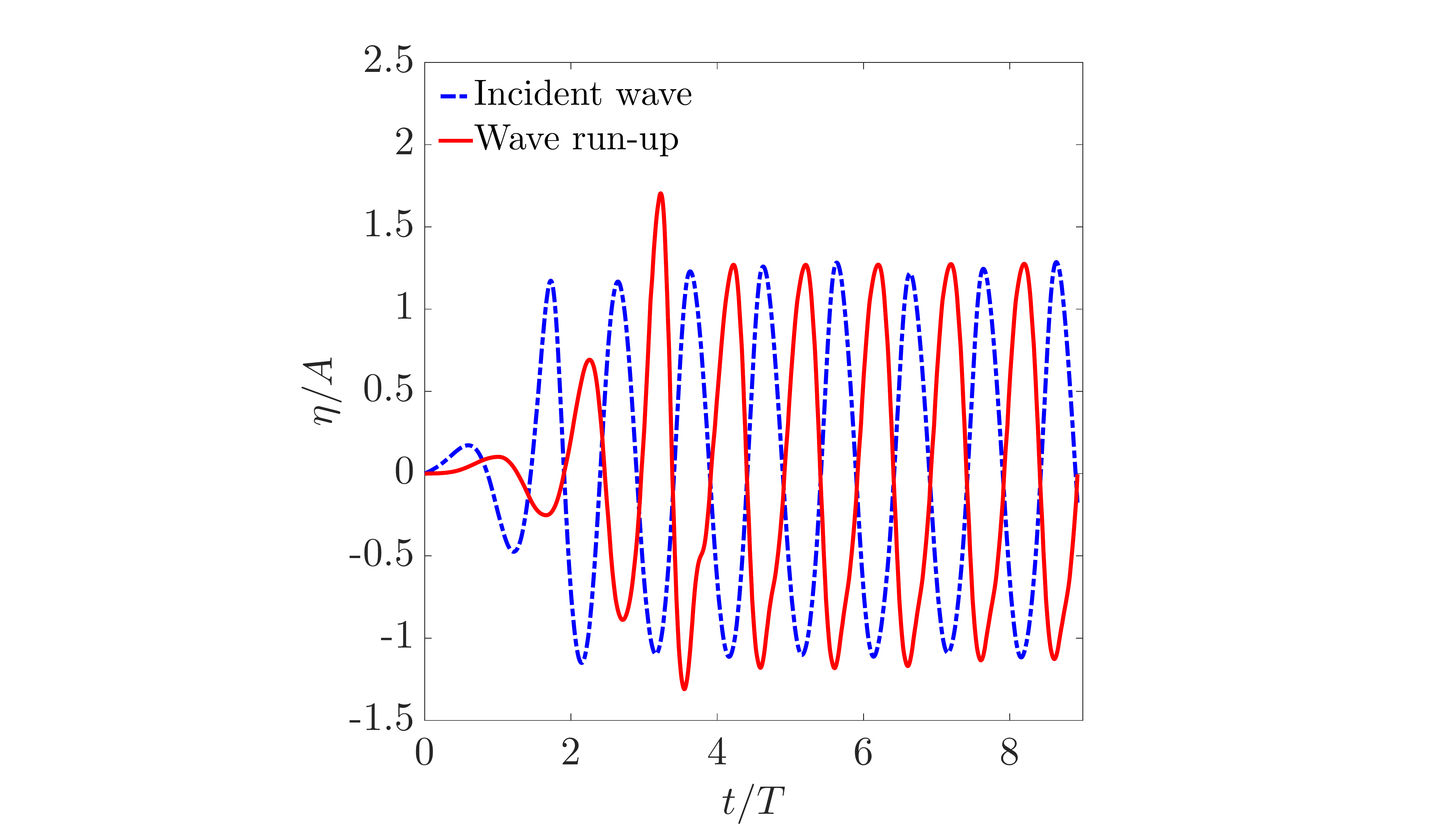}
	\caption{}
	\end{subfigure}

	\begin{subfigure}[b]{0.5\textwidth}
		\centering
		\includegraphics[trim={10cm 0.2cm 10cm 0.2cm},clip,width=7cm]{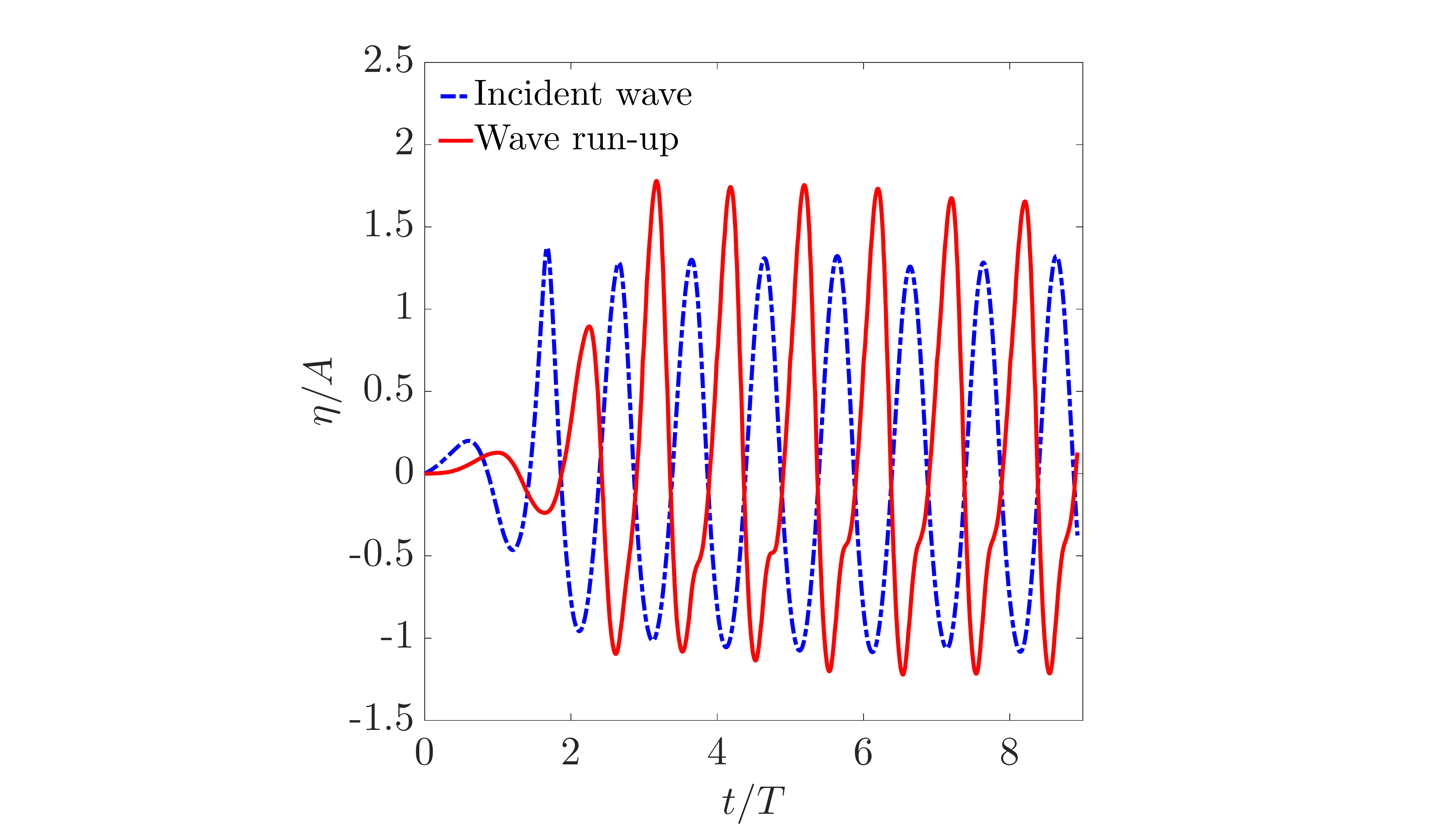}
	\caption{}
	\end{subfigure}%
	\begin{subfigure}[b]{0.5\textwidth}
		\centering
		\includegraphics[trim={10cm 0.2cm 10cm 0.2cm},clip,width=7cm]{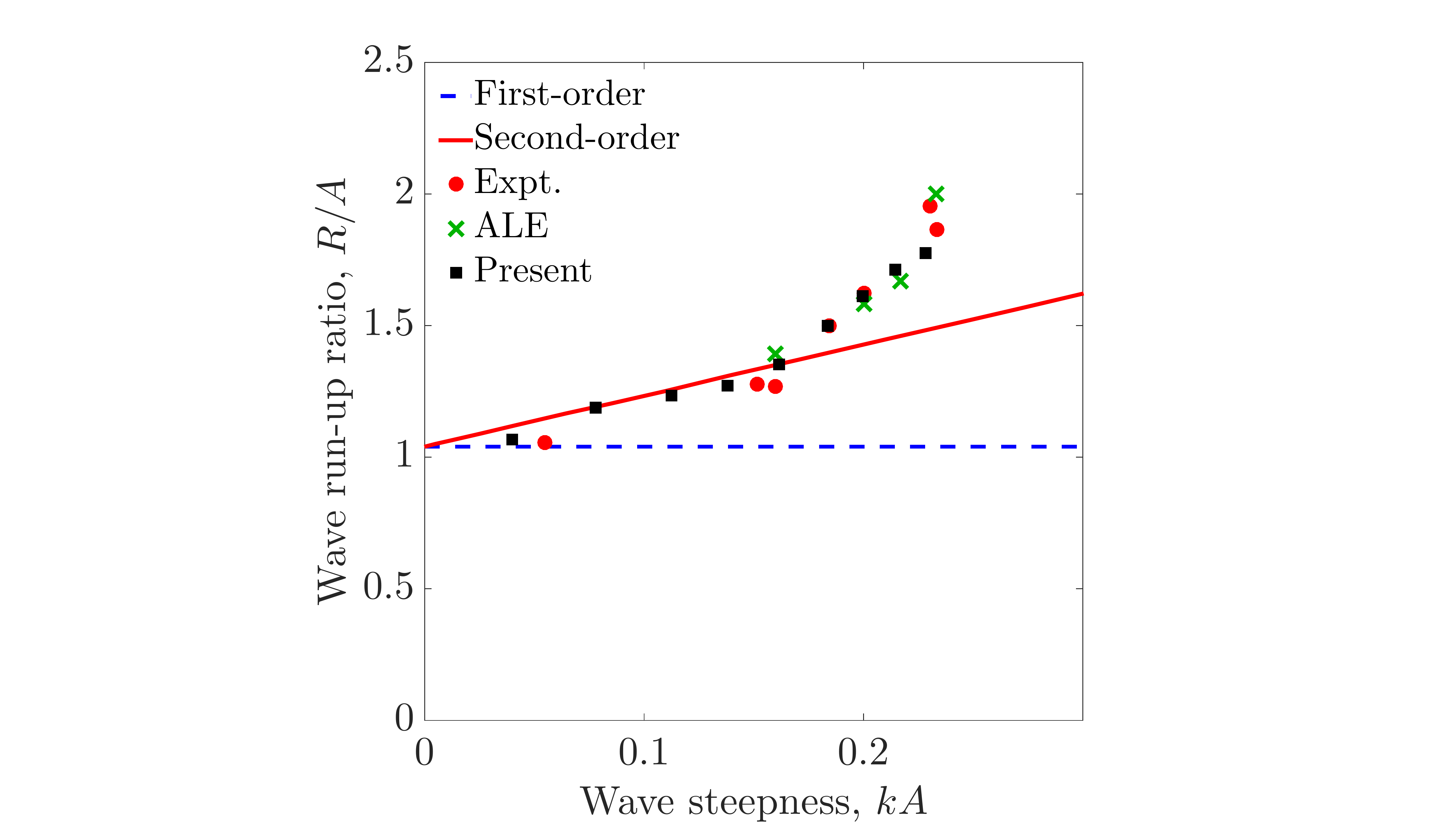}
	\caption{}
	\end{subfigure}
\caption{Water wave run-up on a truncated cylinder: Time histories of the free-surface elevations 
at the probe locations for three representative wave steepness parameters: 
(a) $kA=0.0779$, (b) $0.1381$, (c) $0.2146$, and (d) comparison of the wave run-up ratio with 
the results from the literature at $ka=0.208$: The first- and second-order results are 
from frequency domain potential theory, the experimental results are from the 
campaign in \cite{Morris_thesis} and the ALE results are from the free-surface 
ALE solver in \cite{OMAE_RKJ_NWT}.  
Time and free-surface elevation are non-dimensionalized by time period $T$ and wave amplitude $A$ respectively.} 
\label{Plt_1}
\end{figure}

The free-surface wave elevation is recorded in time at two locations- one at $x=2$m and the other at $x=3.5$m (the front face of the cylinder). The wave elevation is found out by linearly interpolating the order parameter and finding the interface where $\phi=0$. The wave amplitude is calculated as the amplitude of the first harmonic of the free-surface elevation of the incident wave at the first probe point at $x=2$m by taking the Fourier transform of the elevation $\eta$ as $A = A(\omega_1)$, where $\omega_1$ is the first harmonic frequency of the incident wave. The wave run-up on the front-face of the cylinder at $x=3.5$m is calculated as the mean of the maximum amplitude from each wave cycle as $R = \frac{1}{M} \displaystyle\sum_\mathrm{n=1}^M \eta_\mathrm{n}$,
where $\eta_\mathrm{n}$ is the peak amplitude of the free-surface elevation $\eta$ at $x=3.5$m of the $\mathrm{n}$th wave cycle and $M$ is the total number of wave cycles which is equal to $5$ in this case. The wave run-up ratio is evaluated as the ratio of the wave run-up $R$ on the front face of the cylinder and the incident wave amplitude $A$. All the post-processing is performed in the time interval $t/T \in [4,9]$ to exclude any initial transient solutions. The data is post-processed in a similar manner as the experimental campaign \cite{Morris_thesis}. We simulate cases for a range of $kA \in [0.04, 0.2282]$. The iso-contours of the free-surface at $\phi=0$ colored by the free-surface elevation are plotted in Fig. \ref{Plt_2} for two different $kA$ values at $t=12.5$. The time history of the incident wave and run-up for different $kA$ values is plotted in Fig. \ref{Plt_1}(a-c). We observe a secondary kink in the wave run-up at high $kA$ values in the figures similar to the findings in the literature. The wave run-up ratio is compared with the results from the literature in Fig. \ref{Plt_1}(d). Our results are in good agreement with the theoretical and experimental predictions. 
\vspace{-0.5cm}

\section{Conclusions}
\label{conclusion}
In this work, a novel positivity preserving variational method has been developed for the accurate, general and robust phase-field modeling of incompressible two-phase flows involving high density ratios. A one-fluid formulation has been utilized under the diffuse-interface description to represent the interface between the two immiscible phases. The interface is evolved by solving the Allen-Cahn equation under the variational framework which minimizes the free energy functional. The mass conservation and energy stability properties are imparted to the underlying scheme in a relatively simple manner. While the mass is conserved by a Lagrange multiplier term, the discrete energy stability is established by considering the mid-point approximation of the spatial part of the Lagrange multiplier. It is found that the mid-point approximation helps in the energy-stability property of the underlying scheme. We want to emphasize the advantage of the partitioned staggered single-pass coupling which reduces the computational time while providing accuracy and robustness in the solution. Standalone convergence tests of the Allen-Cahn phase-field solver resulted into a second-order accuracy in space and more than first-order for temporal discretization. The positivity preserving property of the scheme is demonstrated by considering the sloshing tank problem. In contrast to the oscillatory solutions from linear stabilized methods, the numerical solution is found to be bounded if we include the PPV-based nonlinear stabilization terms. We also conclude that about 4-5 elements in the equilibrium interfacial region are sufficient to obtain the accurate solution of two-phase flows. Furthermore, a selection of smaller interface thickness parameter $\varepsilon$ increases the accuracy of the solution but this advantage comes with higher computational cost. Tests on the dam break problem demonstrate the advantage of the presented interface capturing technique on the cases with the breaking and merging of the interface. Finally, the ability of the solver to handle a practical problem of wave-structure interaction is demonstrated by analyzing the water run-up on a vertical truncated circular cylinder.
Application of the present formulation to the fluid-structure interaction of floating bodies is a topic of future study.

{\bf{Acknowledgements}}

The first author would like to thank for the financial support from  National Research Foundation through Keppel-NUS Corporate Laboratory. The conclusions put forward reflect the views of the authors alone, and not necessarily those of the institutions.

\appendix
\setcounter{equation}{0} 
\setcounter{figure}{0}

\section{Discrete energy stability proof for the Allen-Cahn equation}
\label{DEL}
Here, we derive the statement of discrete energy stability for our variational phase-field formulation based on the convective Allen-Cahn equation presented in Section \ref{proposed method}. 
We show that our proposed formulation is provably energy stable. We will take one particular case when $\alpha_\mathrm{m}=\gamma = 0.5$. To prove the energy stability, we choose the following approximation
\begin{align}
	\phi^\mathrm{n+\alpha} &= \frac{\phi^\mathrm{n+1} + \phi^\mathrm{n}}{2},\\
	\partial_t\phi^\mathrm{n+\alpha_m} &= \frac{\phi^\mathrm{n+1} - \phi^\mathrm{n}}{\Delta t}.
\end{align}
For simplicity, we assume that $\boldsymbol{u}=\boldsymbol{0}$. The discrete variational formulation is now given as
\begin{align} \label{energy_stable_var}
	&\int_\Omega \bigg( w_\mathrm{h}\partial_t{\phi}_\mathrm{h} + \nabla w_\mathrm{h}\cdot(k\nabla\phi_\mathrm{h} ) + w_\mathrm{h}s\phi_\mathrm{h} - w_\mathrm{h}f \bigg) \mathrm{d}\Omega + \sum_\mathrm{e=1}^\mathrm{n_{el}} \int_{\Omega^\mathrm{e}}\chi \frac{|\mathcal{R}(\phi_\mathrm{h})|}{|\nabla \phi_\mathrm{h}|} k^\mathrm{add}_{c} \nabla w_\mathrm{h} \cdot\mathbf{I} \cdot \nabla\phi_\mathrm{h} \mathrm{d}\Omega = 0.	
\end{align}
Now, substituting $w_\mathrm{h} = \phi^\mathrm{n+1} - \phi^\mathrm{n}$, 
each term of Eq.~(\ref{energy_stable_var})  can be written as follows:
\begin{align}
&\mathrm{Inertia\ term:}\int_\Omega w_\mathrm{h}\partial_t\phi_\mathrm{h} \mathrm{d}\Omega = \int_\Omega (\phi^\mathrm{n+1} - \phi^\mathrm{n})\frac{(\phi^\mathrm{n+1} - \phi^\mathrm{n})}{\Delta t} \mathrm{d}\Omega = \frac{1}{\Delta t} ||\phi^\mathrm{n+1} - \phi^\mathrm{n}||^2_0, \\
&\mathrm{Diffusion\ term:}\int_\Omega \nabla w_\mathrm{h}\cdot (k\nabla\phi_\mathrm{h}) \mathrm{d}\Omega = \int_\Omega \varepsilon^2 \nabla(\phi^\mathrm{n+1} - \phi^\mathrm{n})\cdot \nabla\bigg(\frac{\phi^\mathrm{n+1} + \phi^\mathrm{n}}{2}\bigg) \mathrm{d}\Omega = \frac{\varepsilon^2}{2}||\nabla\phi^\mathrm{n+1}||^2_0 - \frac{\varepsilon^2}{2}||\nabla\phi^\mathrm{n}||^2_0, \\
&\mathrm{Reaction/source\ terms:}\int_\Omega w_\mathrm{h} \big( s \phi_\mathrm{h} - f \big) \mathrm{d}\Omega = \int_\Omega (\phi^\mathrm{n+1} - \phi^\mathrm{n} ) F'(\phi) \mathrm{d}\Omega - \beta(t) \int_\Omega (\phi^\mathrm{n+1} - \phi^\mathrm{n})\sqrt{F(\phi)}\mathrm{d}\Omega\nonumber \\
	&\qquad\qquad\qquad\qquad\qquad= \int_\Omega \big( F(\phi^\mathrm{n+1}) - F(\phi^\mathrm{n}) \big) \mathrm{d}\Omega - \beta(t) \int_\Omega \big( K(\phi^\mathrm{n+1}) - K(\phi^\mathrm{n}) \big) \mathrm{d}\Omega \label{B_6}\nonumber \\
	&\qquad\qquad\qquad\qquad\qquad= \int_\Omega \big( F(\phi^\mathrm{n+1}) - F(\phi^\mathrm{n}) \big) \mathrm{d}\Omega,\\
&\mathrm{PPV\ stabilization\ term:}\sum_\mathrm{e=1}^\mathrm{n_{el}} \int_{\Omega^\mathrm{e}}\chi \frac{|\mathcal{R}(\phi_\mathrm{h})|}{|\nabla \phi_\mathrm{h}|} k^\mathrm{add}_{c} \nabla w_\mathrm{h} \cdot\nabla\phi_\mathrm{h} \mathrm{d}\Omega \nonumber \\
&\qquad\qquad\qquad\qquad\qquad = \int_\Omega \chi \frac{|\mathcal{R}(\phi_\mathrm{h})|}{|\nabla \phi_\mathrm{h}|} k^\mathrm{add}_{c} \nabla (\phi^\mathrm{n+1} - \phi^\mathrm{n} )\cdot\nabla\bigg(\frac{\phi^\mathrm{n+1} + \phi^\mathrm{n}}{2}\bigg) \mathrm{d}\Omega\nonumber \\
&\qquad\qquad\qquad\qquad\qquad=  \frac{\chi}{2} \frac{|\mathcal{R}(\phi_\mathrm{h})|}{|\nabla \phi_\mathrm{h}|} k^\mathrm{add}_{c} ||\nabla\phi^\mathrm{n+1}||^2_0 - \frac{\chi}{2} \frac{|\mathcal{R}(\phi_\mathrm{h})|}{|\nabla \phi_\mathrm{h}|} k^\mathrm{add}_{c} ||\nabla\phi^\mathrm{n}||^2_0.
\end{align}
The second line of Eq.~(\ref{B_6}) is obtained by substituting the corresponding discretizations of $F'(\phi)$ and $\sqrt{F(\phi)}$ which are expressed in Eqs.~(\ref{F_phi}) and (\ref{K_phi}) respectively. Equation~(\ref{K_cons}) further helps to simplify the expression in Eq.~(\ref{B_6}). By substituting the expressions in Eq.~(\ref{energy_stable_var}) and after some algebraic arrangements, we get the following discrete energy statement: 
\begin{align}
	\mathcal{E}(\phi^\mathrm{n+1}) - \mathcal{E}(\phi^\mathrm{n}) = - \frac{1}{\Delta t}||\phi^\mathrm{n+1} - \phi^\mathrm{n}||^2_0 \leq 0,
\end{align}
where $\mathcal{E}(\phi)$ is the modified energy functional given as
\begin{align}
	\mathcal{E}(\phi) = \int_\Omega \bigg( \frac{1}{2} (\varepsilon^2 + \chi \frac{|\mathcal{R}(\phi_\mathrm{h})|}{|\nabla \phi_\mathrm{h}|} k^\mathrm{add}_{c}) |\nabla\phi|^2 + F(\phi) \bigg) \mathrm{d}\Omega.
\end{align}
Hence, the presented scheme is unconditionally energy stable.  In other words, the discrete Allen-Cahn phase-field equation possesses the energy functional that is bounded for 
all time, with the bound depending only on the problem data. Similar proof can be derived for the $\theta$-family of time integration methods (which are specific cases of generalized-$\alpha$ method) with the condition of unconditional energy stability as $\theta\ge 0.5$.

\section{Discrete mass conservation}
\label{DMC}
In this section, we prove the mass conservation property of the discrete Allen-Cahn equation presented in Eq.~(\ref{PPV_AC}). Consider $w_\mathrm{h}=1$ in Eq.~(\ref{PPV_AC}). With the aid of boundary conditions in Eq.~(\ref{CAC}) and the incompressibility condition, we have
\begin{align}
	\int_\Omega \boldsymbol{u}\cdot\nabla\phi_\mathrm{h} \mathrm{d}\Omega = \int_\Gamma (\boldsymbol{u}\phi_\mathrm{h})\cdot \mathbf{n} \mathrm{d}\Gamma - \int_\Omega \phi_\mathrm{h}(\nabla\cdot\boldsymbol{u})\mathrm{d}\Omega = 0.
\end{align} 
Since $\nabla w_\mathrm{h} = 0$, Eq.~(\ref{PPV_AC}) can be rewritten as
\begin{align}
	&\int_\Omega \bigg( \partial_t\phi_\mathrm{h} + s\phi_\mathrm{h} \bigg) \mathrm{d}\Omega = 0,\\
\implies &\int_\Omega \bigg( \partial_t\phi_\mathrm{h} + F'(\phi) - \beta(t)\sqrt{F(\phi)} \bigg) \mathrm{d}\Omega = 0,\\
\implies &\int_\Omega \partial_t\phi_\mathrm{h} \mathrm{d}\Omega + \int_\Omega F'(\phi) \mathrm{d}\Omega - \beta(t)\int_\Omega \sqrt{F(\phi)}\mathrm{d}\Omega = 0.
\end{align}
Using the definition of $\beta(t)$ in Eq.~(\ref{eqn:beta}), we obtain
\begin{align}
	&\int_\Omega \partial_t\phi_\mathrm{h} \mathrm{d}\Omega + \int_\Omega F'(\phi) \mathrm{d}\Omega - \int_\Omega F'(\phi) \mathrm{d}\Omega = 0.
\end{align}
Since $\partial_t\phi_\mathrm{h}$ is evaluated at $\mathrm{n+\alpha_m}$, we can write
\begin{align}
&\int_\Omega \partial_t\phi^\mathrm{n+\alpha_m} \mathrm{d}\Omega = 0,
\end{align}
which can be expressed as follows using Eqs.~(\ref{app_1}) and (\ref{app_2}),
\begin{align}
&\int_\Omega \partial_t\phi^\mathrm{n} - \frac{\alpha_\mathrm{m}}{\gamma}\partial_t\phi^\mathrm{n} + \frac{\alpha_\mathrm{m}}{\gamma\Delta t}(\phi^\mathrm{n+1} - \phi^\mathrm{n} ) \mathrm{d}\Omega = 0,
\end{align}
which implies that the mass will only be conserved when $\alpha_\mathrm{m} = \gamma$. 
This completes the proof of discrete mass conservation.
All the simulations performed in the present study consider $\alpha_\mathrm{m} = \gamma = \alpha = 0.5$.

\section*{References}
\bibliographystyle{unsrt}
\bibliography{citations}

\begin{thebibliography}{10}

\bibitem{Khatavkar}
V.~V. Khatavkar, P.~D. Anderson, P.~C. Duineveld, and H.~H.~E. Meijer.
\newblock Diffuse interface modeling of droplet impact on a pre-patterned solid
  surface.
\newblock {\em Macromolecular Rapid Communications}, 26(4):298--303, 2005.

\bibitem{Fried}
E.~Fried and M.~E. Gurtin.
\newblock Dynamic solid-solid transitions with phase characterized by an order
  parameter.
\newblock {\em Physica D: Nonlinear Phenomena}, 72(4):287 -- 308, 1994.

\bibitem{Sorensen}
R.M. Sorensen.
\newblock {\em Basic Wave Mechanics: {F}or Coastal and Ocean Engineers}.
\newblock A Wiley-Interscience publication. Wiley, 1993.

\bibitem{Chakrabarti}
S.K. Chakrabarti.
\newblock {\em Hydrodynamics of Offshore Structures}.
\newblock Springer Verlag, 1987.

\bibitem{Multiphase_Tryg}
A.~Prosperetti and G.~Tryggvason.
\newblock {\em Computational Methods for Multiphase Flow}.
\newblock Cambridge University Press, 2007.

\bibitem{Unverdi}
S.~O. Unverdi and G.~Tryggvason.
\newblock A front-tracking method for viscous, incompressible, multi-fluid
  flows.
\newblock {\em Journal of Computational Physics}, 100(1):25 -- 37, 1992.

\bibitem{donea}
J.~Donea.
\newblock Arbitrary {L}agrangian-{E}ulerian finite element methods.
\newblock 192:4195--4215, 1983.

\bibitem{Brackbill}
J.~U. Brackbill, D.~B. Kothe, and C.~Zemach.
\newblock A continuum method for modeling surface tension.
\newblock {\em Journal of Computational Physics}, 100(2):335 -- 354, 1992.

\bibitem{Scardovelli}
R.~Scardovelli and S.~Zaleski.
\newblock Direct numerical simulation of free-surface and interfacial flow.
\newblock {\em Annual Review of Fluid Mechanics}, 31(1):567--603, 1999.

\bibitem{Rider}
W.~J. Rider and D.~B. Kothe.
\newblock Reconstructing volume tracking.
\newblock {\em Journal of Computational Physics}, 141(2):112 -- 152, 1998.

\bibitem{Sethian}
J.~A. Sethian and P.~Smereka.
\newblock Level set methods for fluid interfaces.
\newblock {\em Annual Review of Fluid Mechanics}, 35:341, 2003.
\newblock Copyright - Copyright Annual Reviews, Inc. 2003; Last updated -
  2014-05-18.

\bibitem{Lanhao}
L.~Zhao, X.~Bai, T.~Li, and J.~J.~R. Williams.
\newblock Improved conservative level set method.
\newblock {\em International Journal for Numerical Methods in Fluids},
  75(8):575--590, 2014.
\newblock FLD-13-0333.R1.

\bibitem{Sussman_1}
M.~Sussman and E.~Fatemi.
\newblock An efficient, interface-preserving level set redistancing algorithm
  and its application to interfacial incompressible fluid flow.
\newblock {\em SIAM Journal on Scientific Computing}, 20(4):1165--27, 1999.
\newblock Copyright - Copyright] © 1999 Society for Industrial and Applied
  Mathematics; Last updated - 2012-02-15.

\bibitem{Peng}
D.~Peng, B.~Merriman, S.~Osher, H.~Zhao, and M.~Kang.
\newblock A {PDE}-based fast local level set method.
\newblock {\em Journal of Computational Physics}, 155(2):410 -- 438, 1999.

\bibitem{Russo}
G.~Russo and P.~Smereka.
\newblock A remark on computing distance functions.
\newblock {\em Journal of Computational Physics}, 163(1):51 -- 67, 2000.

\bibitem{Enright_1}
D.~Enright, R.~Fedkiw, J.~Ferziger, and I.~Mitchell.
\newblock A hybrid particle level set method for improved interface capturing.
\newblock {\em Journal of Computational Physics}, 183(1):83 -- 116, 2002.

\bibitem{Enright_2}
D.~Enright, F.~Losasso, and R.~Fedkiw.
\newblock A fast and accurate semi-{L}agrangian particle level set method.
\newblock {\em Comput. Struct.}, 83(6-7):479--490, February 2005.

\bibitem{Koh}
C.~G. Koh, M.~Gao, and C.~Luo.
\newblock A new particle method for simulation of incompressible free surface
  flow problems.
\newblock {\em International Journal for Numerical Methods in Engineering},
  89(12):1582--1604, 2012.

\bibitem{Sussman_2}
M.~Sussman and E.~G. Puckett.
\newblock A coupled level set and volume-of-fluid method for computing {3D} and
  axisymmetric incompressible two-phase flows.
\newblock {\em Journal of Computational Physics}, 162(2):301 -- 337, 2000.

\bibitem{Wang}
Z.~Wang, J.~Yang, B.~Koo, and F.~Stern.
\newblock A coupled level set and volume-of-fluid method for sharp interface
  simulation of plunging breaking waves.
\newblock {\em International Journal of Multiphase Flow}, 35(3):227 -- 246,
  2009.

\bibitem{Yang}
X.~Yang, J.~J. Feng, C.~Liu, and J.~Shen.
\newblock Numerical simulations of jet pinching-off and drop formation using an
  energetic variational phase-field method.
\newblock {\em Journal of Computational Physics}, 218(1):417 -- 428, 2006.

\bibitem{Olsson_1}
E.~Olsson and G.~Kreiss.
\newblock A conservative level set method for two phase flow.
\newblock {\em Journal of Computational Physics}, 210(1):225 -- 246, 2005.

\bibitem{Olsson_2}
E.~Olsson, G.~Kreiss, and S.~Zahedi.
\newblock A conservative level set method for two phase flow {II}.
\newblock {\em Journal of Computational Physics}, 225(1):785 -- 807, 2007.

\bibitem{Fries_1}
H.~Sauerland and T-P. Fries.
\newblock The extended finite element method for two-phase and free-surface
  flows: {A} systematic study.
\newblock {\em Journal of Computational Physics}, 230(9):3369 -- 3390, 2011.

\bibitem{Anderson}
D.~M. Anderson, G.~B. McFadden, and A.~A. Wheeler.
\newblock Diffuse-interface methods in fluid mechanics.
\newblock {\em Annual Review of Fluid Mechanics}, 30(1):139--165, 1998.

\bibitem{Anderson_1}
D.~M. Anderson and G.~B. McFadden.
\newblock A diffuse-interface description of fluid systems.
\newblock In {\em NIST IR 5887 (National Institute of Standards and
  Technology)}, 1996.

\bibitem{Allen_cahn}
S.~M. Allen and J.~W. Cahn.
\newblock A microscopic theory for antiphase boundary motion and its
  application to antiphase domain coarsening.
\newblock {\em Acta Metallurgica}, 27(6):1085 -- 1095, 1979.

\bibitem{Cahn_Hilliard}
J.~W. Cahn and J.~E. Hilliard.
\newblock Free energy of a nonuniform system. {I}. {I}nterfacial free energy.
\newblock {\em The Journal of Chemical Physics}, 28(2), 1958.

\bibitem{Rubinstein}
J.~Rubinstein and P.~Sternberg.
\newblock Nonlocal reaction-diffusion equations and nucleation.
\newblock {\em IMA Journal of Applied Mathematics}, 48(3):249--264, 1992.

\bibitem{Bronsard}
L.~Bronsard and B.~Stoth.
\newblock Volume-preserving mean curvature flow as a limit of a nonlocal
  {G}inzburg-{L}andau equation.
\newblock {\em SIAM Journal on Mathematical Analysis}, 28(4):769--39, 07 1997.
\newblock Copyright - Copyright] © 1997 Society for Industrial and Applied
  Mathematics; Last updated - 2012-02-29.

\bibitem{Brassel}
M.~Brassel and E.~Bretin.
\newblock A modified phase field approximation for mean curvature flow with
  conservation of the volume.
\newblock {\em Mathematical Methods in the Applied Sciences},
  34(10):1157--1180, 2011.

\bibitem{Kim_2}
J.~Kim, S.~Lee, and Y.~Choi.
\newblock A conservative {A}llen-{C}ahn equation with a space–time dependent
  {L}agrange multiplier.
\newblock {\em International Journal of Engineering Science}, 84:11 -- 17,
  2014.

\bibitem{H_G_Lee}
Hyun~Geun Lee.
\newblock High-order and mass conservative methods for the conservative
  {A}llen-{C}ahn equation.
\newblock {\em Computers \& Mathematics with Applications}, 72(3):620 -- 631,
  2016.

\bibitem{Tierra}
G.~Tierra and F.~Guill$\mathrm{\acute{e}}$n-Gonz$\mathrm{\acute{a}}$lez.
\newblock Numerical methods for solving the {C}ahn-{H}illiard equation and its
  applicability to related energy-based models.
\newblock {\em Archives of Computational Methods in Engineering},
  22(2):269--289, 04 2015.
\newblock Copyright - CIMNE, Barcelona, Spain 2015; Last updated - 2015-04-04.

\bibitem{Dongsun}
D.~Lee and J.~Kim.
\newblock Comparison study of the conservative {A}llen-{C}ahn and the
  {C}ahn-{H}illiard equations.
\newblock {\em Mathematics and Computers in Simulation}, 119:35 -- 56, 2016.

\bibitem{Jeong}
D.~Jeong, S.~Lee, D.~Lee, J.~Shin, and J.~Kim.
\newblock Comparison study of numerical methods for solving the {A}llen-{C}ahn
  equation.
\newblock {\em Computational Materials Science}, 111:131 -- 136, 2016.

\bibitem{Kim}
J.~Kim.
\newblock Phase-field models for multi-component fluid flows.
\newblock {\em Communications in Computational Physics}, 12(3):613--661, 009
  2012.

\bibitem{Gomez}
H.~Gomez and T.~J.~R. Hughes.
\newblock Provably unconditionally stable, second-order time-accurate, mixed
  variational methods for phase-field models.
\newblock {\em Journal of Computational Physics}, 230(13):5310 -- 5327, 2011.

\bibitem{Zhang}
Z.~Zhang and H.~Tang.
\newblock An adaptive phase field method for the mixture of two incompressible
  fluids.
\newblock {\em Computers \& Fluids}, 36(8):1307 -- 1318, 2007.

\bibitem{Vasconcelos}
D.~F.~M. Vasconcelos, A.~L. Rossa, and A.~L. G.~A. Coutinho.
\newblock A residual-based {A}llen-{C}ahn phase field model for the mixture of
  incompressible fluid flows.
\newblock {\em International Journal for Numerical Methods in Fluids},
  75(9):645--667, 2014.

\bibitem{PPV}
V.~Joshi and R.~K. Jaiman.
\newblock A positivity preserving variational method for multi-dimensional
  convection-diffusion-reaction equation.
\newblock {\em Journal of Computational Physics}, 339:247 -- 284, 2017.

\bibitem{VTI}
A.~Lew, J.~E. Marsden, M.~Ortiz, and M.~West.
\newblock Variational time integrators.
\newblock {\em International Journal for Numerical Methods in Engineering},
  60(1):153--212, 2004.

\bibitem{Gen_alpha}
J.~Chung and G.~M. Hulbert.
\newblock A time integration algorithm for structural dynamics with improved
  numerical dissipation: {T}he generalized-$\alpha$ method.
\newblock {\em Journal of Applied Mechanics}, 60(2):371--375, 1993.

\bibitem{Jansen}
K.~E. Jansen, C.~H. Whiting, and G.~M. Hulbert.
\newblock A generalized-$\alpha$ method for integrating the filtered
  {N}avier-{S}tokes equations with a stabilized finite element method.
\newblock {\em Computer Methods in Applied Mechanics and Engineering},
  190(3–4):305 -- 319, 2000.

\bibitem{Du}
Q.~Du and R.~A. Nicolaides.
\newblock Numerical analysis of a continuum model of phase transition.
\newblock {\em SIAM Journal on Numerical Analysis}, 28(5):1310--1322, 1991.

\bibitem{Hughes_X}
F.~Shakib, T.~J.~R. Hughes, and Z.~Johan.
\newblock A new finite element formulation for computational fluid dynamics:
  {X}. {T}he compressible {E}uler and {N}avier-{S}tokes equations.
\newblock {\em Computer Methods in Applied Mechanics and Engineering},
  89:141--219, 1991.

\bibitem{Akkerman}
I.~Akkerman, Y.~Bazilevs, D.~J. Benson, M.~W. Farthing, and C.~E. Kees.
\newblock Free-surface flow and fluid-object interaction modeling with emphasis
  on ship hydrodynamics.
\newblock {\em Journal of Applied Mechanics}, 79(1):10905, 2012.

\bibitem{Bazilevs_book}
Y.~Bazilevs, K.~Takizawa, and T.~E. Tezduyar.
\newblock {\em Computational {F}luid-{S}tructure {I}nteraction: {M}ethods and
  {A}pplications}.
\newblock Wiley, 2013.

\bibitem{Kim_3}
J.~Kim.
\newblock A continuous surface tension force formulation for diffuse-interface
  models.
\newblock {\em Journal of Computational Physics}, 204(2):784 -- 804, 2005.

\bibitem{Hughes_conserve}
T.~J.~R. Hughes and G.~Wells.
\newblock Conservation properties for the {G}alerkin and stabilised forms of
  the advection-diffusion and incompressible {N}avier-{S}tokes equations.
\newblock {\em Computer Methods in Applied Mechanics and Engineering},
  194:1141--1159, 2005.

\bibitem{Hsu}
M.~Hsu, Y.~Bazilevs, Calo V., T.~Tezduyar, and T.~J.~R. Hughes.
\newblock Improving stability of multiscale formulations of fluid flow at small
  time steps.
\newblock {\em Computer Methods in Applied Mechanics and Engineering},
  199:828--840, 2010.

\bibitem{Brooks}
A.~N. Brooks and T.~J.~R. Hughes.
\newblock Streamline upwind/{P}etrov-{G}alerkin formulations for convection
  dominated flows with particular emphasis on the incompressible
  {N}avier-{S}tokes equations.
\newblock {\em Computer Methods in Applied Mechanics and Engineering}, 32,
  1982.

\bibitem{Tezduyar_1}
T.~E. Tezduyar, S.~Mittal, S.~Ray, and R.~Shih.
\newblock Incompressible flow computations with stabilized bilinear and linear
  equal-order interpolation velocity-pressure elements.
\newblock {\em Computer Methods in Applied Mechanics and Engineering},
  95:221--242, 1992.

\bibitem{France_II}
L.~Franca and S.~Frey.
\newblock Stabilized finite element methods: {II}. {T}he incompressible
  {N}avier-{S}tokes equations.
\newblock {\em Computer Methods in Applied Mechanics and Engineering},
  99:209--233, 1992.

\bibitem{Hughes_inv_est}
I.~Harari and T.~J.~R. Hughes.
\newblock What are {C} and h?: {I}nequalities for the analysis and design of
  finite element methods.
\newblock {\em Computer Methods in Applied Mechanics and Engineering},
  97(2):157--192, 1992.

\bibitem{Johnson}
C.~Johnson.
\newblock {\em Numerical solutions of partial differential equations by the
  finite element method}.
\newblock Cambridge {U}niversity {P}ress, 1987.

\bibitem{Hughes_V}
T.~J.~R. Hughes, L.~P. Franca, and M.~A. Balestra.
\newblock A new finite element formulation for computational fluid dynamics:
  {V}. {C}ircumventing the {B}abuska-{B}rezzi condition: {A} stable
  {P}etrov-{G}alerkin formulation of the {S}tokes problem accommodating
  equal-order interpolations.
\newblock {\em Computer Methods in Applied Mechanics and Engineering},
  59:85--99, 1986.

\bibitem{Tezduyar_stab}
T.~E. Tezduyar.
\newblock Finite elements in fluids: {S}tabilized formulations and moving
  boundaries and interfaces.
\newblock {\em Computers and Fluids}, 36:191--206, 2007.

\bibitem{saad1986}
Y.~Saad and M.~H. Schultz.
\newblock {GMRES}: A generalized minimal residual algorithm for solving
  nonsymmetric linear systems.
\newblock {\em SIAM J. Sci. Stat. Comput.}, 7(3):856--869, July 1986.

\bibitem{mpi}
{MPI} webpage (www.mpi-formum.org).
\newblock Technical report, 2009.

\bibitem{Martin_Moyce}
J.~C. Martin and W.~J. Moyce.
\newblock Part {IV}. {A}n experimental study of the collapse of liquid columns
  on a rigid horizontal plane.
\newblock {\em Philosophical Transactions of the Royal Society of London.
  Series A, Mathematical and Physical Sciences}, 244(882):312--324, 1952.

\bibitem{Ubbink_thesis}
O.~Ubbink.
\newblock {\em Numerical prediction of two fluid systems with sharp
  interfaces}.
\newblock PhD thesis, Imperial College of Science, Technology \& Medicine,
  1997.

\bibitem{Walhorn_thesis}
E.~Walhorn.
\newblock {\em Ein simultanes Berechnungsverfahren f${\ddot{u}}$r
  Fluid-Struktur-Wechselwirkungen mit finiten Raum-Zeit-Elementen}.
\newblock PhD thesis, Technische Universit${\ddot{a}}$t Braunschweig, 2002.

\bibitem{Issa}
R.~Issa and D.~Violeau.
\newblock Test-case 2, {3D} dambreaking, {R}elease 1.1
  (http://app.spheric-sph.org/sites/spheric/tests/test-2).
\newblock Technical report, Electricit$\mathrm{\acute{e}}$ De France,
  Laboratoire National d'Hydraulique et Environnement, 2006.

\bibitem{Elias}
R.~N. Elias and A.~L. G.~A. Coutinho.
\newblock Stabilized edge-based finite element simulation of free-surface
  flows.
\newblock {\em International Journal for Numerical Methods in Fluids},
  54(6-8):965--993, 2007.

\bibitem{Akkerman_LS}
I.~Akkerman, Y.~Bazilevs, C.~E. Kees, and M.~W. Farthing.
\newblock Isogeometric analysis of free-surface flow.
\newblock {\em Journal of Computational Physics}, 230(11):4137 -- 4152, 2011.
\newblock Special issue High Order Methods for CFD Problems.

\bibitem{Kleefsman}
K.~M.~T. Kleefsman, G.~Fekken, A.~E.~P. Veldman, B.~Iwanowski, and B.~Buchner.
\newblock A volume-of-fluid based simulation method for wave impact problems.
\newblock {\em Journal of Computational Physics}, 206(1):363 -- 393, 2005.

\bibitem{Stokes}
George~Gabriel Stokes.
\newblock {\em On the Theory of Oscillatory Waves}, volume~1 of {\em Cambridge
  Library Collection - Mathematics}, pages 197--229.
\newblock Cambridge University Press, 2009.

\bibitem{Morris_thesis}
M.~Morris-Thomas.
\newblock {\em An investigation into wave run-up on vertical surface piercing
  cylinders in monochromatic waves}.
\newblock PhD thesis, The University of Western Australia, 2003.

\bibitem{Thiagarajan}
K.~P. Thiagarajan and N.~Repalle.
\newblock Wave run-up on columns of deepwater platforms.
\newblock {\em Proceedings of the Institution of Mechanical Engineers, Part M:
  Journal of Engineering for the Maritime Environment}, 227(3):256--265, 2013.

\bibitem{gmsh}
C.~Geuzaine and J.~F. Remacle.
\newblock Gmsh: {A} 3-{D} finite element mesh generator with built-in pre- and
  post-processing facilities.
\newblock {\em International Journal for Numerical Methods in Engineering},
  79(11):1309--1331, 2009.

\bibitem{OMAE_RKJ_NWT}
J.~Kim, R.~Jaiman, S.~Cosgrove, and J.~O'Sullivan.
\newblock Numerical wave tank analysis of wave run-up on a truncated vertical
  cylinder.
\newblock In {\em ASME 30th International Conference on Ocean, Offshore and
  Arctic Engineering (OMAE)}, Rotterdam, The Netherlands, 2011.

\end{thebibliography}

\end{document}